\documentclass[a4paper,11pt]{article}
\pdfoutput=1

\usepackage{jcappub}
\usepackage{amsmath}
\usepackage{amssymb}
\usepackage{color}
\usepackage{graphicx}

\usepackage[utf8]{inputenc}

\graphicspath{{figures/}}

\begin{document}

\title{\boldmath Reconstructing the dark sector interaction with LISA}

\author[a,b]{Rong-Gen Cai}
\emailAdd{cairg@itp.ac.cn}

\author[c]{Nicola Tamanini}

\emailAdd{nicola.tamanini@cea.fr}

\author[a,b]{Tao Yang}
\emailAdd{yangtao@itp.ac.cn}

\affiliation[a]{CAS Key Laboratory of Theoretical Physics, Institute of Theoretical Physics, Chinese Academy of Sciences, P.O. Box 2735, Beijing 100190, China}
\affiliation[b]{School of Physical Sciences, University of Chinese Academy of Sciences, No.19A Yuquan Road, Beijing 100049, China}
\affiliation[c]{Institut de Physique Th\'eorique, CEA-Saclay, CNRS UMR 3681, Universit\'e Paris-Saclay, F-91191 Gif-sur-Yvette, France}

\date{\today}

\abstract{
We perform a forecast analysis of the ability of the LISA space-based interferometer to reconstruct the dark sector interaction using gravitational wave standard sirens at high redshift.
We employ Gaussian process methods to reconstruct the distance-redshift relation in a model independent way.
We adopt simulated catalogues of standard sirens given by merging massive black hole binaries visible by LISA, with an electromagnetic counterpart detectable by future telescopes.
The catalogues are constructed considering three different astrophysical scenarios for the evolution of massive black hole mergers based on the semi-analytic model of E.~Barausse, \textit{Mon.~Not.~Roy.~Astron.~Soc.}~{\bf 423} (2012) 2533.
We first use these standard siren datasets to assess the potential of LISA in reconstructing a possible interaction between vacuum dark energy and dark matter.
Then we combine the LISA cosmological data with supernovae data simulated for the Dark Energy Survey.
We consider two scenarios distinguished by the time duration of the LISA mission: 5 and 10 years.
Using only LISA standard siren data, the dark sector interaction can be well reconstructed from redshift $z\sim1$ to $z\sim3$ (for a 5 years mission) and $z\sim1$ up to $z\sim5$ (for a 10 years mission), though the reconstruction is inefficient at lower redshift.
When combined with the DES datasets, the interaction is well reconstructed in the whole redshift region from $z\sim0$ to $z\sim3$ (5 yr) and $z\sim0$ to $z\sim5$ (10 yr), respectively.
Massive black hole binary standard sirens can thus be used to constrain the dark sector interaction at redshift ranges not reachable by usual supernovae datasets which probe only the $z\lesssim 1.5$ range.
Gravitational wave standard sirens will not only constitute a complementary and alternative way, with respect to familiar electromagnetic observations, to probe the cosmic expansion, but will also provide new tests to constrain possible deviations from the standard $\Lambda$CDM dynamics, especially at high redshift.
}

\maketitle

\section{Introduction} 
\label{sec:introduction}
Current experiments suggest that our Universe is presently undergoing an accelerated phase of expansion~\cite{Riess:1998cb,Perlmutter:1998np}, driven by a mysterious entity named dark energy (DE).
The simplest model which fits the current observational data is called $\Lambda$CDM: within this scenario the dynamics of the Universe is described by the combination of a simple cosmological constant, playing the role of DE, and a cold dark matter (DM) component.
In spite of its observational success, the $\Lambda$CDM model is however faced with some unsolved theoretical issues: specifically the fine-tuning problem and the cosmic coincidence problem \cite{Weinberg:1988cp,Martin:2012bt}.
To alleviate these problems a plethora of alternative solutions has been proposed in the literature, where DE is identified as a dynamical component arising either from new matter degrees of freedom or from modifications of the standard cosmological equations.
Among all these models, the sub-class admitting an interaction between DE and DM is particularly suited for dealing with the cosmic coincidence problem \cite{Wetterich:1994bg,Amendola:1999er}, although at the cosmological perturbation and fully covariant level some issues must still be properly addressed \cite{Li:2014eha,Faraoni:2014vra,Tamanini:2015iia,Skordis:2015yra,Valiviita:2008iv,He:2008si}.
Mild indications of a late-time interaction in the dark sector are present in some current observational datasets \cite{Salvatelli:2014zta,Cai:2015zoa}; see also \cite{Wang:2016lxa} and references therein.
All these datasets are built using information collected through observations of electromagnetic (EM) radiation: for example type Ia supernova (SNIa), baryon acoustic oscillations (BAO) or the temperature and polarization anisotropies of the cosmic microwave background (CMB) radiation.
There is however a new observational window that one can now exploit to obtain independent astronomical information: gravitational waves (GWs).

On 11 February 2016, the Laser Interferometer Gravitational Wave Observatory (LIGO) collaboration reported the first direct detection of the gravitational wave source GW150914~\cite{Abbott:2016blz}, marking the beginning of the GW astronomy era.
GW experiments can be used not only to obtain astrophysical information on the emitting sources, but also to probe cosmology.
In 1986, Schutz showed that it is possible to determine the Hubble constant from GW observations, by using the fact that GWs from stellar binary systems encode distance information~\cite{Schutz:1986gp}.
Thus the inspiralling and merging compact binaries consisting of neutron stars (NSs) and black holes (BHs), can be considered as standard distance rulers, named standard sirens in analogy to standard candles, namely SNIa.
In the last decades, several papers have studied the possibility of using GW as standard sirens~\cite{Holz:2005df,Dalal:2006qt,MacLeod:2007jd,Nissanke:2009kt,Taylor:2011fs,DelPozzo:2011yh,DelPozzo:2015bna}, especially in connection with future GW detectors such as the Einstein Telescope (ET)~\cite{Sathyaprakash:2009xt,Zhao:2010sz,Cai:2016sby}, a third-generation ground-based interferometers, and the Laser Interferometric Space Antenna (LISA)~\cite{Petiteau:2011we,Tamanini:2016zlh,Kyutoku:2016zxn,Tamanini:2016uin,DelPozzo:2017kme}, a space mission designed to probe the GW landscape around the mHz region where the signal produced by massive black hole binaries (MBHBs) is expected to be the loudest~\cite{elisaweb}.
In the most recent configuration for LISA, conceived to answer the call from ESA for its L3 mission~\cite{2017arXiv170200786A}, the observatory will be based on three arms with six active laser links, between three identical spacecraft in a triangular formation separated by 2.5 million km.
The proposed nominal mission duration in science mode is of 4 years, but the mission should be designed with consumables and orbital stability to facilitate a total mission duration of up to 10 years.

The aim of the present work is to reconstruct the dark sector interaction using simulated MBHB standard siren events with LISA.
The simulated datasets of MBHB standard sirens used in what follows are taken from \cite{Tamanini:2016zlh}, where realistic catalogues of GW events detected by LISA and for which a EM counterpart can be identified by future telescopes were constructed.
The statistical analysis is instead based on the Gaussian process method~\cite{Holsclaw:2010nb,Holsclaw:2010sk,Holsclaw:2011wi,Seikel:2012uu} which allows for a time, or better redshift, reconstruction of the interaction without assuming any {\it a priori} parametrization.
A similar study has been performed in~\cite{Cai:2015zoa} where supernova Union 2.1 data and data simulated for the Dark Energy Survey (DES) have been employed to reconstruct the interaction between dark energy and dark matter in a model-independent fashion.
However SNIa data can cover the redshift range up to $z \sim 2$ at most, while the MBHB standard siren datasets constructed with LISA contain events that could be detected up to $z\sim 10$.
It is thus expected that GW information can extend the reconstructed dark interaction to a much larger redshift.
To show this result in what follows the interaction between DE and DM will first be reconstructed using LISA MBHB standard siren data alone, and then combining GW information with the simulated data from DES.

The outline of the paper is as follows.
In Sec.~\ref{sec:massive_black_hole_binary_mergers_as_standard_sirens_for_lisa}, the basics of MBHB mergers as standard sirens for LISA are discussed.
In Sec.~\ref{sec:the_dark_sector_interaction} the cosmological models where an interaction between DE and DM is presented, and in Sec.~\ref{sec:the_reconstruction_method} the reconstruction method is introduced.
All results are exposed in Sec.~\ref{sec:results}, while discussions and conclusions are considered in Sec.~\ref{sec:discussion_and_conclusion}.


\section{Massive black hole binary mergers as standard sirens for LISA} 
\label{sec:massive_black_hole_binary_mergers_as_standard_sirens_for_lisa}

In this section we briefly introduce the concept of standard siren and summarise the method of \cite{Tamanini:2016zlh} employed to simulate observations of MBHB standard siren events with LISA.
The data produced in \cite{Tamanini:2016zlh} have been used to forecast constraints on cosmological parameters of both standard and alternative models \cite{Tamanini:2016zlh,Caprini:2016qxs,Tamanini:2016uin}, and will be used in Sec.~\ref{sec:results} to reconstruct the interaction between DM and DE.
Here we will not present the details regarding the production of such data, referring the interested reader to \cite{Tamanini:2016zlh} for the explanation of the whole process.

Standard sirens are GW sources which can be used as accurate cosmological distance indicators \cite{Schutz:1986gp,Holz:2005df}.
They are the GW equivalence of EM standard candles, namely type-Ia supernovae (SNIa).
Compact binary inspirals are excellent examples of standard sirens, irrespective of the masses of the inspiralling objects.
In fact the waveform $h$ produced by these systems during the inspiral phase, is theoretically well described by the analytical solution
\begin{align}
	h_\times = \frac{4}{d_L(z)} \left( \frac{G \mathcal{M}_c(z)}{c^2} \right)^{5/3} \left( \frac{\pi f}{c} \right)^{2/3} \cos\iota \sin[\Phi(f)] \,,
	\label{eq:gwf}
\end{align}
which is valid at the lowest (Newtonian) order for the ``cross'' GW polarization (the same expression holds for the ``plus'' polarization with a difference dependence on the orientation of the binary's orbital plane $\iota$).
Here $\mathcal{M}_c(z)$ is the (redshifted) chirp mass, $f$ the GW frequency at the observer and $\Phi(f)$ its phase.
For our purposes the luminosity distance $d_L(z)$ is the most important parameter entering the waveform \eqref{eq:gwf}.
Parameter estimation over the observed GW signal \eqref{eq:gwf} directly yields the value of the luminosity distance of the GW source, together with an uncertainty due to the detector noise.
This implies that once the signal is detected and characterised, the luminosity distance of the source can be extracted, without the need of cross calibrating with other distance indicators or unwanted systematics, as in the case of SNIa.
If also the redshift $z$ of the GW source (or the one of its hosting galaxy) is measured, one obtains a data point in the distant-redshift diagram which can be used to constrain the distance-redshift relation
\begin{equation}
	d_L(z) = c\left(1+z\right)\int_0^z \frac{dz'}{H(z')} \,,
	\label{eq:dL_z}
\end{equation}
where $H(z)$ is the Hubble rate depending on the parameters characterizing the background evolution of the cosmological model at hand.
With a sufficient amount of standard siren data, the theoretical distance-redshift relation can be fitted and constraints on the cosmological parameters can be derived statistically.
Again the analysis is analogue to the one performed for SNIa.

The main drawback of GW standard sirens is the need of measuring the redshift of the GW source.
The most direct way of achieving this goal consists in identifying an EM counterpart to the GW signal.
One can then obtain the redshift with standard spectroscopic or photometric techniques directly from its observation (or from its hosting galaxy).
The problem is that EM counterparts are not guaranteed to be produced by any GW source, and only neutron star and MBHBs are expected to emit detectable EM radiation at merger.
Moreover to effectively identify EM counterparts an accurate and fast sky localisation by the GW detector is necessary, otherwise telescopes cannot be promptly pointed towards the right direction.
These arguments suggest that MBHB mergers will probably constitute one of the most interesting classes of standard sirens, especially in association with LISA, which might provide an accurate sky localisation error for several of such sources.
Furthermore MBHBs produce GWs at very high redshift, up to $z\sim 20$, and EM counterpart could be detected up to $z\sim 10$ with planned future telescopes \cite{Tamanini:2016zlh}.
This implies that MBHB standard sirens can be used to map the expansion of the universe at redshift ranges not reachable by standard candles or other distance probes.

The point now is to predict how many MBHB standard sirens will be detected by LISA and what kind of cosmological data they will provide.
To address this issue, in \cite{Tamanini:2016zlh} the rate and redshift distribution of MBHB merger events have been reproduced using the results of semi-analytical simulations based on \cite{mymodel}; see also \cite{spin_model,letter,newpaper}.
Three different competing scenarios for the initial conditions of the massive BH population at high redshift have been taken into account:
\begin{itemize}
	\item {\bf popIII}: a ``light-seeds'' scenario where the massive BHs form from the remnants of population III stars;
	\item {\bf Q3d}: a ``heavy-seeds'' scenario where massive BHs form from the collapse of protogalactic disks, but a time delay between the merger of the hosting galaxies and the BHs is present;
	\item {\bf Q3nod}: another ``heavy-seeds'' scenario with no time delay between the merger of hosting galaxies and BHs.
\end{itemize}
For each of these scenarios the simulations produce catalogues of MBHB merger events, storing all the relevant information about the massive BHs and their hosting galaxy (redshift, masses, amount of gas, ...).
The parameters of each MBHB merger event are then used as input for the waveform template used in a LISA Fisher matrix code yielding the signal to noise ratio (SNR) and the 1$\sigma$ errors on the waveform parameters (cf.~\cite{Klein:2015hvg}), including in particular the ones on the luminosity distance $\Delta d_L$ and the sky location $\Delta\Omega$ (the code takes into account all the phases of the GW signal: inspiral, merger and ringdown).
The LISA configuration considered for the code coincides with the one labelled as N2A2M5L6 in \cite{Tamanini:2016zlh,Klein:2015hvg}.
The design is characterized by six active laser links (three active arms), an armlength of 2 million km, a mission duration of 5 years and the low frequency noise level coinciding with the one tested by LISA pathfinder \cite{Armano:2016bkm} (N2).
This choice is made in order to work with the configuration, among the ones analysed in \cite{Tamanini:2016zlh}, more similar to the one proposed by the LISA consortium to answer ESA's call for its L3 mission \cite{2017arXiv170200786A}: six laser links, 2.5 million km arms, nominal mission duration of 4 years and low frequency noise comparable with N2.

Among all the GW signals analysed by the LISA Fisher matrix code, the ones with SNR~$> 8$ (confirmed detections) and $\Delta\Omega< 10\, {\rm deg}^2$ (sufficient sky location accuracy) are selected.
Then the events for which an EM counterpart can be detected are further identified.
For this purpose a general MBHB merger astrophysical model predicting an optical quasar-like luminosity and magnetic field induced radio flare and jet is considered.
The characteristics of future planned telescopes, such as LSST, ELT and SKA, are then used to estimate the number of EM counterparts detected taking into account the amount of EM radiation emitted in this way.
The whole process provides catalogues of standard sirens listing $z$, $d_L$ and $\Delta d_L$ for each event.
The uncertainty on $d_L$ is not only provided by the 1$\sigma$ ``instrumental'' error of LISA, but also peculiar velocity and lensing contributions are added (we also consider a de-lensing factor of 2 in agreement with the ``optimistic'' scenario in \cite{Tamanini:2016zlh}).
Moreover the uncertainty on the redshift measurements, which is non negligible only for photometric observations, is also propagated and added to $\Delta d_L$ assuming $\Lambda$CDM.
In \cite{Tamanini:2016zlh} 118 catalogues of 5 years where produced for each MBHB formation model to reduce the statistical uncertainty due to the low number of events in each catalogue.
In the investigation that follows we select a representative catalogue among all 118 of them.
To do this for each catalogue we compute the figure of merit
\begin{equation}
	{\rm FoM} = \det(F_{ij})^{\frac{1}{4}} \,,
	\label{eq:FoM}
\end{equation}
as defined in \cite{Tamanini:2016zlh}, where $F_{ij}$ is the Fisher matrix over the $\Lambda$CDM parameters.
We then select the representative catalogue simply choosing the one associated with the median of the distribution of the FoM.
This will allow us to use the reconstruction method (see Sec.~\ref{sec:the_reconstruction_method}) using the representative catalogue, which provides a fairly realistic realisation of MBHB standard sirens data detected by LISA\footnote{A more accurate statistical analysis should be performed taking into account all 118 catalogues, as done e.g.~in \cite{Tamanini:2016zlh,Caprini:2016qxs}. However the reconstruction method does not perform well for those catalogues whose number of events is not large enough, and averaging the results obtained over those catalogues would provide inconsistent outcomes. This is the reason why we prefer to first select a representative catalogue and then perform the reconstruction method only over its data (which always provide reliable results, at least up to the considered redshift).}.
We will also repeat the analysis considering a LISA mission duration of 10 years, for which the number of standard sirens events on average doubles with respect to a 5 years mission.
For this analysis we will combine together the two 5 years catalogues having their FoMs closest  the median of the distribution of the FoMs, producing in this way a representative 10 years catalogue of standard siren data.


\section{The dark sector interaction} 
\label{sec:the_dark_sector_interaction}

We consider a minimal extension of the $\Lambda$CDM model where the vacuum energy (defined as DE with $w=-1$) is allowed to interact with DM, without 
introducing any additional degrees of freedom.
A more general model of DE whose equation of state (EoS) is dynamical has been considered in~\cite{Cai:2015zoa}.
Here we only assume the interacting vacuum model which allows energy-momentum transfer between CDM and the vacuum as for example in~\cite{Salvatelli:2014zta}.
From now on we will set physical units such that $8\pi G=1$.

In a flat universe, the Friedmann equation describing the evolution of the universe is given by
\begin{equation}
3 {H^2} = {\rho _m} + {\rho _v},
\label{eq:friedmann}
\end{equation}
where $\rho_m$ and $\rho_v$ are the energy densities of dark matter and the vacuum, respectively.
Considering the interaction between dark matter and the vacuum, the conservation equations can be written as
\begin{equation}
\dot\rho_m+3H\rho_m =  - Q,
\label{rhom}
\end{equation}
and
\begin{equation}
\dot\rho_v = Q,
\label{rhov}
\end{equation}
where $H$ is the expansion rate of the universe, and $Q$ describes the strength of the interaction between dark matter and dark energy. Obviously, $Q=0$ recovers the $\Lambda$CDM model where $\rho_v$ is simply the cosmological constant.
Following~\cite{Cai:2015zoa}, we do not parametrize the interaction term $Q$ but we will reconstruct it using a model-independent method.

From Eqs.~\eqref{eq:dL_z}, \eqref{eq:friedmann}--\eqref{rhov}, we can obtain~\cite{Cai:2015zoa}
\begin{align}
q \equiv \frac{Q}{H_0^3} = &~~2\left(\frac{{3D'{'^2}}}{{D{'^5}}} - \frac{{D'''}}{{D{'^4}}}\right){(1 + z)^2} + 4\frac{{D''}}{{D{'^4}}}(1 + z),
\label{equa:qD}
\end{align}
where
\begin{equation}
	D(z)=\frac{H_0}{c} \frac{1}{1+z} d_L(z) \,,
\end{equation}
is the normalized comoving distance, and $q$ is the dimensionless interacting term.
Using \eqref{equa:qD}, we can reconstruct the interaction from the observed distance-redshift relationship datasets simulated for LISA.
In what follows we do not make any additional assumption about the coupling between the vacuum DE and DM, i.e.~we do not assume any particular parametrization for $Q(z)$, but let it be an arbitrary function of $z$.
The Gaussian process method, presented in Sec.~\ref{sec:the_reconstruction_method}, will allow us to reconstruct the distance $d_L(z)$ in a model-independent way and to set uncertainties on its value at any considered redshift.
Note that this analysis is more general than other investigations where an explicit redshift dependency of $Q$, usually through the quantities $H(z)$, $\rho_m(z)$ and $\rho_v(z)$, is assumed; see e.g.~\cite{Salvatelli:2014zta,Ade:2015rim,Costa:2016tpb,Nunes:2016dlj,Li:2015vla,Sharov:2017iue}.
The results presented in Sec.~\ref{sec:results} thus nicely generalize the work performed in \cite{Caprini:2016qxs} where LISA MBHB standard sirens were shown to effectively constrain specific interacting DE models, with $Q\propto H\rho_m$ or $Q\propto H\rho_v$, in the redshift range $1 \lesssim z \lesssim 10$.


\section{The reconstruction method} 
\label{sec:the_reconstruction_method}

Gaussian processes (GPs)~\cite{Holsclaw:2010nb,Holsclaw:2010sk,Holsclaw:2011wi,Seikel:2012uu} allow one to reconstruct a function from a set of data without assuming any parametrization for it.
We use GPs in Python (GaPP)~\cite{Seikel:2012uu} to derive our GP reconstruction results.
The distribution over functions  provided by the GP is suitable to describe the observed data.
At each point $z$, the reconstructed function $f(z)$ is also a Gaussian distribution with a mean value and Gaussian error. The functions at different points $z$ and $\tilde{z}$ are related by a covariance function $k(z,\tilde{z})$, which only depends on a set of hyperparameters $\ell$ and $\sigma_f$. Here $\ell$ gives a measure of the coherence length of the correlation in the $x$-direction and $\sigma_f$ denotes the overall amplitude of the correlation in the $y$-direction. Both of them are optimized by the GP with the observed dataset.
In contrast to actual parameters, the GP does not specify the form of the reconstructed function. Instead it characterizes the typical changes of the function.

The different choices of the covariance function may affect the reconstruction to some extent. The covariance function usually takes  the squared exponential form as~\cite{Seikel:2012uu}
\begin{equation}
k(z,\tilde{z})={\sigma_f}^2 \exp\Big(-\frac{(z-\tilde{z})}{2\ell^2}\Big).
\end{equation}
But this is not always a suitable choice. Here we take the Mat\'{e}rn ($\nu = 9/2$) covariance function
\begin{align}
k(z,\tilde z) = &~{\sigma _f}^2\exp\left( - \frac{{3\left| {z - \tilde z} \right|}}{\ell }\right) \times\Big[1 + \frac{{3\left| {z - \tilde z} \right|}}{\ell } + \frac{{27{{(z - \tilde z)}^2}}}{{7{\ell ^2}}} + \frac{{18{{\left| {z - \tilde z} \right|}^3}}}{{7{\ell ^3}}} + \frac{{27{{(z - \tilde z)}^4}}}{{35{\ell ^4}}}\Big],
\end{align}
according to the analysis made in~\cite{Seikel:2013fda}, where the authors considered various assumed models and many realizations of mock datasets for a test and concluded that the Mat\'{e}rn ($\nu=9/2$) covariance function can lead to more reliable results than all others when applying GP to reconstructions using $D$ measurements.
The detailed analysis and description of the GP method can be found in~\cite{Seikel:2012uu,Seikel:2013fda}, where the authors studied the use of the GP method to reconstruct dark energy dynamics from supernovae data. Some of the GPs applications can also be found in~\cite{Cai:2015zoa,Cai:2015pia,Cai:2016vmn}.


\section{Results} 
\label{sec:results}

Using Gaussian processes, we can obtain the reconstructed $D(z)$ and its derivatives, as well as the covariance matrix between them.
Then we can apply Monte Carlo samplings to determine the reconstructed $q(z)$ from Eq.~\eqref{equa:qD} at every redshift $z$.
Since the reconstructions involve the third-order derivative of the distance, the errors from the data increase as the redshift gets large, implying that the reconstructed errors of the distance will become very large in the high redshift region.
This leads to somehow uncontrollable large errors at high redshift when applied to the Monte Carlo sampling in~\eqref{equa:qD}.
For this reason, following~\cite{Cai:2015zoa}, for a better exposition of the results, we introduce a prefactor $(1+z)^{-6}$ to plot the figures of the interaction term $q$, that is we define $\tilde{q}(z)=q(z)(1+z)^{-6}$.
Such a prefactor is just considered as a scale transformation with respect to the redshift, introduced to better expose the results in the following plots.
The main scope of our analysis is in fact to compare the results with the ones obtained in~\cite{Cai:2015zoa}, where the same rescaling has been adopted.
For this reason choosing to plot $\tilde{q}(z)$ guarantees a direct comparison with~\cite{Cai:2015zoa}.
One must keep in mind however that the actual errors on $Q(z)$ will be larger especially at higher redshift.
To check consistency of a specific interacting model with the results showed below one should compute $\tilde{q}(z)$ and compare its value at all redshift with the uncertainty confidence level provided in the plots below.
In any case since we are only interested in testing the compatibility of the reconstructed $Q$ with its null hypothesis, i.e.~with $Q=0$, for what concerns our scope we are free to consider this rescaling without loss of generality; cf.~\cite{Yahya:2013xma}.

Fig.~2 of~\cite{Cai:2015zoa} shows that using the DES simulated data one can reconstruct the interaction well from redshift 0 to 1.2.
At higher redshift the errors become out of control as the lack of data points make them rapidly diverge.
We can conclude that the DES will have the ability to reconstruct the dark sector interaction up to redshift 1.2, though the errors do get very large in the high redshift region.
In this paper, we want to use the simulated MBHB standard siren data for LISA to repeat the same test.
Given the fact that the bulk of LISA cosmological data will appear at redshift from 1 to 5~\cite{Tamanini:2016zlh}, we expect that LISA will be able to extend the result obtained only with DES to larger redshift values.
As we are going to see the dark sector interaction will be well reconstructed up to higher redshift, depending on the number of standard sirens available, i.e.~on the duration of the LISA mission.
The following analysis is divided into two parts.
In the first part we use the LISA MBHB standard siren data alone to check what LISA can do without being combined with other probes.
In the second part we add the DES and LISA data to see how the DES result obtained in~\cite{Cai:2015zoa} can improve.
In both cases we do the analysis for a LISA mission duration of 5 years (conservative) and 10 years (optimistic); cf.~\cite{2017arXiv170200786A}.

\subsection{LISA MBHB standard sirens alone} 
\label{sub:lisa_mbhb_standard_sirens_alone}

\begin{figure}
	\centering
	\includegraphics[width=.31\textwidth]{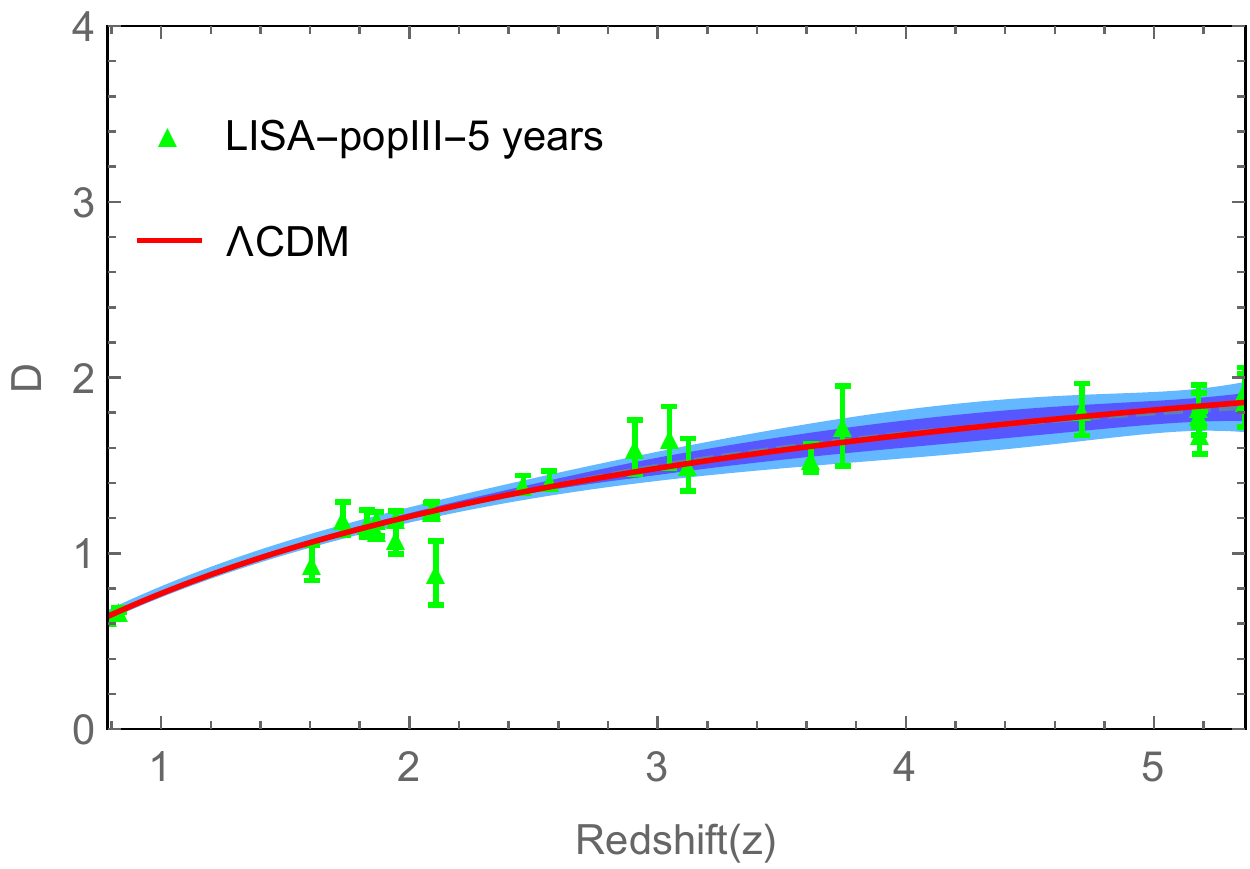}
	\includegraphics[width=.31\textwidth]{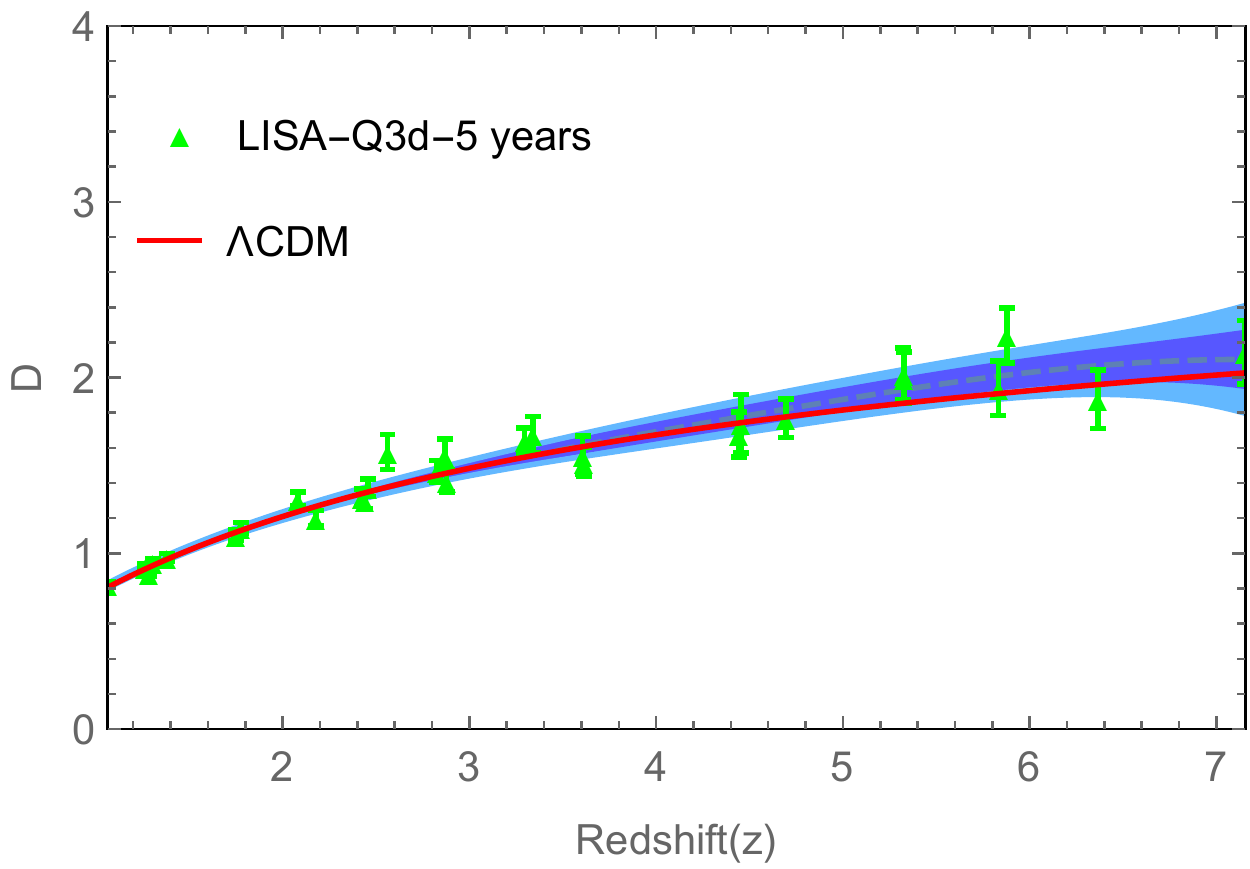}
	\includegraphics[width=.31\textwidth]{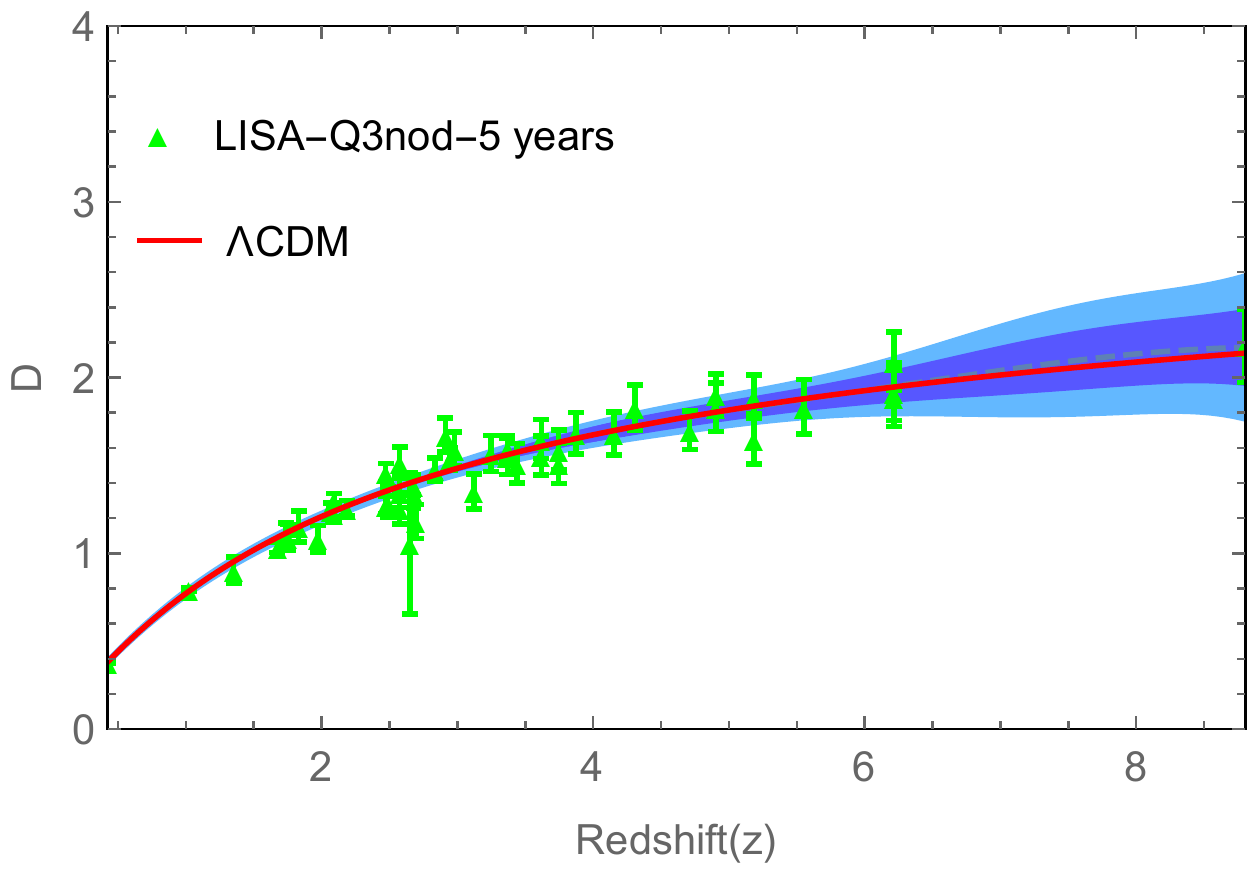}\\
	\caption{Reconstruction of the distance $D(z)$ using LISA alone for a 5 years mission. From left to right each column reports the results for popIII, Q3d, Q3nod. The shaded blue regions are the 68\% and 95\% C.L.~of the reconstructed function.}
\label{fig:5yD_LISA}
\end{figure}

\begin{figure}
	\centering
	\includegraphics[width=.31\textwidth]{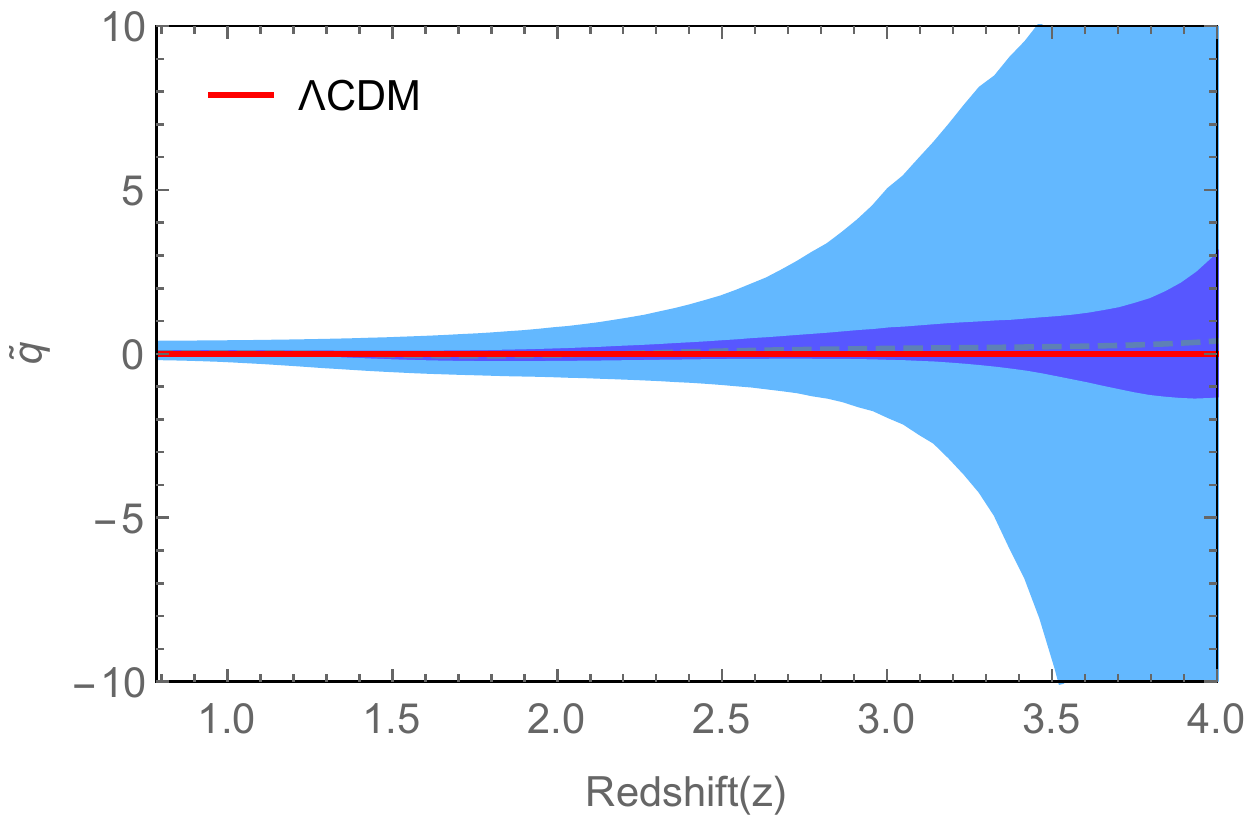}
	\includegraphics[width=.31\textwidth]{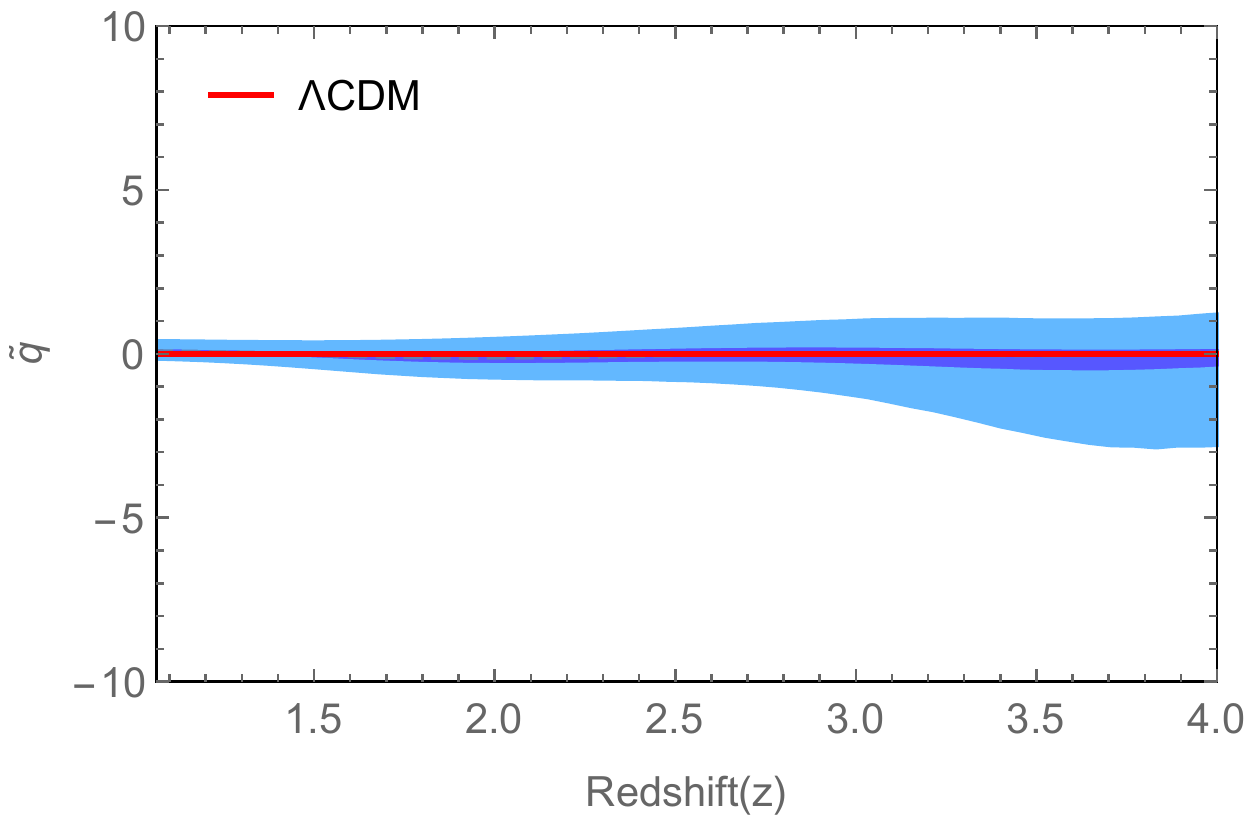}
	\includegraphics[width=.31\textwidth]{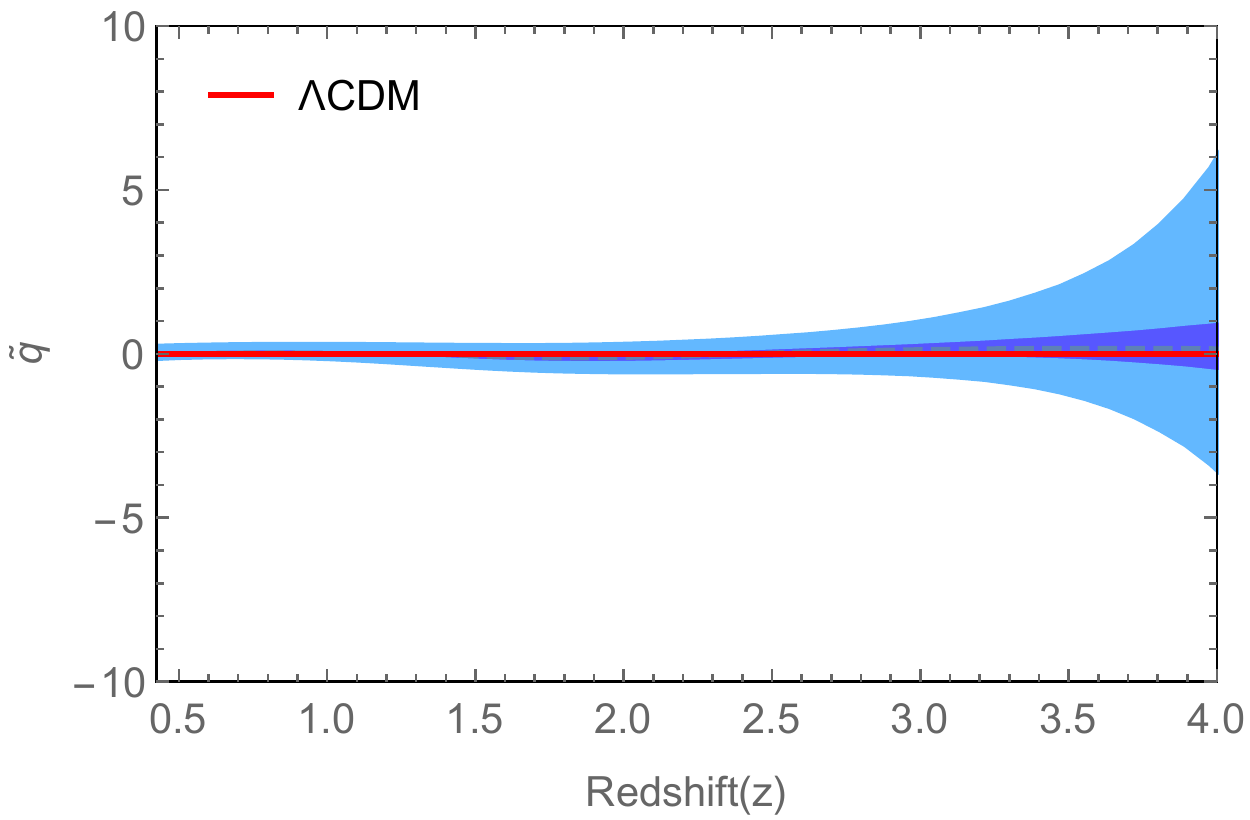}
\caption{Reconstructions of the interaction using LISA alone for a 5 years mission. From left to right each column reports the results for popIII, Q3d, Q3nod.}
\label{fig:5yq_LISA}
\end{figure}

Since LISA MBHB standard siren data cover the redshift mostly from 1 to 8 (see Appendix~A of~\cite{Caprini:2016qxs}), we just reconstruct the distance in the redshift region covered by the data.
The reconstructed errors outside such a region, including the late time $z<1$ epoch, would quickly diverge due to the lack of data points.
We assume the $\Lambda$CDM model with $\Omega_m=0.3$ as our fiducial model (note that $H_0$ do not appear in our analysis, so we are not biased by its value ).
Figs.~\ref{fig:5yD_LISA} and~\ref{fig:5yq_LISA} show the reconstructions of the distance $D(z)$ and the final interaction term $\tilde{q}(z)$ using LISA alone for a 5 years mission.
The reconstruction of the derivatives of $D(z)$ is reported in Fig.~\ref{fig:5yDprime_LISA} at the end of the paper.
In the redshift region covered by the data, the simulated MBHB standard siren data for LISA can reconstruct the dark sector interaction well from redshift about $z\sim1$ to $z\sim3$ and even higher.
The results of the three different competing scenarios for the initial conditions of the massive BH population are all shown.
We can conclude that the reconstruction for the Q3d and Q3nod models are a litter better than the first popIII model, whose errors diverge at lower redshift.

\begin{figure}
	\centering
	\includegraphics[width=.31\textwidth]{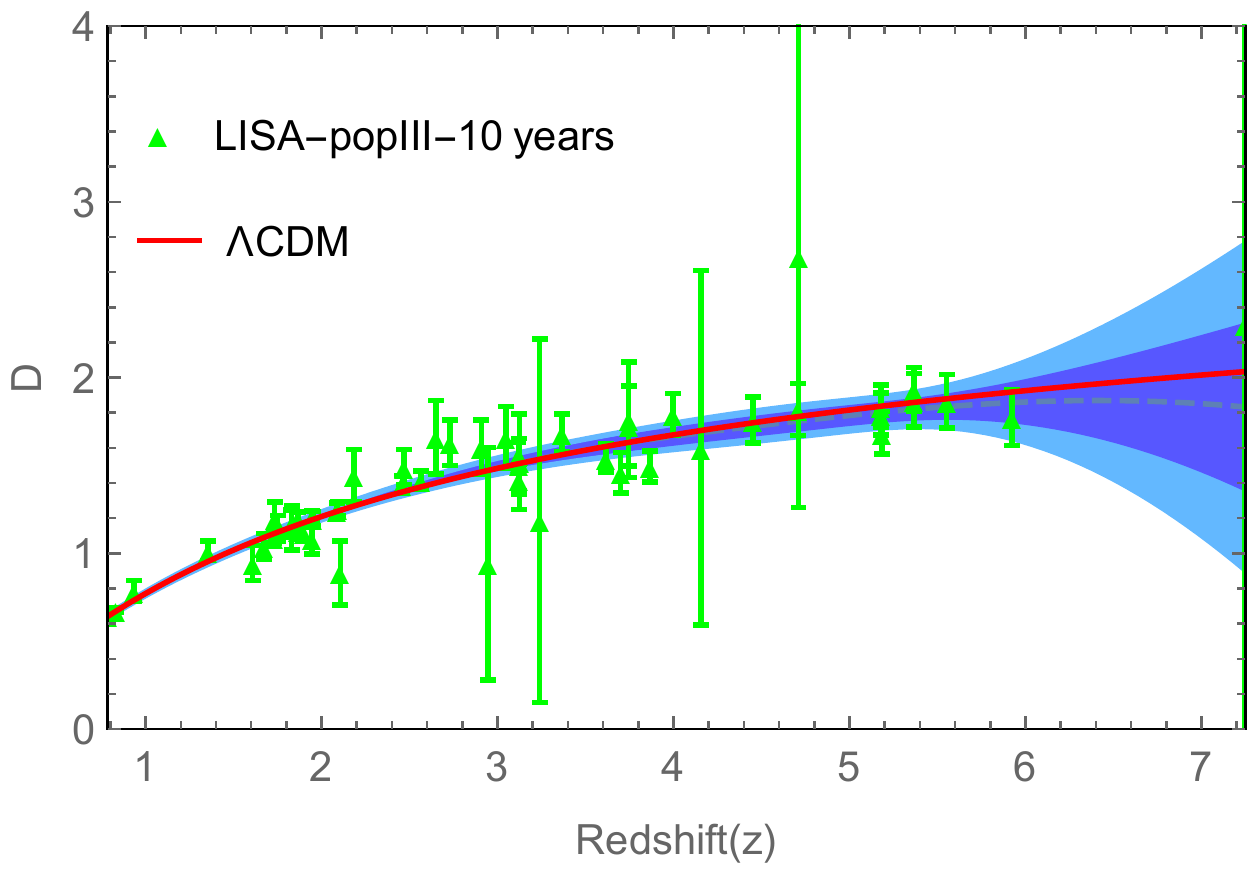}
	\includegraphics[width=.31\textwidth]{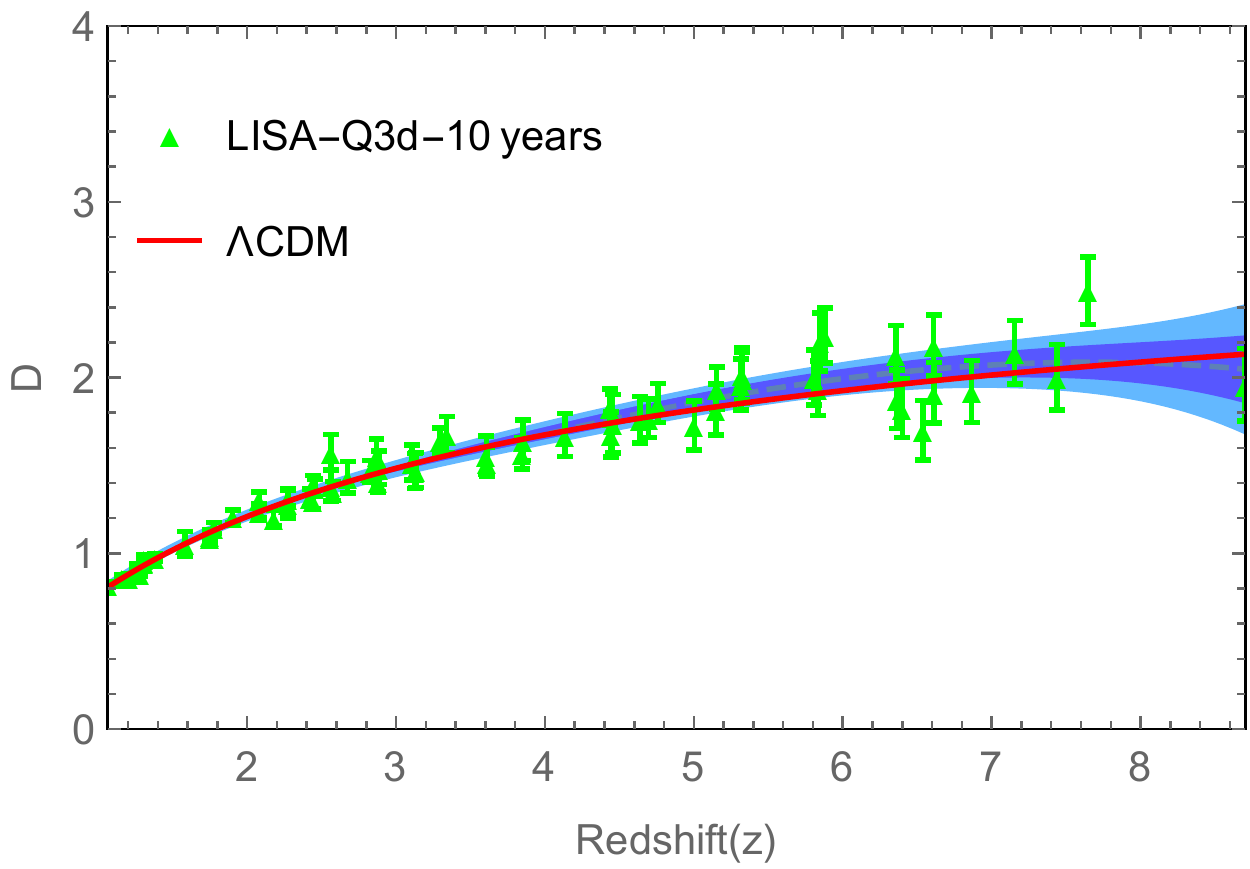}
	\includegraphics[width=.31\textwidth]{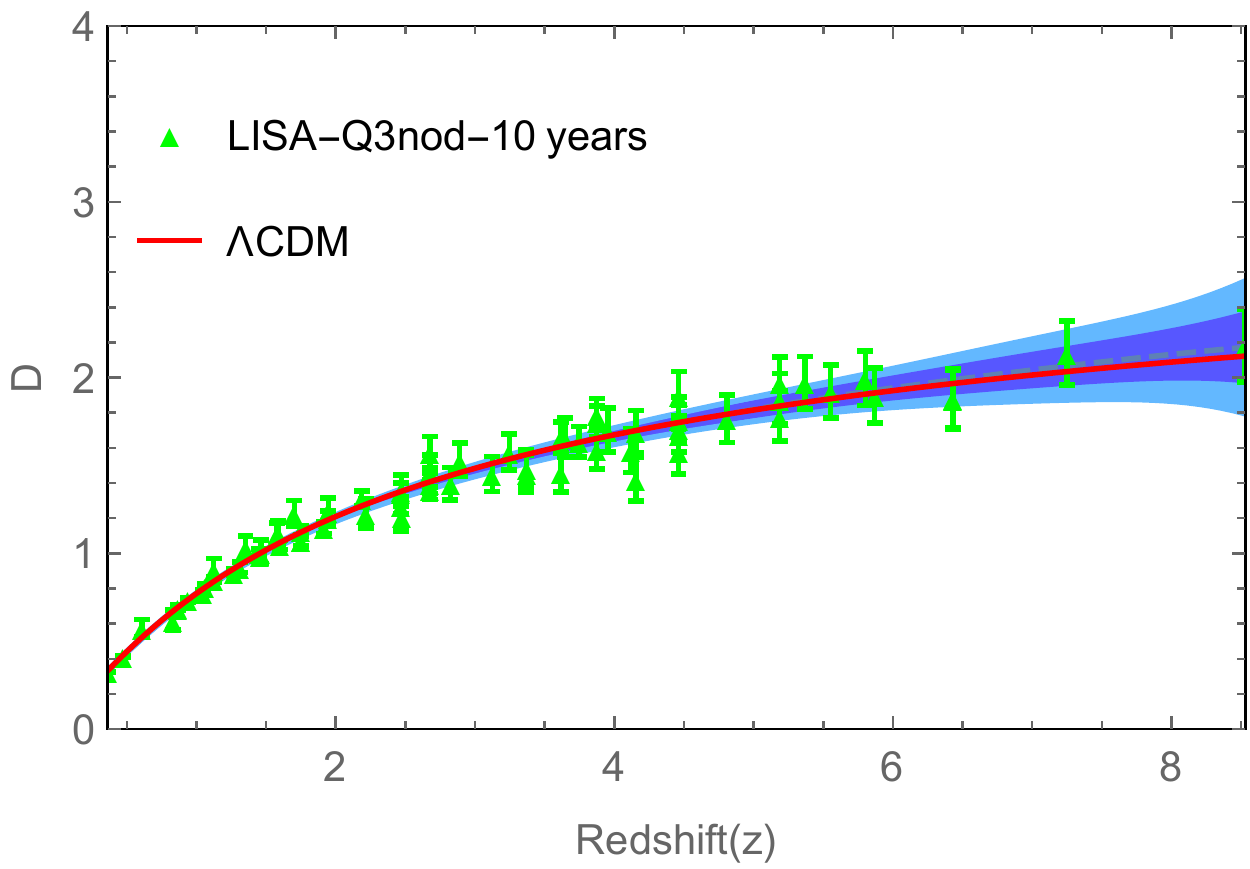}\\
	\caption{Reconstruction of the distance $D(z)$ using LISA alone for a 10 years mission. From left to right each column reports the results for popIII, Q3d, Q3nod. The shaded blue regions are the 68\% and 95\% C.L.~of the reconstruction.}
\label{fig:10yD_LISA}
\end{figure}

\begin{figure}
	\centering
	\includegraphics[width=.31\textwidth]{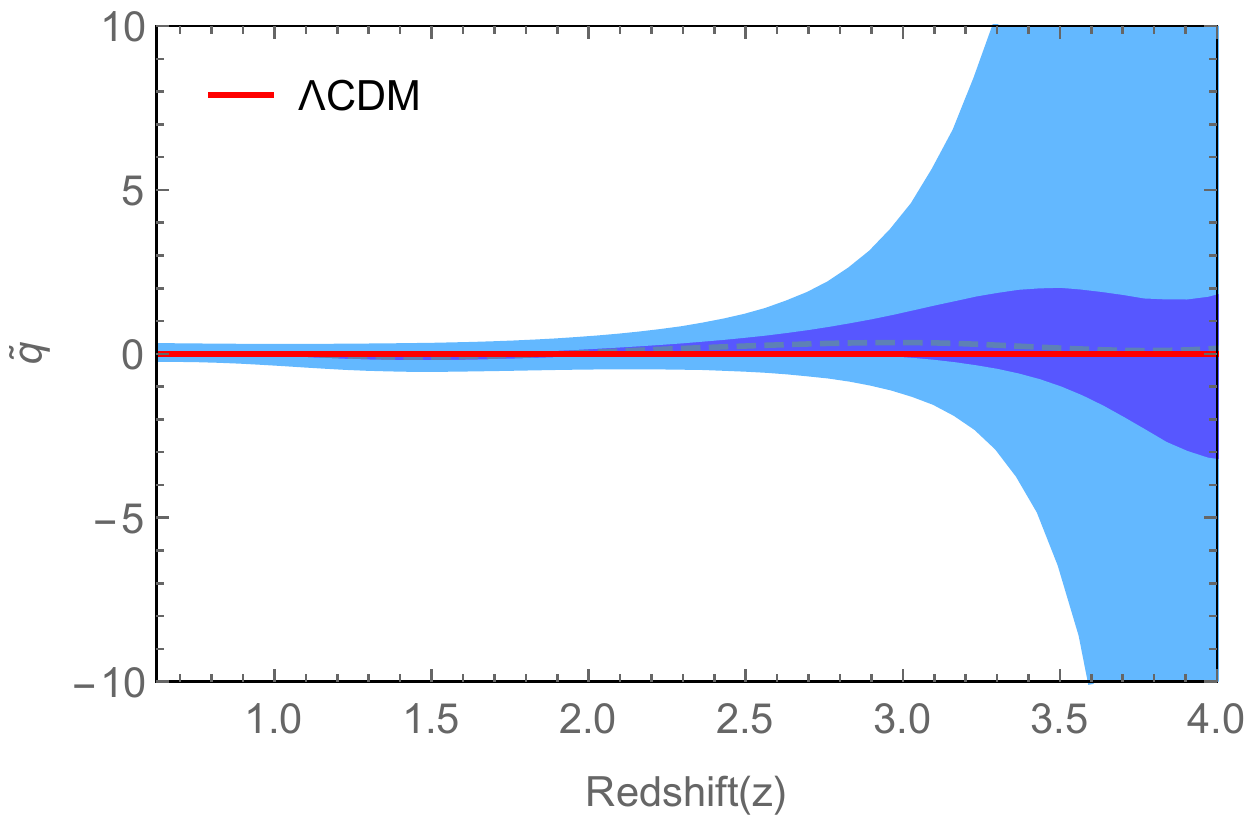}
	\includegraphics[width=.31\textwidth]{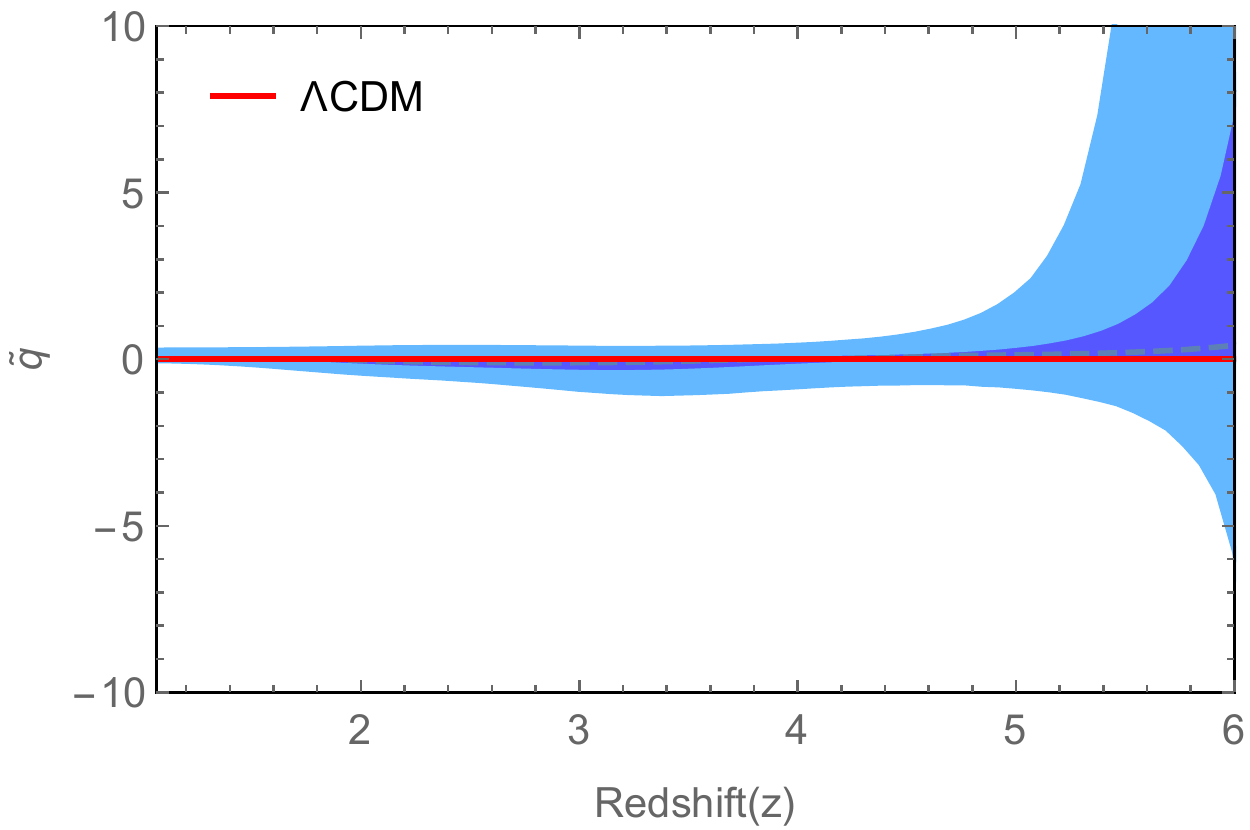}
	\includegraphics[width=.31\textwidth]{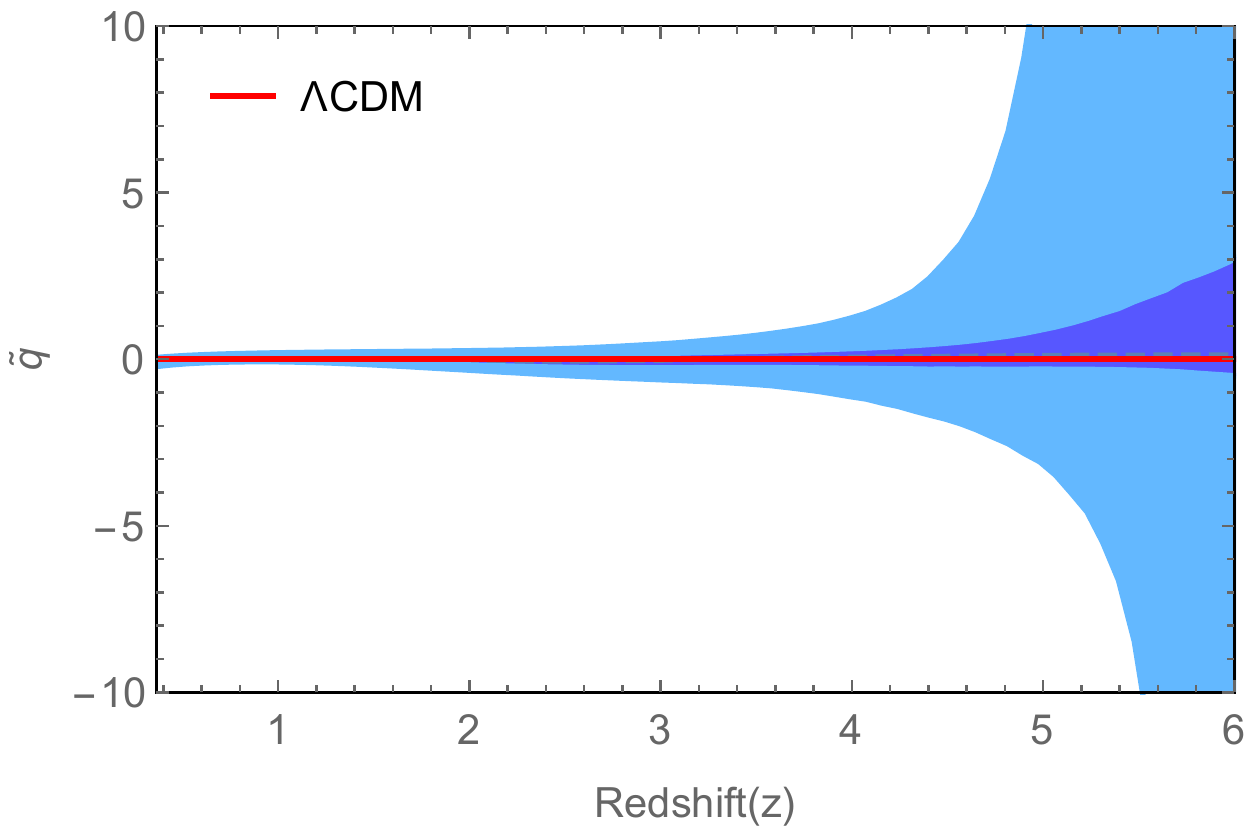}
\caption{Reconstructions of the interaction using LISA alone for a 10 years mission. From left to right each column reports the results for popIII, Q3d, Q3nod.}
\label{fig:10yq_LISA}
\end{figure}

Figs.~\ref{fig:10yD_LISA} and~\ref{fig:10yq_LISA} show the reconstruction of the distance $D(z)$ and the rescaled interaction term $\tilde{q}(z)$ using LISA data alone for a 10 years mission.
The reconstruction of the derivatives of $D(z)$ for this case is plotted in Fig.~\ref{fig:10yDprime_LISA} at the end of the paper.
With a 10 years mission, we can obtain a larger number, roughly double, of MBHB standard sirens data.
The 10 years mission results show that we can extend the well reconstructed redshift range to more than $z\sim5$, although the reconstruction is still inefficient at redshift lower than one due to the absence of MBHB data points.
Again, the results for Q3d and Q3nod MBHB models are better than the popIII one; see Fig.~\ref{fig:10yq_LISA}.
The results for a 5 years and a 10 years LISA mission show that in the redshift region covered by the MBHB standard siren data, we will be able to reconstruct the interaction of vacuum dark energy and dark matter well up to redshift $z\sim3$ (for 5 years mission) and $z\sim5$ (for 10 years mission), although the reconstruction errors at low redshift become uncontrollable for both cases.

\subsection{LISA MBHB standard sirens plus DES standard candles} 
\label{sub:lisa_mbhb_standard_sirens_plus_des_standard_candles}

\begin{figure}
	\centering
	\includegraphics[width=.31\textwidth]{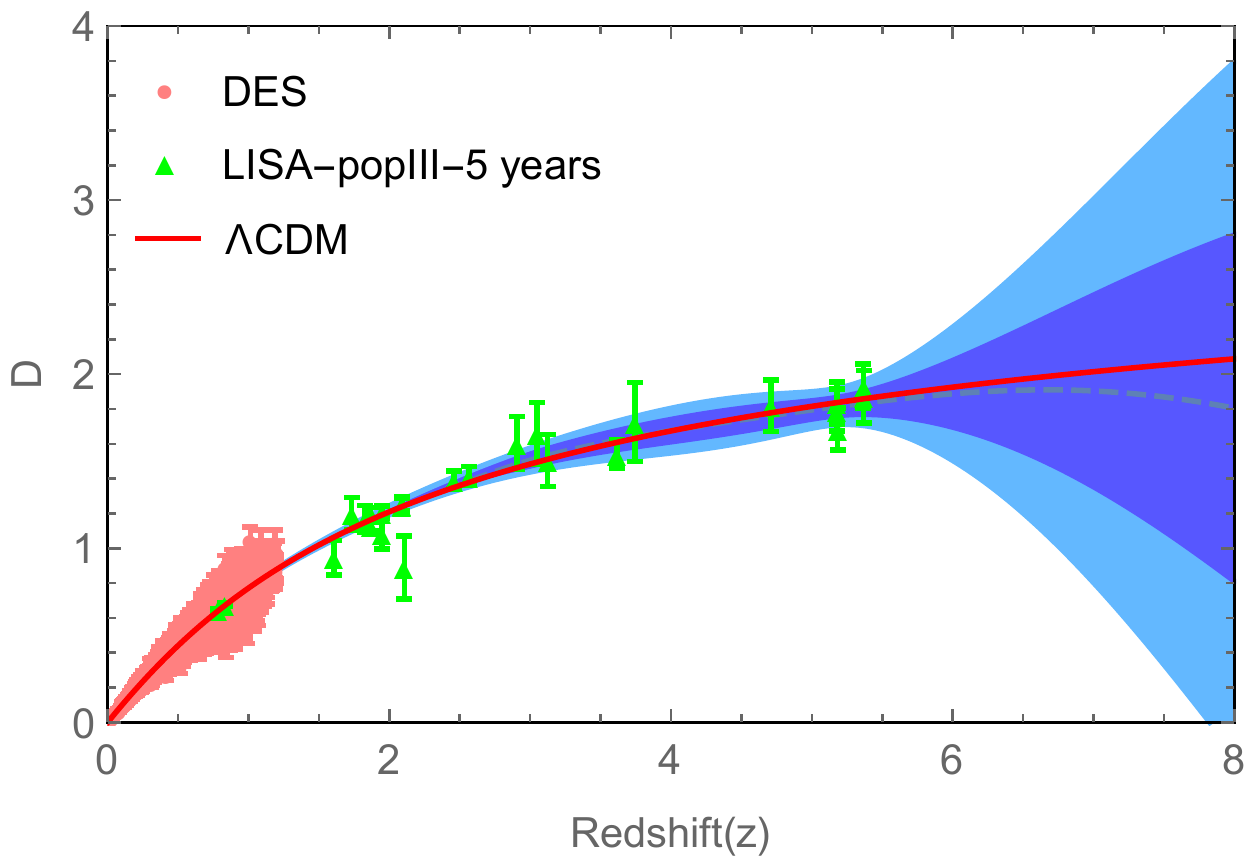}
	\includegraphics[width=.31\textwidth]{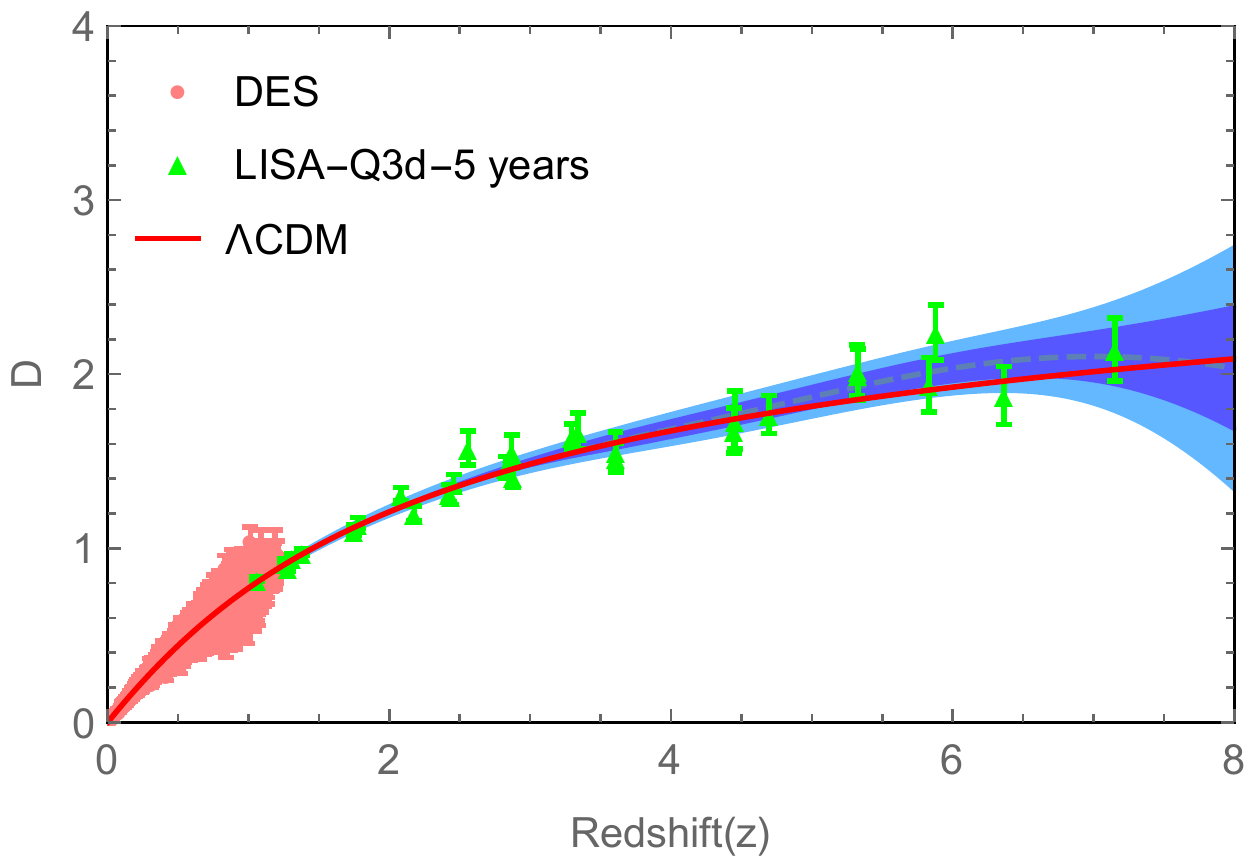}
	\includegraphics[width=.31\textwidth]{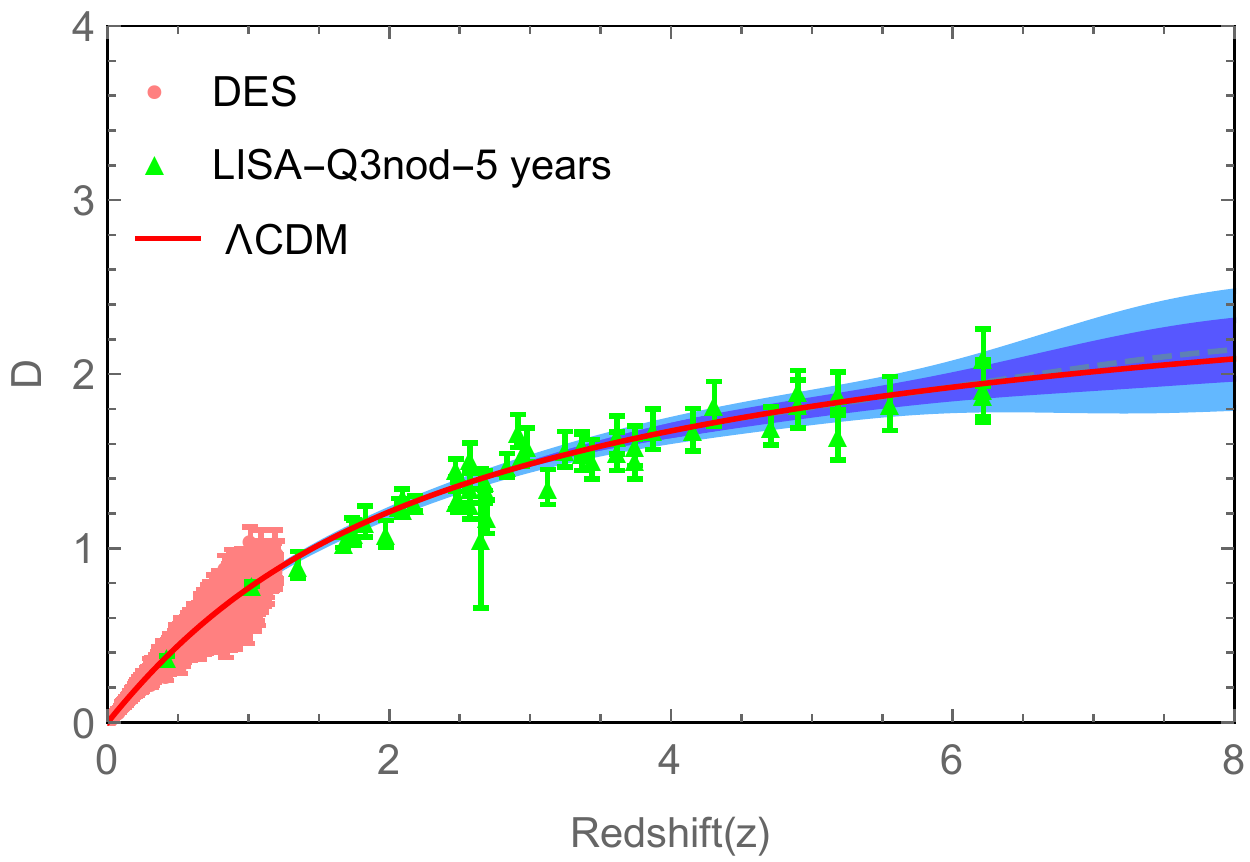}\\
	\caption{Reconstruction of the distance $D(z)$ using DES$+$LISA for a 5 years mission. From left to right each column reports the results for popIII, Q3d, Q3nod. The shaded blue regions are the 68\% and 95\% C.L.~of the reconstruction.}
\label{fig:5yD_LISAplusDES}
\end{figure}

\begin{figure}
	\centering
	\includegraphics[width=.31\textwidth]{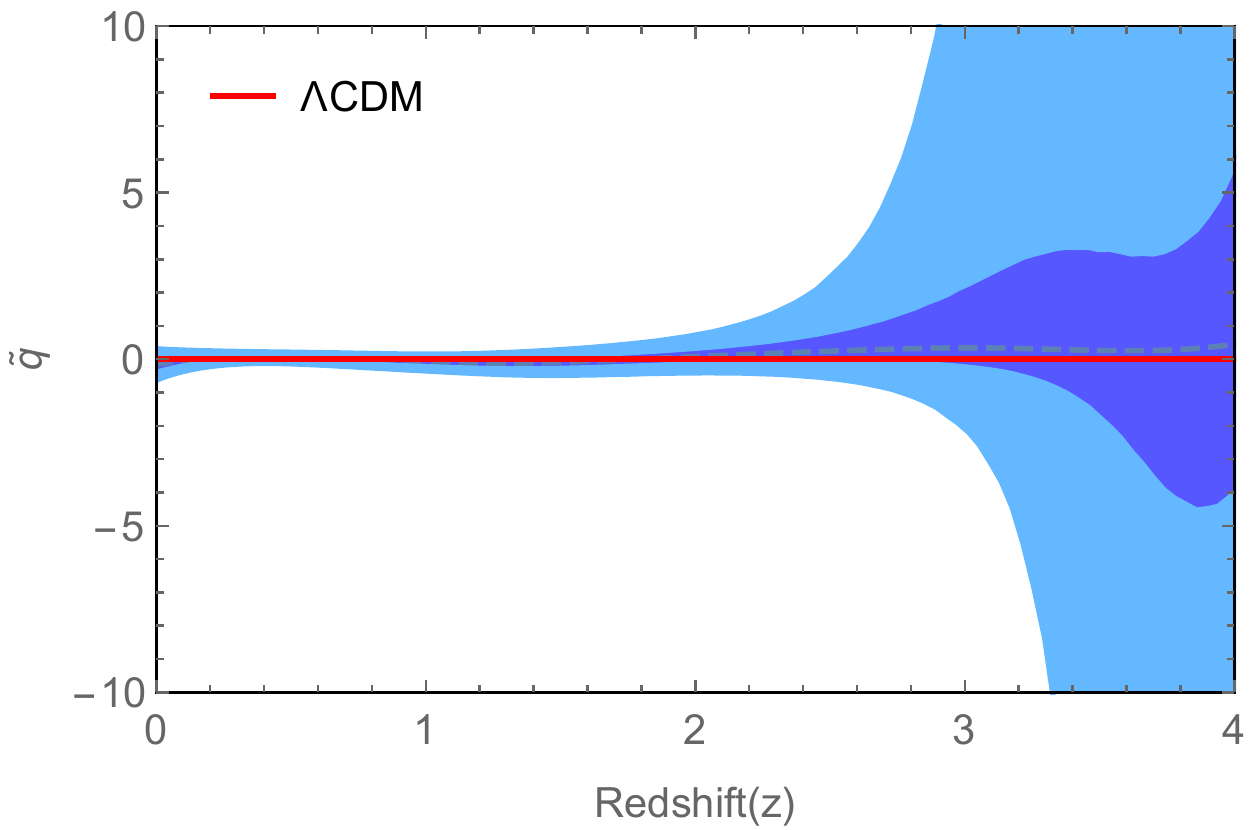}
	\includegraphics[width=.31\textwidth]{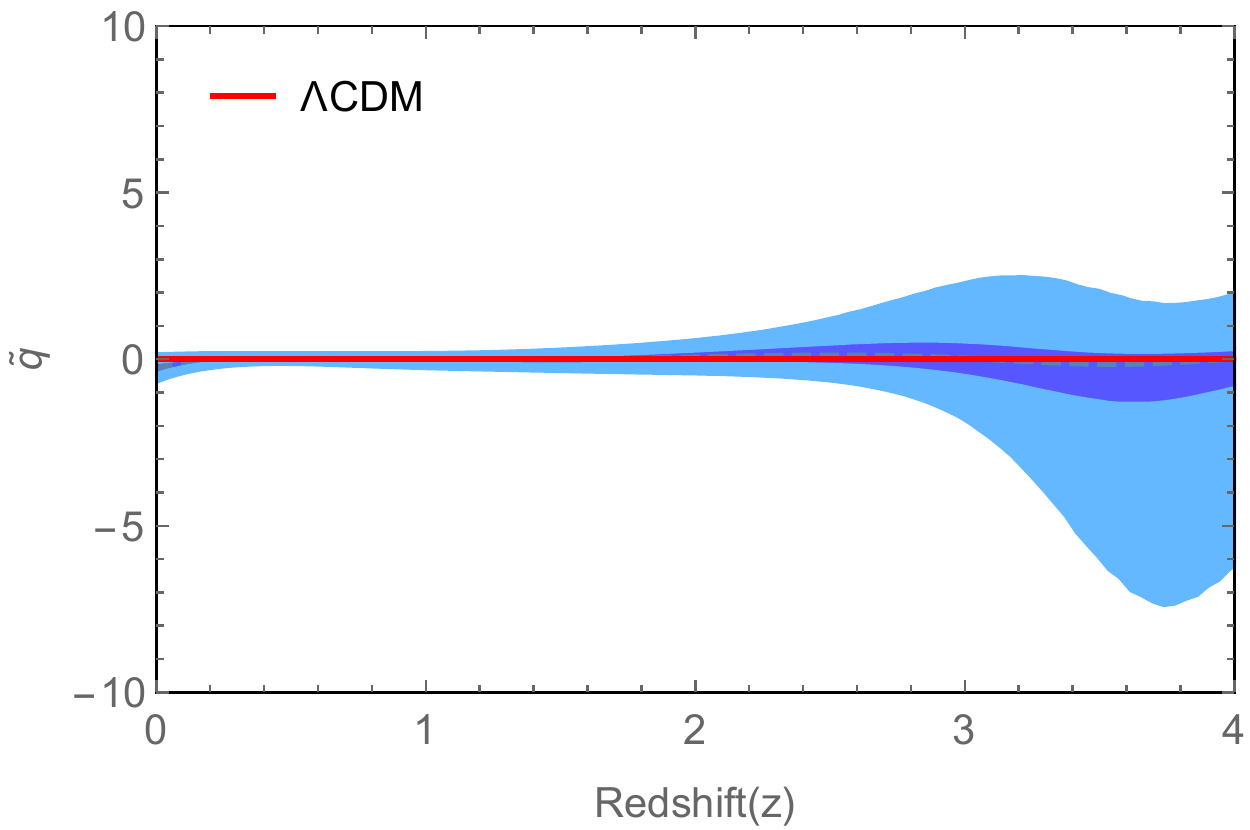}
	\includegraphics[width=.31\textwidth]{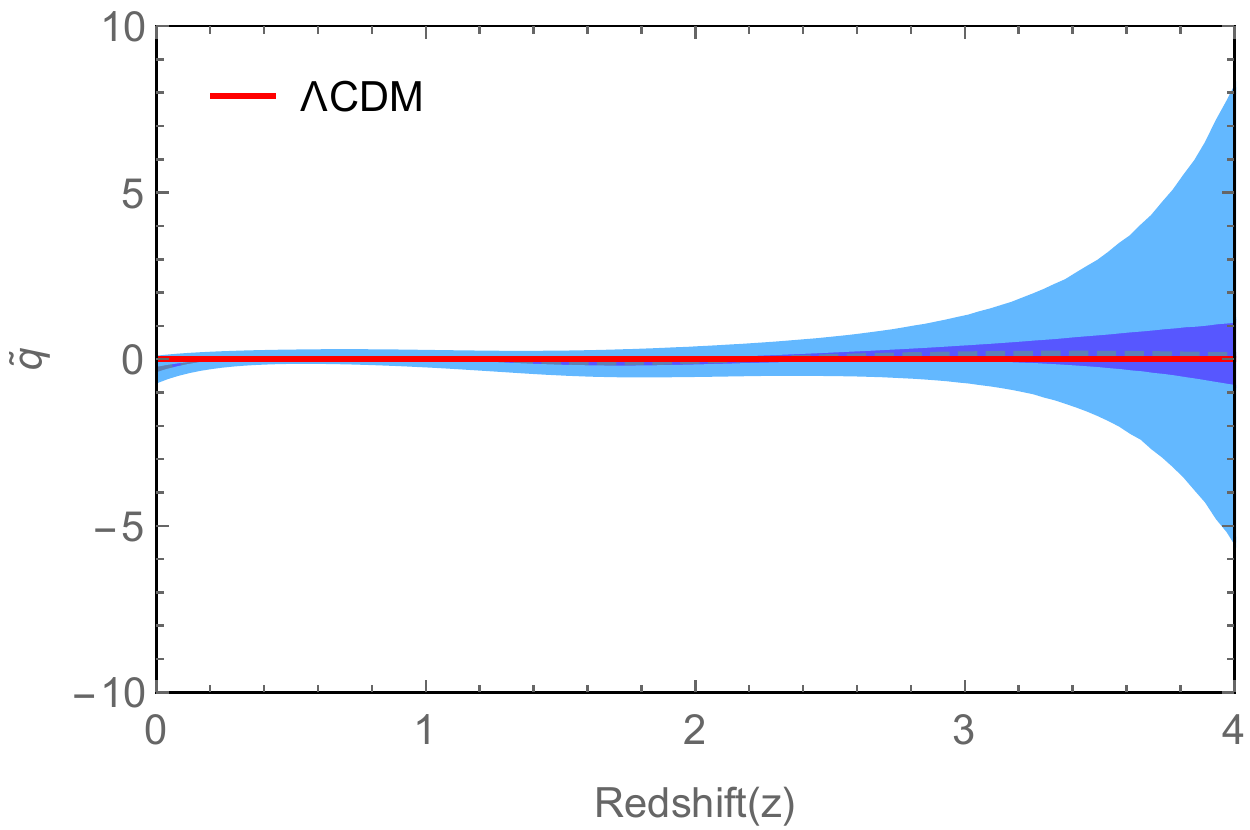}
\caption{Reconstruction of the interacting term $\tilde{q}(z)$ using DES$+$LISA (5 years) data. From left to right each column reports the results for popIII, Q3d, Q3nod.}
\label{fig:5yq_LISAplusDES}
\end{figure}

Having obtained the results using LISA cosmological data alone, we want now to go a step further, that is, combining the GW data of LISA with the traditional EM signal of DES, to check how these can reconstruct the interaction of vacuum DE and DM.
The DES is expected to obtain high quality light curves for about 4000 SNe Ia from $z=0.05$ to $z=1.2$ in the near future~\cite{Bernstein:2011zf}.
We produce mock data for DES following the same procedure explained in~\cite{Cai:2015zoa}, to which we refer for more details.
Combining the LISA and DES datasets, which together cover the whole redshift region from 0 to about 8, we expect to be able to reconstruct the dark interaction from today up to high redshift and consequently to extend the results of~\cite{Cai:2015zoa} to higher redshift.
Figs.~\ref{fig:5yD_LISAplusDES} and~\ref{fig:5yq_LISAplusDES} show the reconstructions of the distance $D(z)$ and the rescaled interaction $\tilde{q}(z)$ using DES data combined with MBHB standard sirens obtained by LISA with a 5 years mission.
The reconstructions of the derivatives of $D(z)$ are reported in Fig.~\ref{fig:5yDprime_LISAplusDES} at the end of the paper.
We can see that the two datasets can reconstruct the interaction well from redshift $z\sim0$ to more than $z\sim3$, extending the results of Fig.~\ref{fig:5yq_LISA} to the low redshift region.
Again, similarly to the results using LISA data alone (cf.~Fig.~\ref{fig:5yq_LISA}), the MBHB scenarios of Q3d and Q3nod are better than popIII in the results.

\begin{figure}
	\centering
	\includegraphics[width=.31\textwidth]{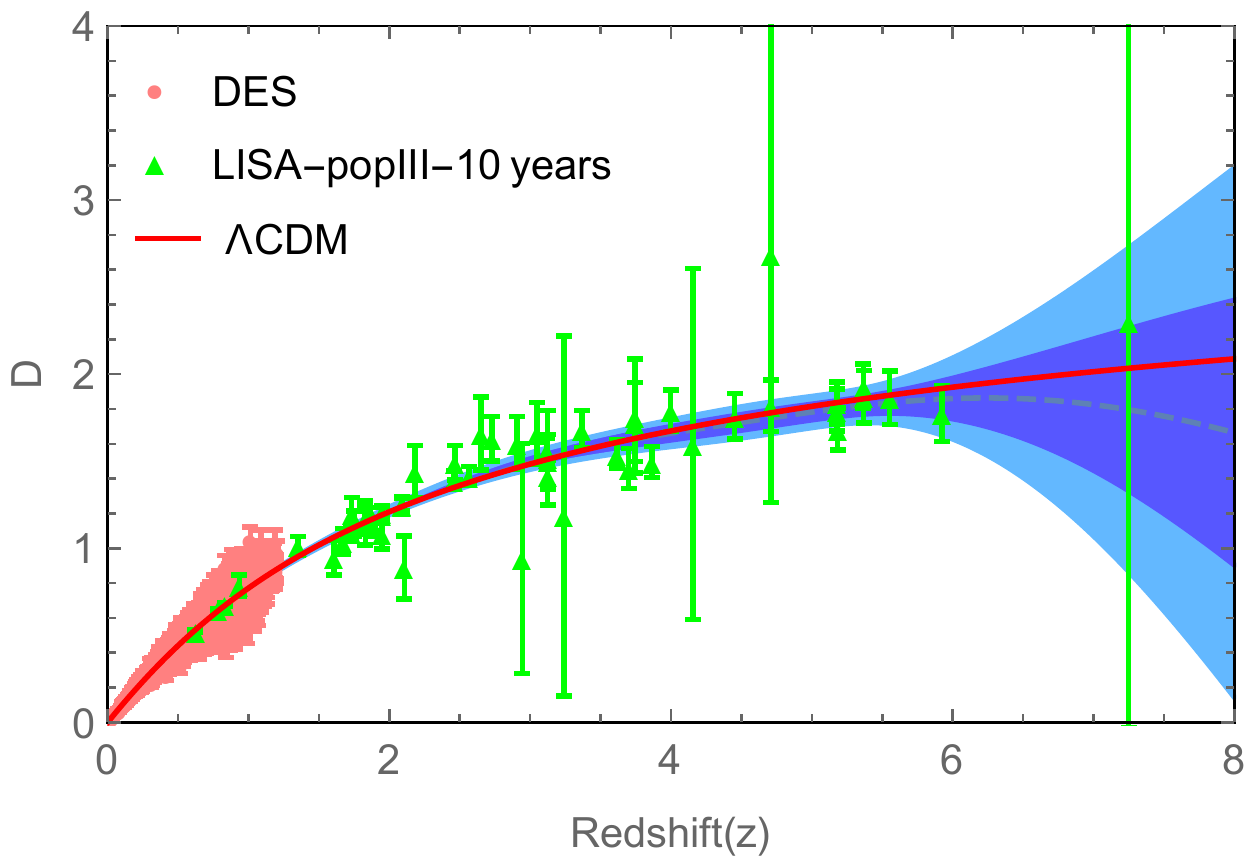}
	\includegraphics[width=.31\textwidth]{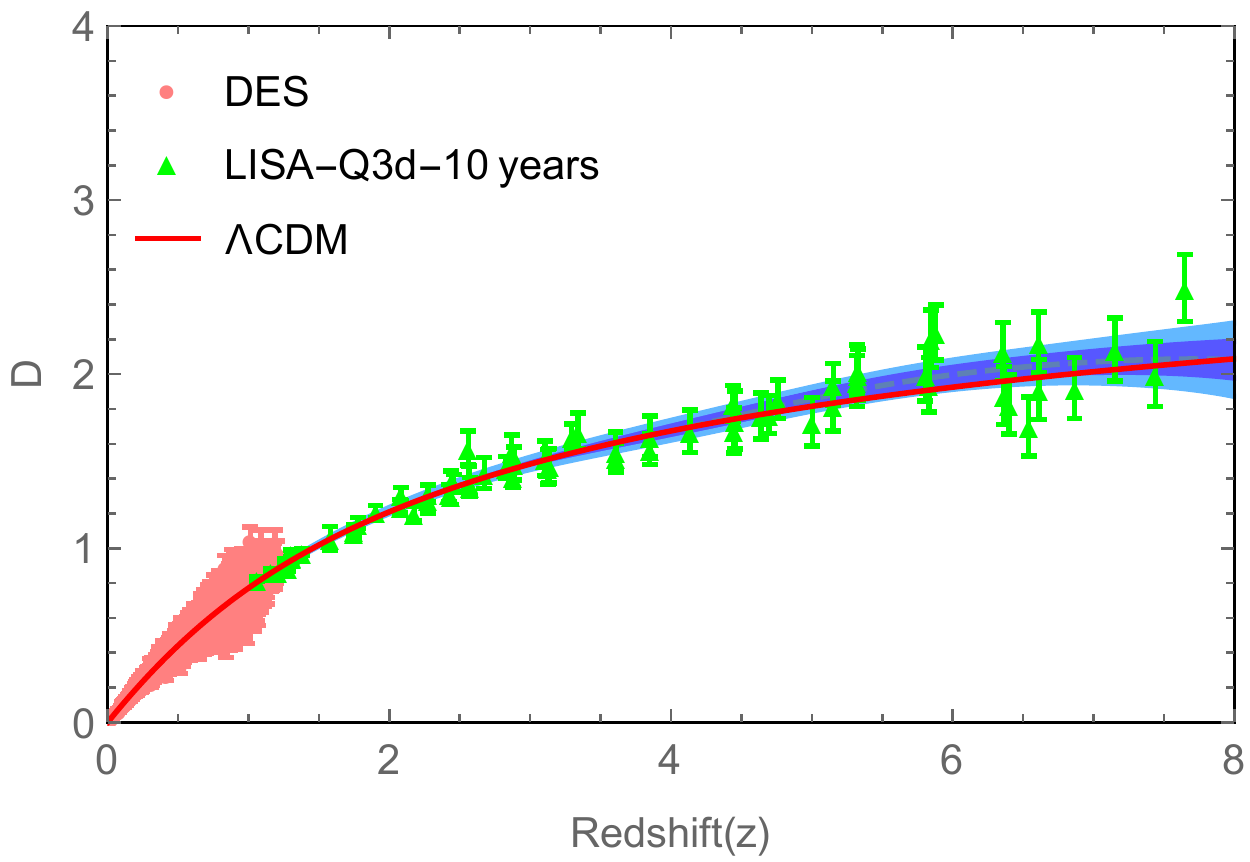}
	\includegraphics[width=.31\textwidth]{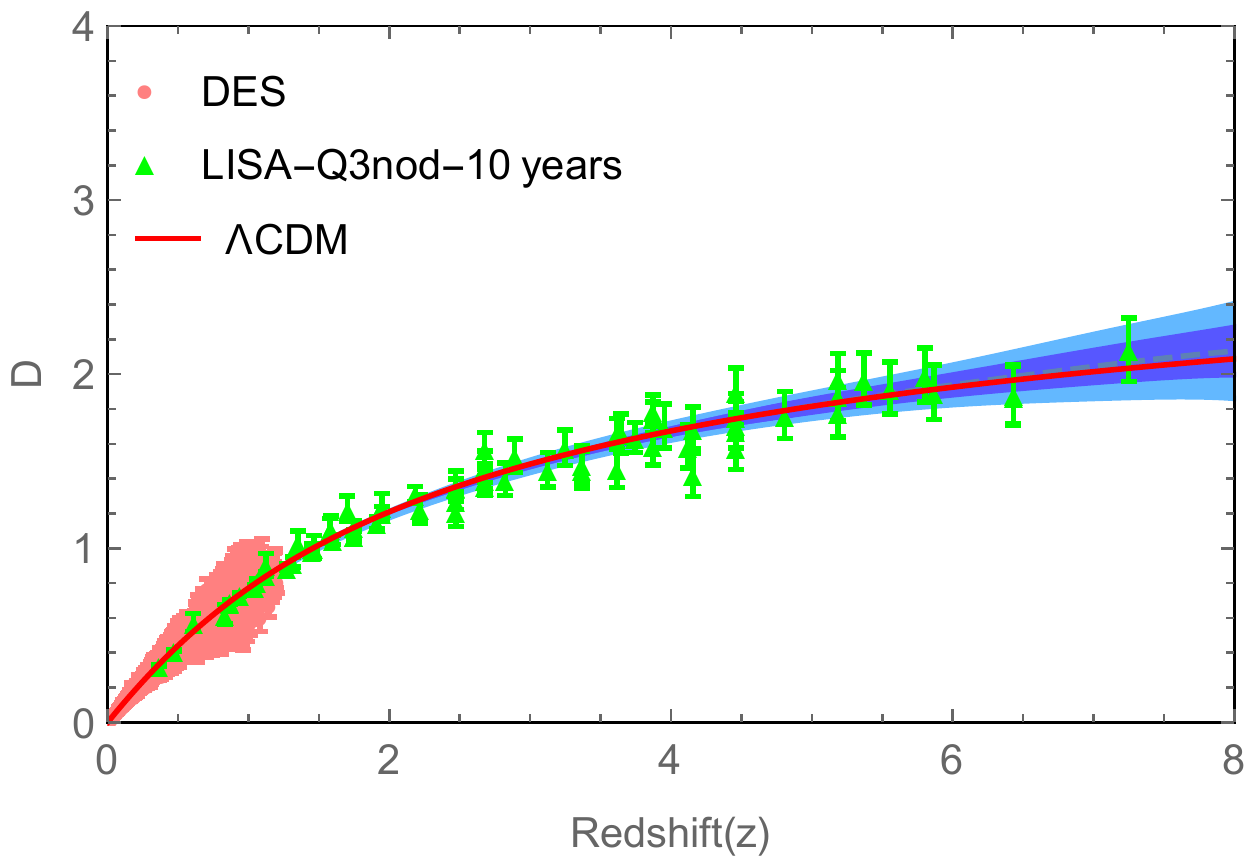}\\
	\caption{Reconstruction of the distance $D(z)$ using DES$+$LISA for a 10 years mission. From left to right each column reports the results for popIII, Q3d, Q3nod. The shaded blue regions are the 68\% and 95\% C.L.~of the reconstruction.}
\label{fig:10yD_LISAplusDES}
\end{figure}

\begin{figure}
	\centering
	\includegraphics[width=.31\textwidth]{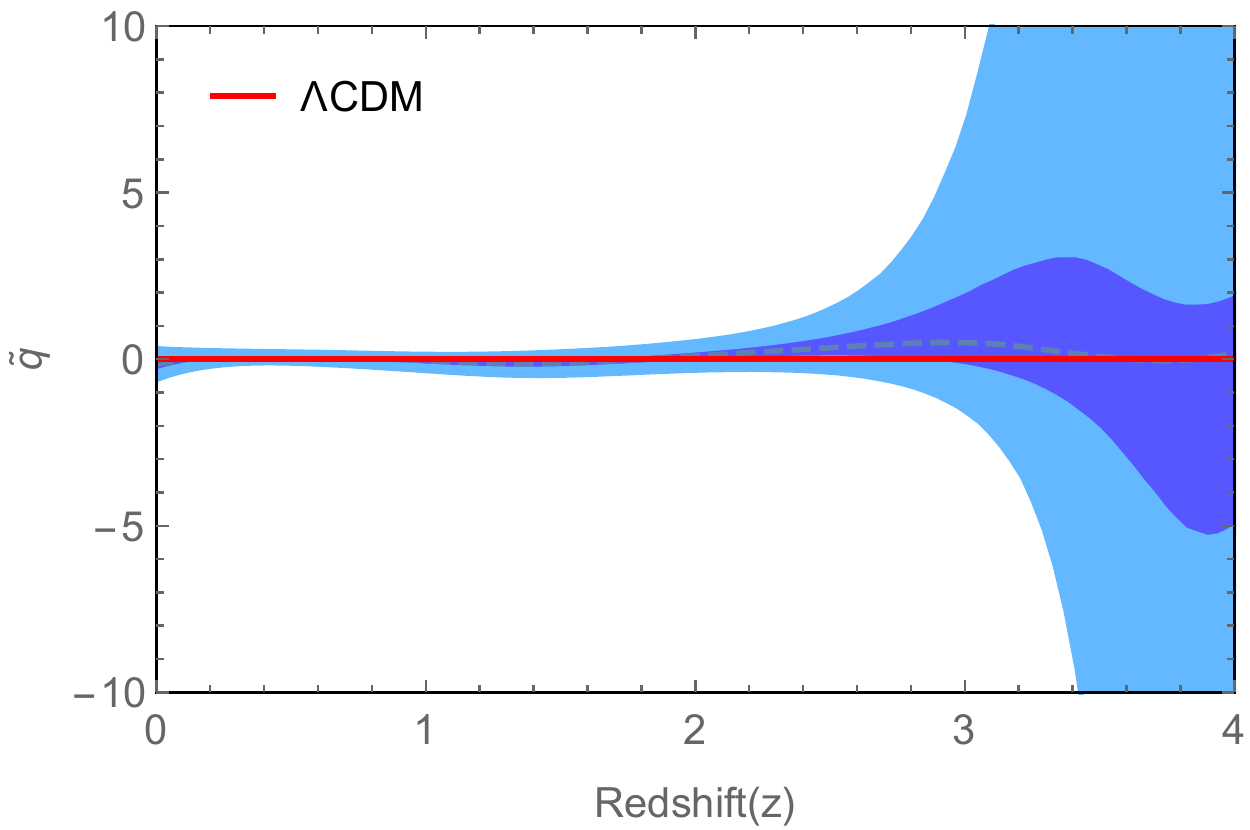}
	\includegraphics[width=.31\textwidth]{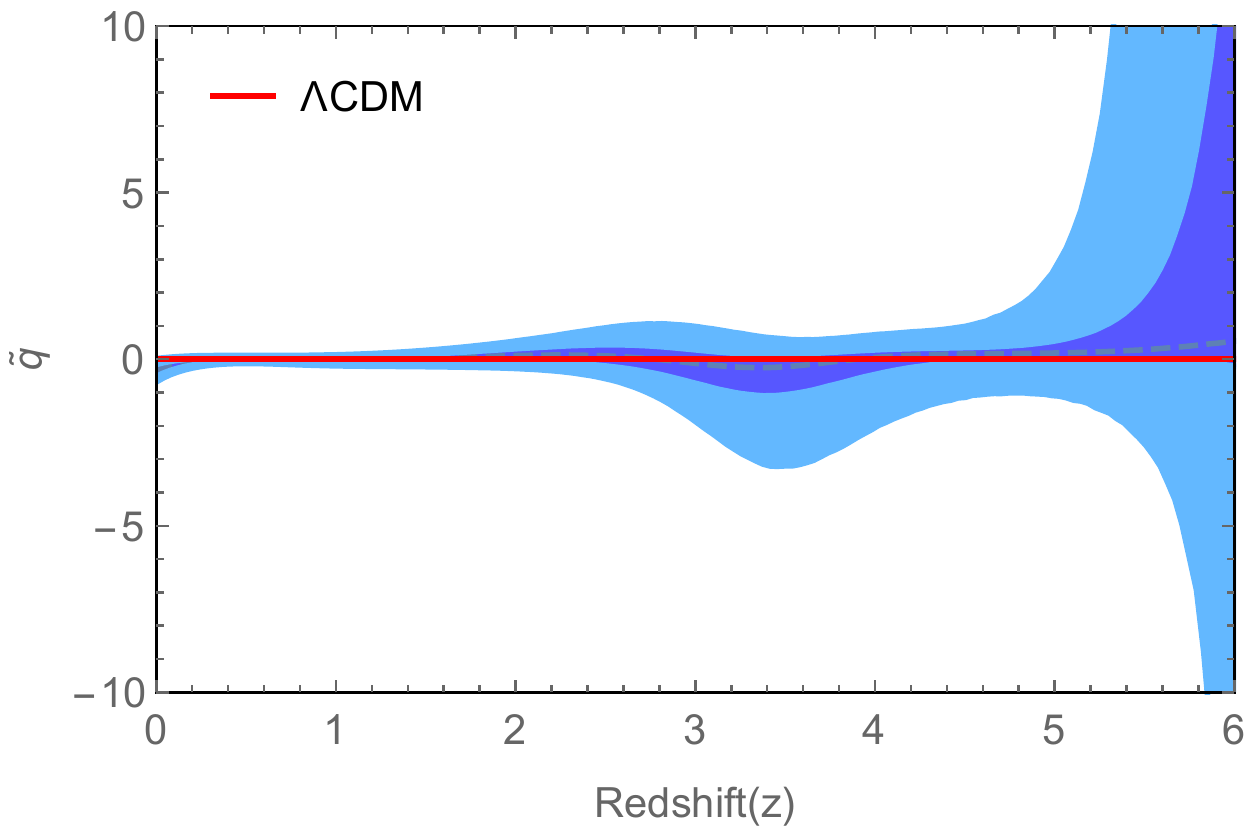}
	\includegraphics[width=.31\textwidth]{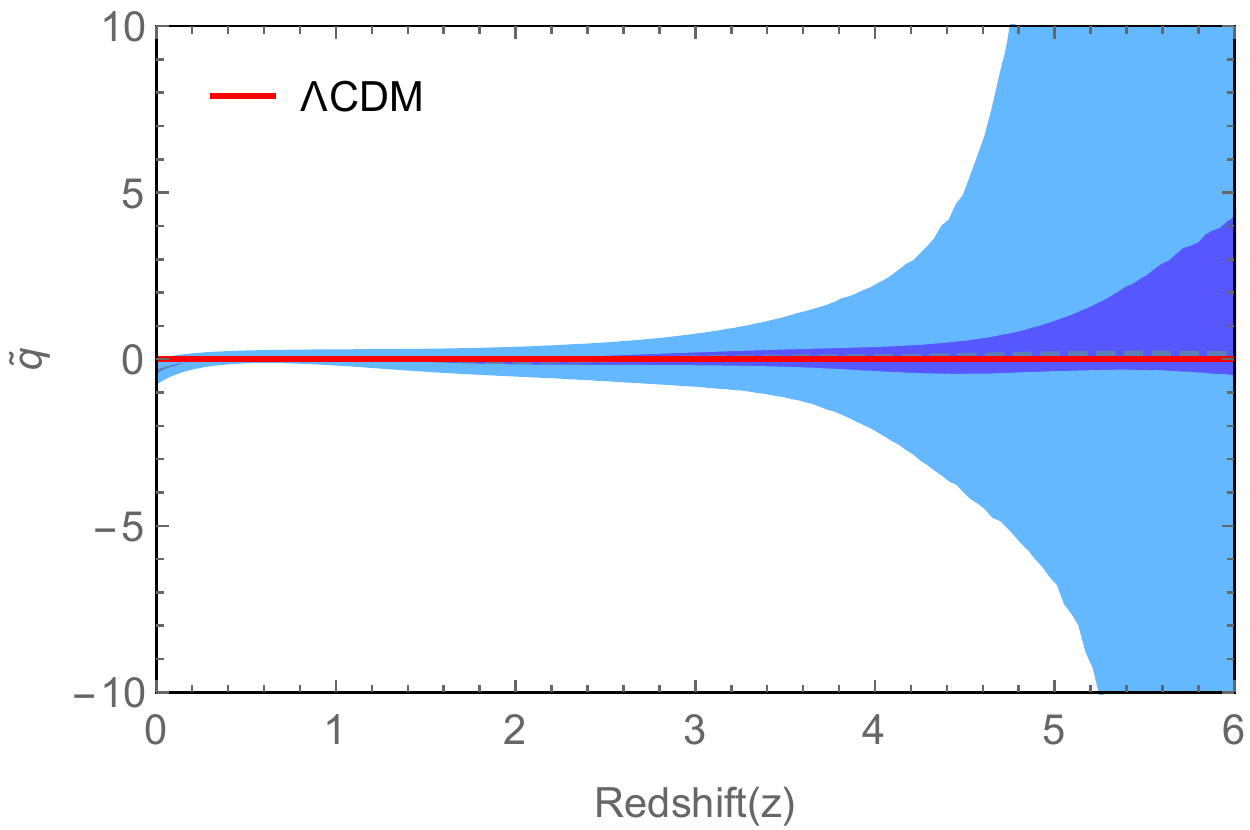}
\caption{Reconstructions of the interaction using DES $+$ LISA for 10 years mission. From left to right each column reports the results for popIII, Q3d, Q3nod.}
\label{fig:10yq_LISAplusDES}
\end{figure}

Figs.~\ref{fig:10yD_LISAplusDES} and~\ref{fig:10yq_LISAplusDES} show the reconstructions of the distance $D(z)$ and the rescaled interacting term $\tilde{q}(z)$ using data from DES plus LISA for a 10 years mission.
The derivatives of $D(z)$ are given in Fig.~\ref{fig:10yDprime_LISAplusDES} after the bibliography.
These two datasets can reconstruct the dark interaction well from redshift $z\sim0$ to more than $z\sim5$, extending the results obtained with a 5 years LISA mission to even higher redshift.
As usual the MBHB scenarios of Q3d and Q3nod are better than popIII in the results.
The results for 5 years and 10 years mission of LISA plus DES show that the combined LISA and DES data can reconstruct the interaction of vacuum DE and DM well in the whole redshift region from $z\sim0$ to $z\sim3$ for a 5 years mission, and from $z\sim0$ to $z\sim5$ for a 10 years mission.
In both cases the low redshift region below $z\simeq 1$ is well covered by the DES datasets, showing that the reconstructions obtained by LISA alone at high redshift (cf.~Sec.~\ref{sub:lisa_mbhb_standard_sirens_alone}) can be integrated with the one obtained by DES data at low redshift.

\section{Discussion and conclusion} 
\label{sec:discussion_and_conclusion}

The main scope of this paper was to repeat and extend the analysis performed in~\cite{Cai:2015zoa} to reconstruct the dark sector interaction, by using simulated standard siren data for LISA.
Taking simulated catalogues from the analysis of~\cite{Tamanini:2016zlh}, we have considered MBHB LISA events whose redshift can be determined through the observation of an EM counterpart.
These data points constrain the distance-redshift relation at high redshift (up to $z\sim 10$) and turn out useful to probe deviations from the standard $\Lambda$CDM dynamics deep into the matter dominated era.
Specifically an interaction in the dark sector can lead to a different expansion rate at high redshift which can be efficiently constrained using LISA MBHB standard siren data, as shown before in~\cite{Caprini:2016qxs} for some specific interacting models.
Following~\cite{Cai:2015zoa} in this paper we have reconstructed the dark sector interaction in a completely model independent way using Gaussian processes.
The analysis has been performed first using LISA datasets alone (based on three different BH formation models) and then combining with simulated data for DES.
In the first case we looked at the potential of LISA to constrain the interaction between DE and DM at higher redshift alone.
In the second case we wanted to find out how the addition of LISA high redshift data improves the results obtained in~\cite{Cai:2015zoa} using only DES simulated data at redshift $z<1.5$.

Using MBHB standard siren alone, the simulated LISA data can reconstruct the interaction well from about $z\sim1$ to $z\sim3$ (for a 5 years mission) and $z\sim4$ or even $z\sim5$ (for a 10 years mission), as shown in Figs.~\ref{fig:5yq_LISA} and \ref{fig:10yq_LISA}.
However the reconstruction is not efficient at redshift lower than $z\sim 1$ as MBHB events are much rare at late times.
When combined with the simulated DES datasets, Gaussian processes can reconstruct the interaction well in the whole redshift region from $z\sim0$ to $z\sim3$ for 5 years LISA and from $z\sim0$ to $z\sim5$ for 10 years LISA, respectively, as shown in Figs.~\ref{fig:5yq_LISAplusDES} and \ref{fig:10yq_LISAplusDES}.
Thus the addition of DES data points, which cover the $0<z<1.5$ range, to the LISA MBHB standard siren catalogues, covering the range $1<z<10$, effectively reconstruct the dark interaction at all accessible redshift values, nicely integrating with each other.
We can also note that in all our results, the MBHB scenarios of Q3d and Q3nod are better than popIII at reconstructing the dark sector interaction, in agreement with the results of~\cite{Caprini:2016qxs}.

\begin{figure}
	\centering
	\includegraphics[width=.48\textwidth]{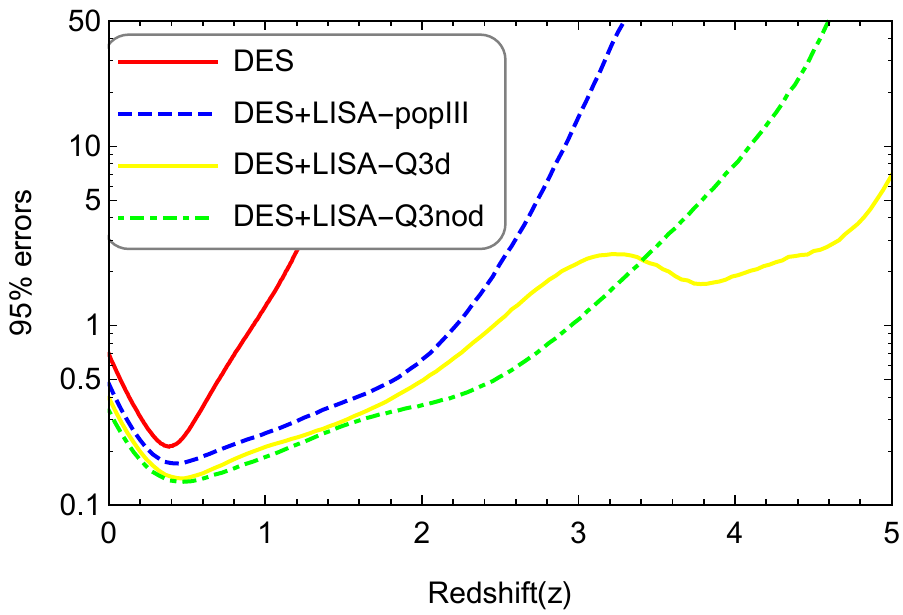}
	\includegraphics[width=.48\textwidth]{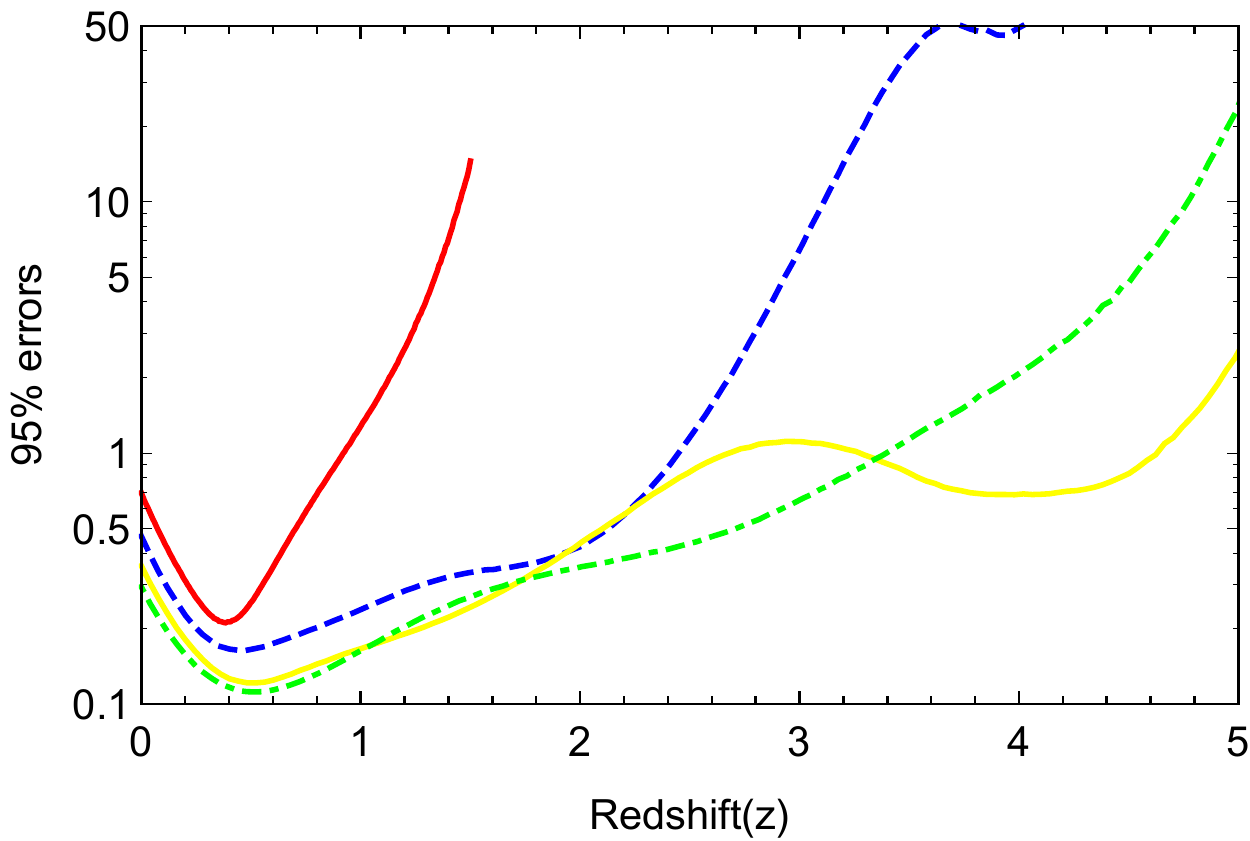}
\caption{95\% confidence errors on $\tilde{q}(z)$ given by DES data alone (red solid), and DES+LISA data for the three different BH formation models considered: popIII (blue dashed), Q3d (yellow solid) and Q3nod (green dot-dashed). The left and right panels report results for a 5 and 10 years LISA mission duration, respectively. The error given by DES data alone quickly diverge after $z\sim 1.5$ due to the absence of SNIa data points at higher redshift.}
\label{fig:95errors}
\end{figure}

In order to understand how MBHB cosmological data will improve the constraints provided by traditional EM signal such as SNIa (standard candles), we can compare the reconstruction uncertainty provided by~\cite{Cai:2015zoa}, i.e.~using DES only data, with the ones obtained by DES+LISA data in Fig.~\ref{fig:5yq_LISAplusDES} and \ref{fig:10yq_LISAplusDES}.
This comparison is presented in Fig.~\ref{fig:95errors} where the 95\% C.L.~errors on $\tilde{q}(z)$ are reported for both DES and DES+LISA, including all three BH formation models.
The reconstruction error from DES quickly diverge after $z\sim 1.5$ due to the absence of SNIa data points at higher redshift.
However when the contribution of LISA MBHB standard siren is added the errors remain small up to $z\sim 3$ (for popIII) or $z\sim 5$ (for Q3), where they reach levels comparable to the value at $z\sim 1.5$ given by DES only.
Finally Fig.~\ref{fig:95errors} shows also that a 10 years LISA mission does not consistently improve the results obtained with a 5 years mission at redshift $z\lesssim 3$, while the errors at high redshift become smaller (up to an order of magnitude) due to the larger number of data points.

We can thus conclude that LISA will extend the ability of SNIa data to constrain the dark sector interaction in a model independent manner up to higher redshift.
These results suggest that GW standard sirens will not only constitute a complementary and alternative way, with respect to familiar EM observations, to probe the cosmic expansion, but will also provide new tests to constrain possible deviations from the standard $\Lambda$CDM dynamics, especially at high redshift.


\subsection*{Acknowledgements} 
\label{sub:acknowledgements}

The authors would like to thank Enrico Barausse, Chiara Caprini, Antoine Klein, Alberto Sesana and Antoine Petiteau for useful discussions and for their original contribution in the production of the LISA cosmological data used in this work.
RGC and TY are supported in part by the National Natural Science Foundation of China Grants No.11690022,  No.11375247, No.11435006, and No. 11647601, and by the Strategic Priority Research Program of CAS Grant No.XDB23030100 and by the Key Research Program of Frontier Sciences of CAS Grant No.QYZDJ-SSWSYS006.
NT acknowledges support from the Labex P2IO and an Enhanced Eurotalents Fellowship.



\bibliographystyle{unsrt}
\bibliography{IDE_biblio}

\begin{thebibliography}{10}

\bibitem{Riess:1998cb}
Adam~G. Riess et~al.
\newblock {Observational evidence from supernovae for an accelerating universe
  and a cosmological constant}.
\newblock {\em Astron. J.}, 116:1009--1038, 1998.
\newblock arXiv:astro-ph/9805201.

\bibitem{Perlmutter:1998np}
S.~Perlmutter et~al.
\newblock {Measurements of Omega and Lambda from 42 high redshift supernovae}.
\newblock {\em Astrophys. J.}, 517:565--586, 1999.
\newblock arXiv:astro-ph/9812133.

\bibitem{Weinberg:1988cp}
Steven Weinberg.
\newblock {The Cosmological Constant Problem}.
\newblock {\em Rev. Mod. Phys.}, 61:1--23, 1989.

\bibitem{Martin:2012bt}
Jerome Martin.
\newblock {Everything You Always Wanted To Know About The Cosmological Constant
  Problem (But Were Afraid To Ask)}.
\newblock {\em Comptes Rendus Physique}, 13:566--665, 2012.

\bibitem{Wetterich:1994bg}
Christof Wetterich.
\newblock {The Cosmon model for an asymptotically vanishing time dependent
  cosmological 'constant'}.
\newblock {\em Astron. Astrophys.}, 301:321--328, 1995.
\newblock arXiv:hep-th/9408025.

\bibitem{Amendola:1999er}
Luca Amendola.
\newblock {Coupled quintessence}.
\newblock {\em Phys. Rev.}, D62:043511, 2000.
\newblock arXiv:astro-ph/9908023.

\bibitem{Li:2014eha}
Yun-He Li, Jing-Fei Zhang, and Xin Zhang.
\newblock {Parametrized Post-Friedmann Framework for Interacting Dark Energy}.
\newblock {\em Phys. Rev.}, D90(6):063005, 2014.
\newblock arXiv:1404.5220.

\bibitem{Faraoni:2014vra}
Valerio Faraoni, James~B. Dent, and Emmanuel~N. Saridakis.
\newblock {Covariantizing the interaction between dark energy and dark matter}.
\newblock {\em Phys. Rev.}, D90(6):063510, 2014.
\newblock arXiv:1405.7288.

\bibitem{Tamanini:2015iia}
Nicola Tamanini.
\newblock {Phenomenological models of dark energy interacting with dark
  matter}.
\newblock {\em Phys. Rev.}, D92(4):043524, 2015.
\newblock arXiv:1504.07397.

\bibitem{Skordis:2015yra}
C.~Skordis, A.~Pourtsidou, and E.~J. Copeland.
\newblock {Parametrized post-Friedmannian framework for interacting dark energy
  theories}.
\newblock {\em Phys. Rev.}, D91(8):083537, 2015.
\newblock arXiv:1502.07297.

\bibitem{Valiviita:2008iv}
Jussi Valiviita, Elisabetta Majerotto, and Roy Maartens.
\newblock {Instability in interacting dark energy and dark matter fluids}.
\newblock {\em JCAP}, 0807:020, 2008.
\newblock arXiv:0804.0232.

\bibitem{He:2008si}
Jian-Hua He, Bin Wang, and Elcio Abdalla.
\newblock {Stability of the curvature perturbation in dark sectors' mutual
  interacting models}.
\newblock {\em Phys. Lett.}, B671:139--145, 2009.
\newblock arXiv:0807.3471.

\bibitem{Salvatelli:2014zta}
Valentina Salvatelli, Najla Said, Marco Bruni, Alessandro Melchiorri, and David
  Wands.
\newblock {Indications of a late-time interaction in the dark sector}.
\newblock {\em Phys. Rev. Lett.}, 113(18):181301, 2014.
\newblock arXiv:1406.7297.

\bibitem{Cai:2015zoa}
Tao Yang, Zong-Kuan Guo, and Rong-Gen Cai.
\newblock {Reconstructing the interaction between dark energy and dark matter
  using Gaussian Processes}.
\newblock {\em Phys. Rev.}, D91(12):123533, 2015.
\newblock arXiv:1505.04443.

\bibitem{Wang:2016lxa}
B.~Wang, E.~Abdalla, F.~Atrio-Barandela, and D.~Pavon.
\newblock {Dark Matter and Dark Energy Interactions: Theoretical Challenges,
  Cosmological Implications and Observational Signatures}.
\newblock {\em Rept. Prog. Phys.}, 79(9):096901, 2016.

\bibitem{Abbott:2016blz}
B.~P. Abbott et~al.
\newblock {Observation of Gravitational Waves from a Binary Black Hole Merger}.
\newblock {\em Phys. Rev. Lett.}, 116(6):061102, 2016.
\newblock arXiv:1602.03837.

\bibitem{Schutz:1986gp}
Bernard~F. Schutz.
\newblock {Determining the Hubble Constant from Gravitational Wave
  Observations}.
\newblock {\em Nature}, 323:310--311, 1986.

\bibitem{Holz:2005df}
Daniel~E. Holz and Scott~A. Hughes.
\newblock {Using gravitational-wave standard sirens}.
\newblock {\em Astrophys. J.}, 629:15--22, 2005.
\newblock arXiv:astro-ph/0504616.

\bibitem{Dalal:2006qt}
Neal Dalal, Daniel~E. Holz, Scott~A. Hughes, and Bhuvnesh Jain.
\newblock {Short grb and binary black hole standard sirens as a probe of dark
  energy}.
\newblock {\em Phys. Rev.}, D74:063006, 2006.
\newblock arXiv:astro-ph/0601275.

\bibitem{MacLeod:2007jd}
Chelsea~L. MacLeod and Craig~J. Hogan.
\newblock {Precision of Hubble constant derived using black hole binary
  absolute distances and statistical redshift information}.
\newblock {\em Phys. Rev.}, D77:043512, 2008.
\newblock arXiv:0712.0618.

\bibitem{Nissanke:2009kt}
Samaya Nissanke, Daniel~E. Holz, Scott~A. Hughes, Neal Dalal, and Jonathan~L.
  Sievers.
\newblock {Exploring short gamma-ray bursts as gravitational-wave standard
  sirens}.
\newblock {\em Astrophys. J.}, 725:496--514, 2010.
\newblock arXiv:0904.1017.

\bibitem{Taylor:2011fs}
Stephen~R. Taylor, Jonathan~R. Gair, and Ilya Mandel.
\newblock {Hubble without the Hubble: Cosmology using advanced
  gravitational-wave detectors alone}.
\newblock {\em Phys. Rev.}, D85:023535, 2012.
\newblock arXiv:1108.5161.

\bibitem{DelPozzo:2011yh}
Walter Del~Pozzo.
\newblock {Inference of the cosmological parameters from gravitational waves:
  application to second generation interferometers}.
\newblock {\em Phys. Rev.}, D86:043011, 2012.

\bibitem{DelPozzo:2015bna}
Walter Del~Pozzo, Tjonnie G.~F. Li, and Chris Messenger.
\newblock {Cosmological inference using only gravitational wave observations of
  binary neutron stars}.
\newblock {\em Phys. Rev.}, D95(4):043502, 2017.
\newblock arXiv:1506.06590.

\bibitem{Sathyaprakash:2009xt}
B.~S. Sathyaprakash, B.~F. Schutz, and C.~Van Den~Broeck.
\newblock {Cosmography with the Einstein Telescope}.
\newblock {\em Class. Quant. Grav.}, 27:215006, 2010.
\newblock arXiv:0906.4151.

\bibitem{Zhao:2010sz}
W.~Zhao, C.~Van Den~Broeck, D.~Baskaran, and T.~G.~F. Li.
\newblock {Determination of Dark Energy by the Einstein Telescope: Comparing
  with CMB, BAO and SNIa Observations}.
\newblock {\em Phys. Rev.}, D83:023005, 2011.
\newblock arXiv:1009.0206.

\bibitem{Cai:2016sby}
Rong-Gen Cai and Tao Yang.
\newblock {Estimating cosmological parameters by the simulated data of
  gravitational waves from the Einstein Telescope}.
\newblock {\em Phys. Rev.}, D95:044024, 2017.
\newblock arXiv:1608.08008.

\bibitem{Petiteau:2011we}
Antoine Petiteau, Stanislav Babak, and Alberto Sesana.
\newblock {Constraining the dark energy equation of state using LISA
  observations of spinning Massive Black Hole binaries}.
\newblock {\em Astrophys. J.}, 732:82, 2011.

\bibitem{Tamanini:2016zlh}
Nicola Tamanini, Chiara Caprini, Enrico Barausse, Alberto Sesana, Antoine
  Klein, and Antoine Petiteau.
\newblock {Science with the space-based interferometer eLISA. III: Probing the
  expansion of the Universe using gravitational wave standard sirens}.
\newblock {\em JCAP}, 1604(04):002, 2016.
\newblock arXiv:1601.07112.

\bibitem{Kyutoku:2016zxn}
Koutarou Kyutoku and Naoki Seto.
\newblock {Gravitational-wave cosmography with eLISA and the Hubble tension}.
\newblock 2016.

\bibitem{Tamanini:2016uin}
Nicola Tamanini.
\newblock {Late time cosmology with LISA: probing the cosmic expansion with
  massive black hole binary mergers as standard sirens}.
\newblock In {\em {11th International LISA Symposium Zurich, Switzerland,
  September 5-9, 2016}}, 2016.

\bibitem{DelPozzo:2017kme}
Walter Del~Pozzo, Alberto Sesana, and Antoine Klein.
\newblock {Stellar binary black holes in the LISA band: a new class of standard
  sirens}.
\newblock 2017.

\bibitem{elisaweb}
{LISA}.
\newblock {\tt www.elisascience.org}.

\bibitem{2017arXiv170200786A}
P.~Amaro-Seoane et~al.
\newblock {Laser Interferometer Space Antenna}.
\newblock arXiv:1702.00786.

\bibitem{Holsclaw:2010nb}
Tracy Holsclaw, Ujjaini Alam, Bruno Sanso, Herbert Lee, Katrin Heitmann, Salman
  Habib, and David Higdon.
\newblock {Nonparametric Reconstruction of the Dark Energy Equation of State}.
\newblock {\em Phys. Rev.}, D82:103502, 2010.
\newblock arXiv:1009.5443.

\bibitem{Holsclaw:2010sk}
Tracy Holsclaw, Ujjaini Alam, Bruno Sanso, Herbert Lee, Katrin Heitmann, Salman
  Habib, and David Higdon.
\newblock {Nonparametric Dark Energy Reconstruction from Supernova Data}.
\newblock {\em Phys. Rev. Lett.}, 105:241302, 2010.
\newblock arXiv:1011.3079.

\bibitem{Holsclaw:2011wi}
Tracy Holsclaw, Ujjaini Alam, Bruno Sanso, Herbie Lee, Katrin Heitmann, Salman
  Habib, and David Higdon.
\newblock {Nonparametric Reconstruction of the Dark Energy Equation of State
  from Diverse Data Sets}.
\newblock {\em Phys. Rev.}, D84:083501, 2011.
\newblock arXiv:1104.2041.

\bibitem{Seikel:2012uu}
Marina Seikel, Chris Clarkson, and Mathew Smith.
\newblock {Reconstruction of dark energy and expansion dynamics using Gaussian
  processes}.
\newblock {\em JCAP}, 1206:036, 2012.
\newblock arXiv:1204.2832.

\bibitem{Caprini:2016qxs}
Chiara Caprini and Nicola Tamanini.
\newblock {Constraining early and interacting dark energy with gravitational
  wave standard sirens: the potential of the eLISA mission}.
\newblock {\em JCAP}, 1610(10):006, 2016.

\bibitem{mymodel}
Enrico Barausse.
\newblock {The evolution of massive black holes and their spins in their
  galactic hosts}.
\newblock {\em Mon.~Not.~Roy.~Astron.~Soc.}, 423:2533--2557, 2012.
\newblock arXiv:1201.5888.

\bibitem{spin_model}
A.~Sesana, E.~Barausse, M.~Dotti, and E.M. Rossi.
\newblock {Linking the spin evolution of massive black holes to galaxy
  kinematics}.
\newblock {\em Astrophys.J.}, 794(2):104, 2014.
\newblock arXiv:1402.7088.

\bibitem{letter}
Fabio Antonini, Enrico Barausse, and Joseph Silk.
\newblock {The imprint of massive black-hole mergers on the correlation between
  nuclear star clusters and their host galaxies}.
\newblock {\em Astrophys. J.}, 806(1):L8, 2015.
\newblock arXiv:1504.04033.

\bibitem{newpaper}
Fabio Antonini, Enrico Barausse, and Joseph Silk.
\newblock {The Coevolution of Nuclear Star Clusters, Massive Black Holes, and
  their Host Galaxies}.
\newblock {\em Astrophys. J.}, 812(1):72, 2015.
\newblock arXiv:1506.02050.

\bibitem{Klein:2015hvg}
Antoine Klein et~al.
\newblock {Science with the space-based interferometer eLISA: Supermassive
  black hole binaries}.
\newblock {\em Phys. Rev.}, D93(2):024003, 2016.
\newblock arXiv:1511.05581.

\bibitem{Armano:2016bkm}
M.~Armano et~al.
\newblock {Sub-Femto- g Free Fall for Space-Based Gravitational Wave
  Observatories: LISA Pathfinder Results}.
\newblock {\em Phys. Rev. Lett.}, 116(23):231101, 2016.

\bibitem{Ade:2015rim}
P.~A.~R. Ade et~al.
\newblock {Planck 2015 results. XIV. Dark energy and modified gravity}.
\newblock 2015.
\newblock arXiv:1502.01590.

\bibitem{Costa:2016tpb}
Andre~A. Costa, Xiao-Dong Xu, Bin Wang, and E.~Abdalla.
\newblock {Constraints on interacting dark energy models from Planck 2015 and
  redshift-space distortion data}.
\newblock 2016.
\newblock arXiv:1605.04138.

\bibitem{Nunes:2016dlj}
Rafael~C. Nunes, Supriya Pan, and Emmanuel~N. Saridakis.
\newblock {New constraints on interacting dark energy from cosmic
  chronometers}.
\newblock {\em Phys. Rev.}, D94(2):023508, 2016.
\newblock arXiv:1605.01712.

\bibitem{Li:2015vla}
Yun-He Li, Jing-Fei Zhang, and Xin Zhang.
\newblock {Testing models of vacuum energy interacting with cold dark matter}.
\newblock {\em Phys. Rev.}, D93(2):023002, 2016.
\newblock arXiv:1506.06349.

\bibitem{Sharov:2017iue}
German~S. Sharov, Subhra Bhattacharya, Supriya Pan, Rafael~C. Nunes, and
  Subenoy Chakraborty.
\newblock {A new interacting two fluid model and its consequences}.
\newblock {\em Mon. Not. Roy. Astron. Soc.}, 466:3497--3506, 2017.

\bibitem{Seikel:2013fda}
Marina Seikel and Chris Clarkson.
\newblock {Optimising Gaussian processes for reconstructing dark energy
  dynamics from supernovae}.
\newblock 2013.
\newblock arXiv:1311.6678.

\bibitem{Cai:2015pia}
Rong-Gen Cai, Zong-Kuan Guo, and Tao Yang.
\newblock {Null test of the cosmic curvature using $H(z)$ and supernovae data}.
\newblock {\em Phys. Rev.}, D93(4):043517, 2016.
\newblock arXiv:1509.06283.

\bibitem{Cai:2016vmn}
Rong-Gen Cai, Zong-Kuan Guo, and Tao Yang.
\newblock {Dodging the cosmic curvature to probe the constancy of the speed of
  light}.
\newblock {\em JCAP}, 1608(08):016, 2016.
\newblock arXiv:1601.05497.

\bibitem{Yahya:2013xma}
Sahba Yahya, Marina Seikel, Chris Clarkson, Roy Maartens, and Mathew Smith.
\newblock {Null tests of the cosmological constant using supernovae}.
\newblock {\em Phys. Rev.}, D89(2):023503, 2014.

\bibitem{Bernstein:2011zf}
J.~P. Bernstein et~al.
\newblock {Supernova Simulations and Strategies For the Dark Energy Survey}.
\newblock {\em Astrophys. J.}, 753:152, 2012.
\newblock arXiv:1111.1969.

\end{thebibliography}

\newpage

\begin{figure}
	\centering
	\includegraphics[width=.31\textwidth]{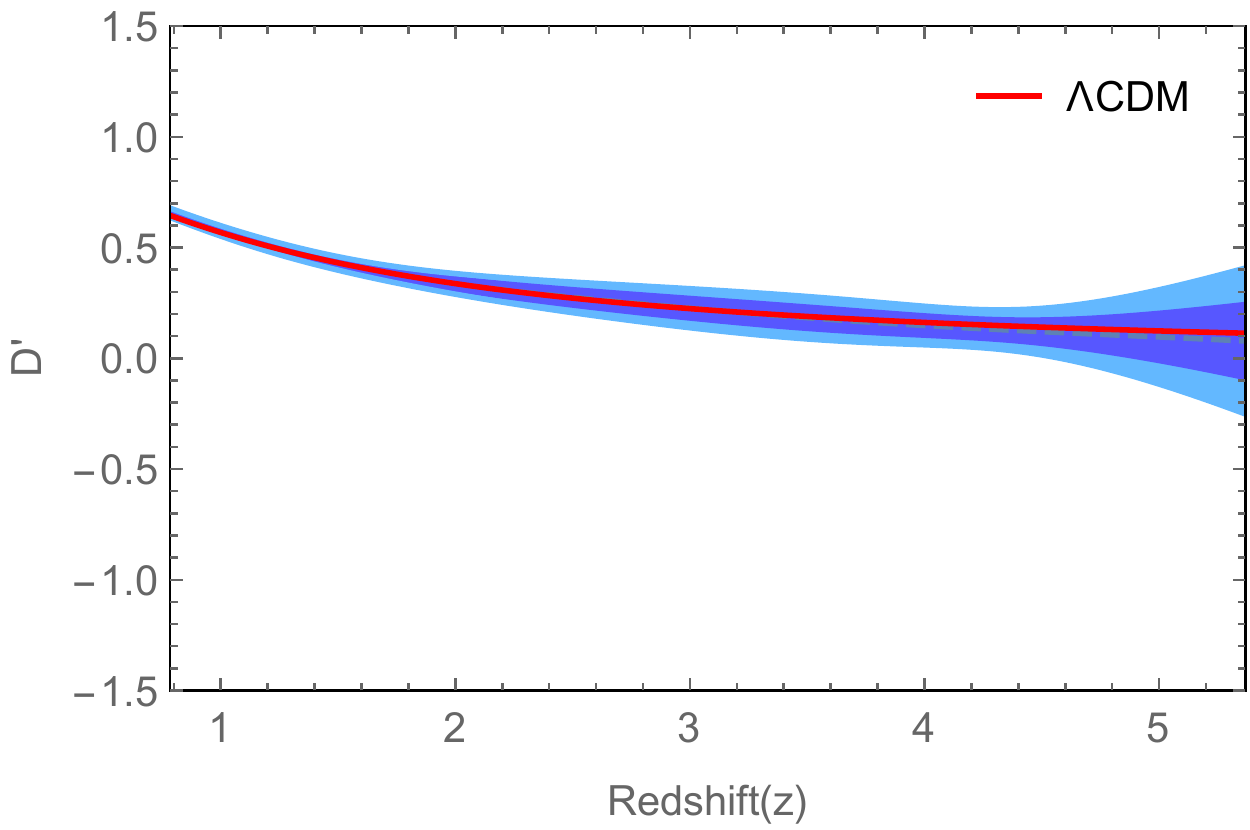}
	\includegraphics[width=.31\textwidth]{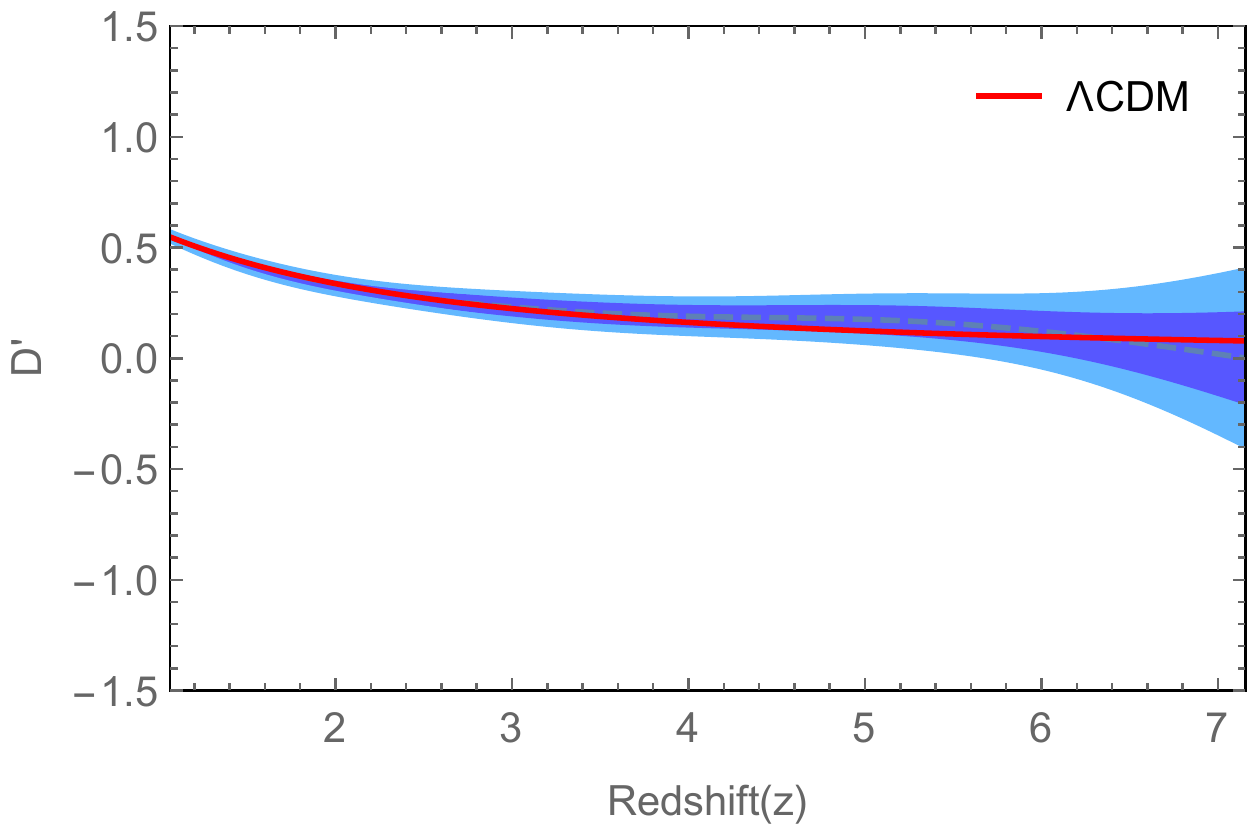}
	\includegraphics[width=.31\textwidth]{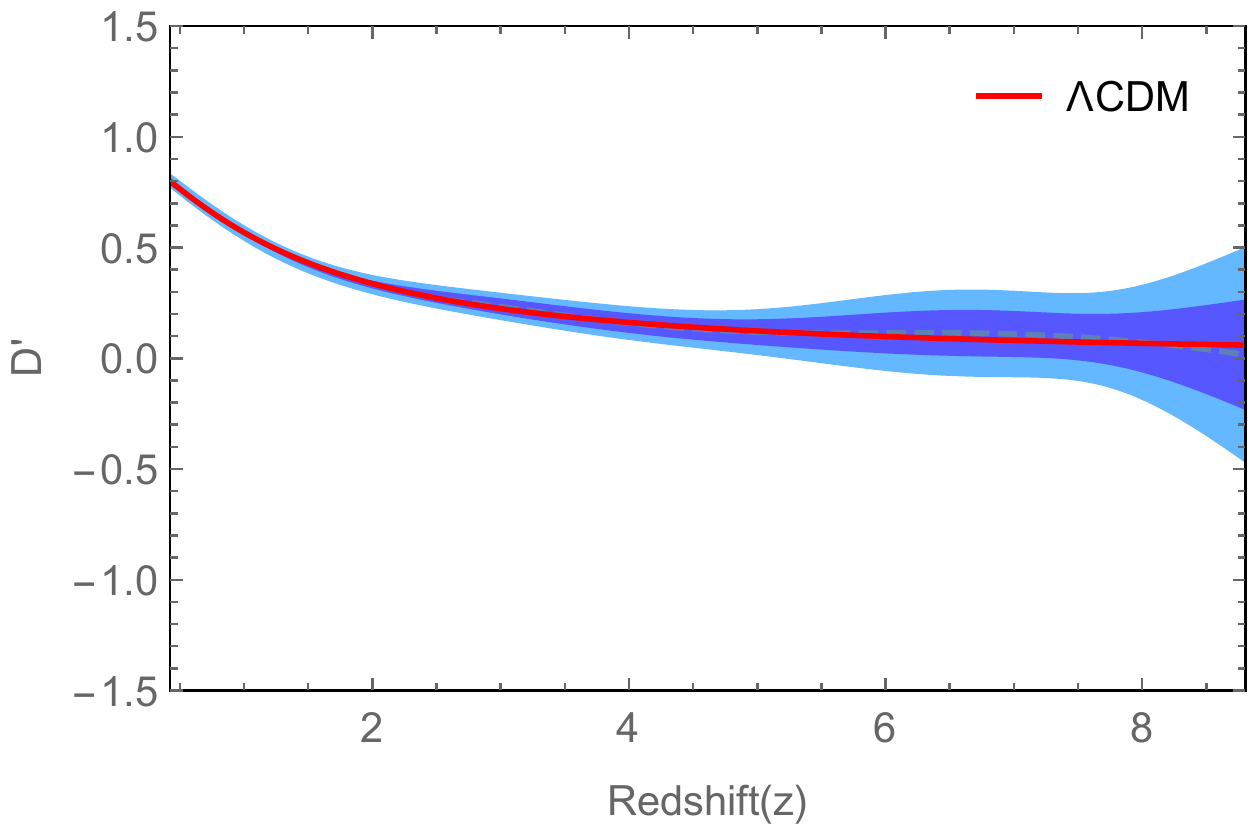}\\
	\includegraphics[width=.31\textwidth]{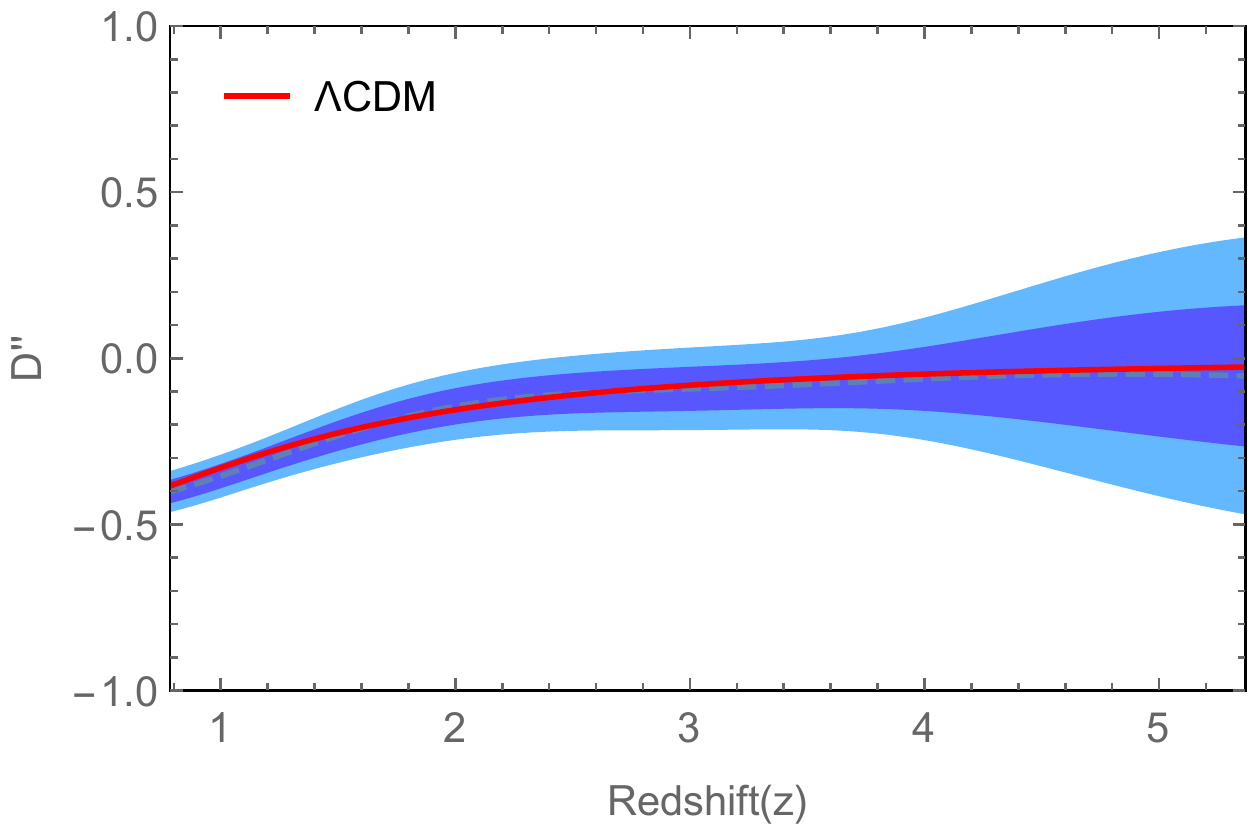}
	\includegraphics[width=.31\textwidth]{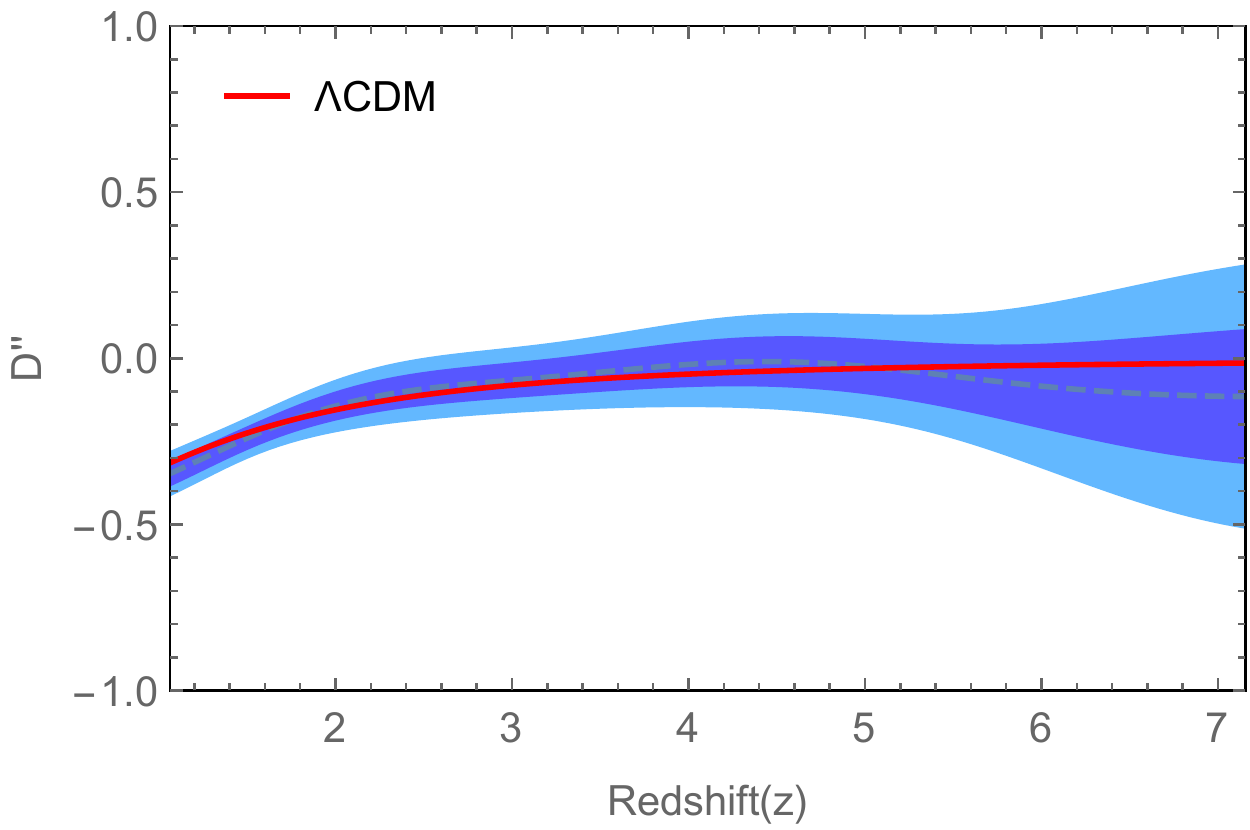}
	\includegraphics[width=.31\textwidth]{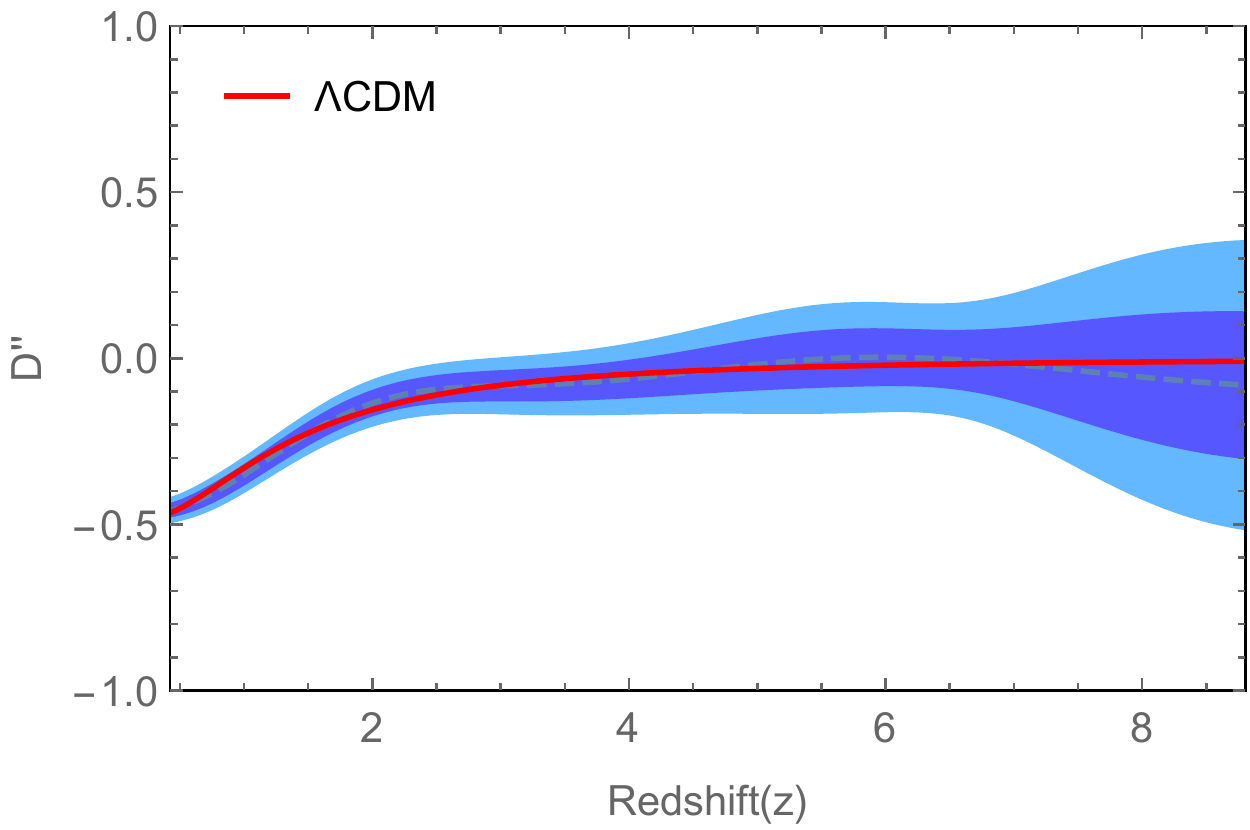}\\
	\includegraphics[width=.31\textwidth]{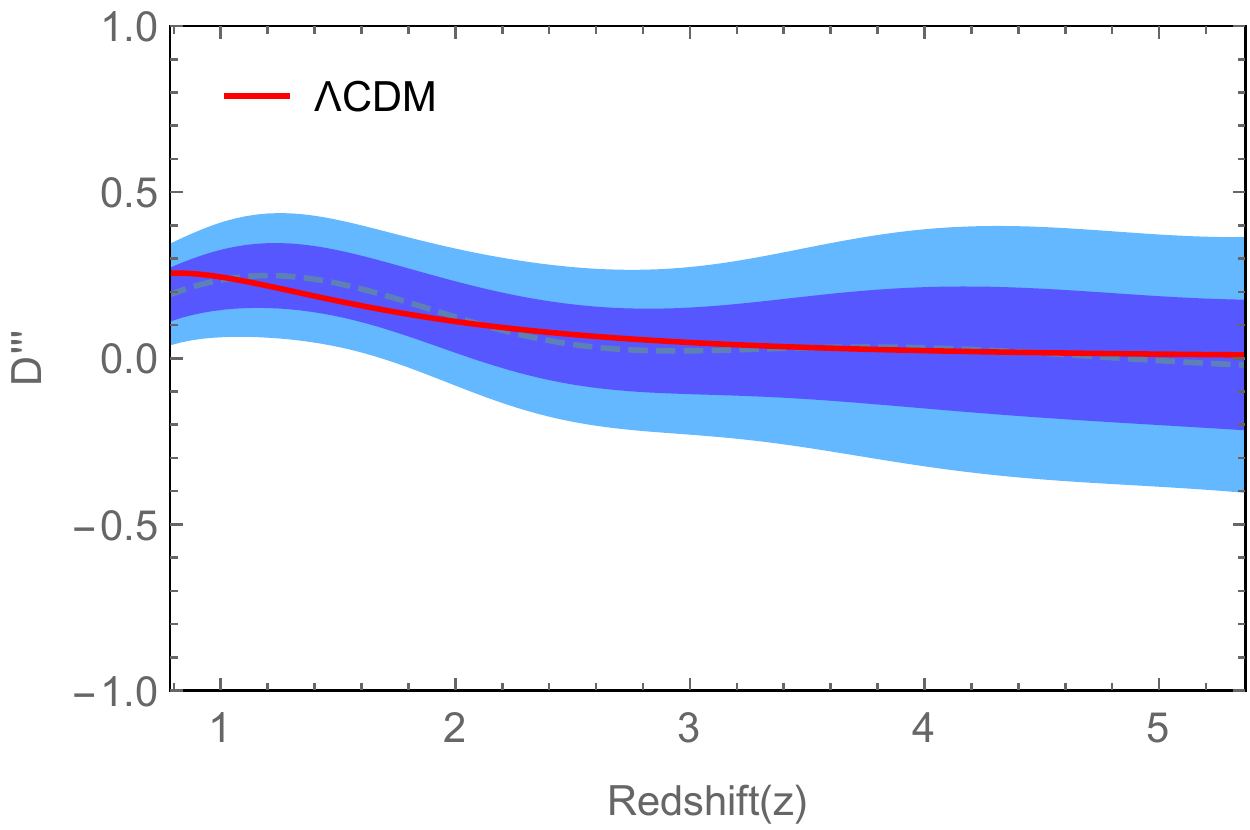}
	\includegraphics[width=.31\textwidth]{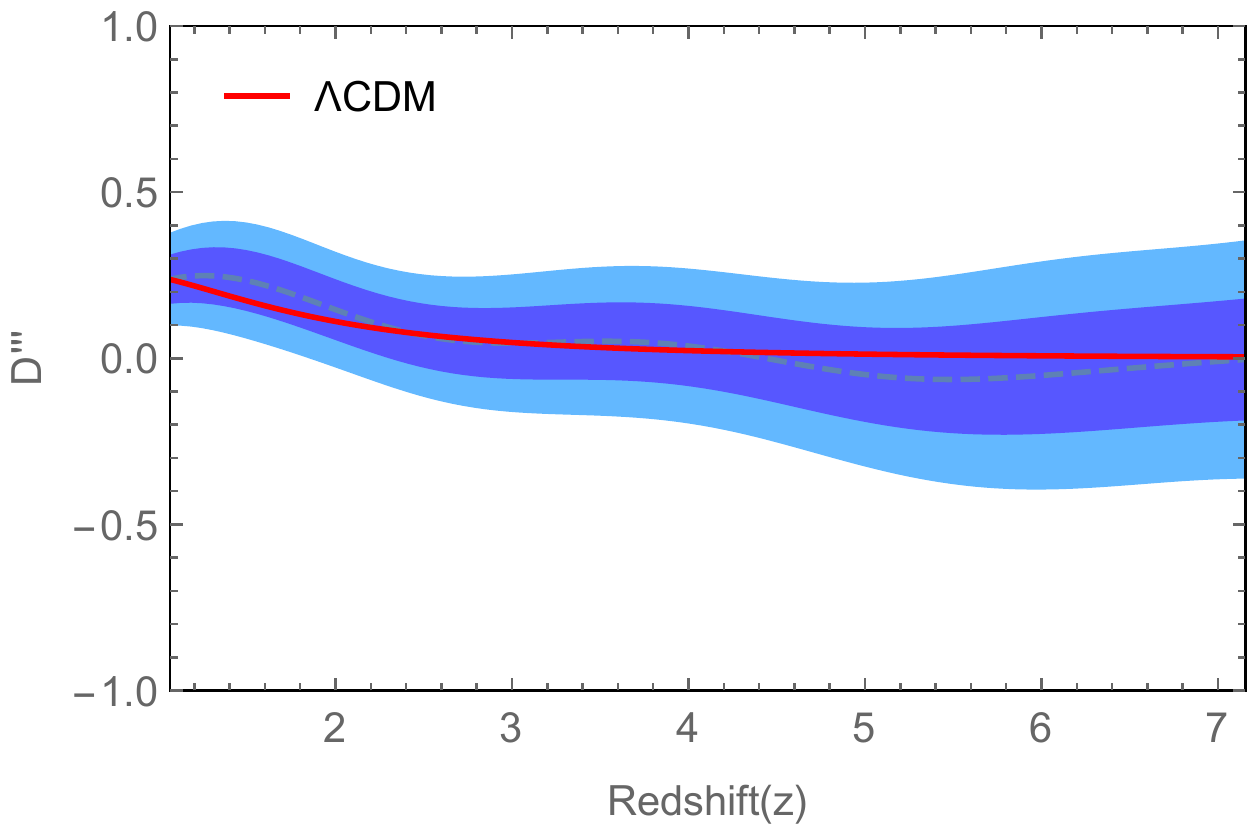}
	\includegraphics[width=.31\textwidth]{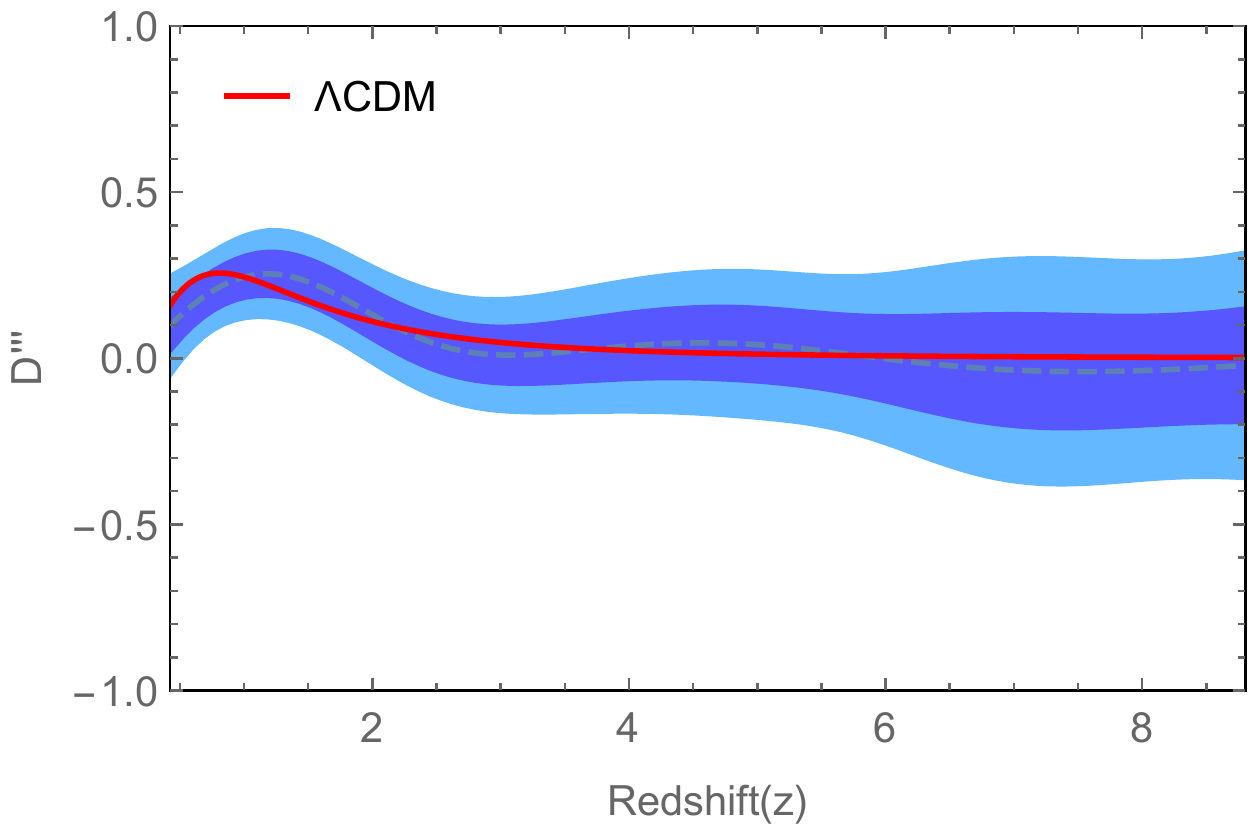}
	\caption{Reconstructions of the derivatives of the distance $D(z)$ using LISA alone for a 5 years mission. From left to right each column reports the results for popIII, Q3d, Q3nod. The shaded blue regions are the 68\% and 95\% C.L.~of the reconstruction.}
\label{fig:5yDprime_LISA}
\end{figure}

\begin{figure}
	\centering
	\includegraphics[width=.31\textwidth]{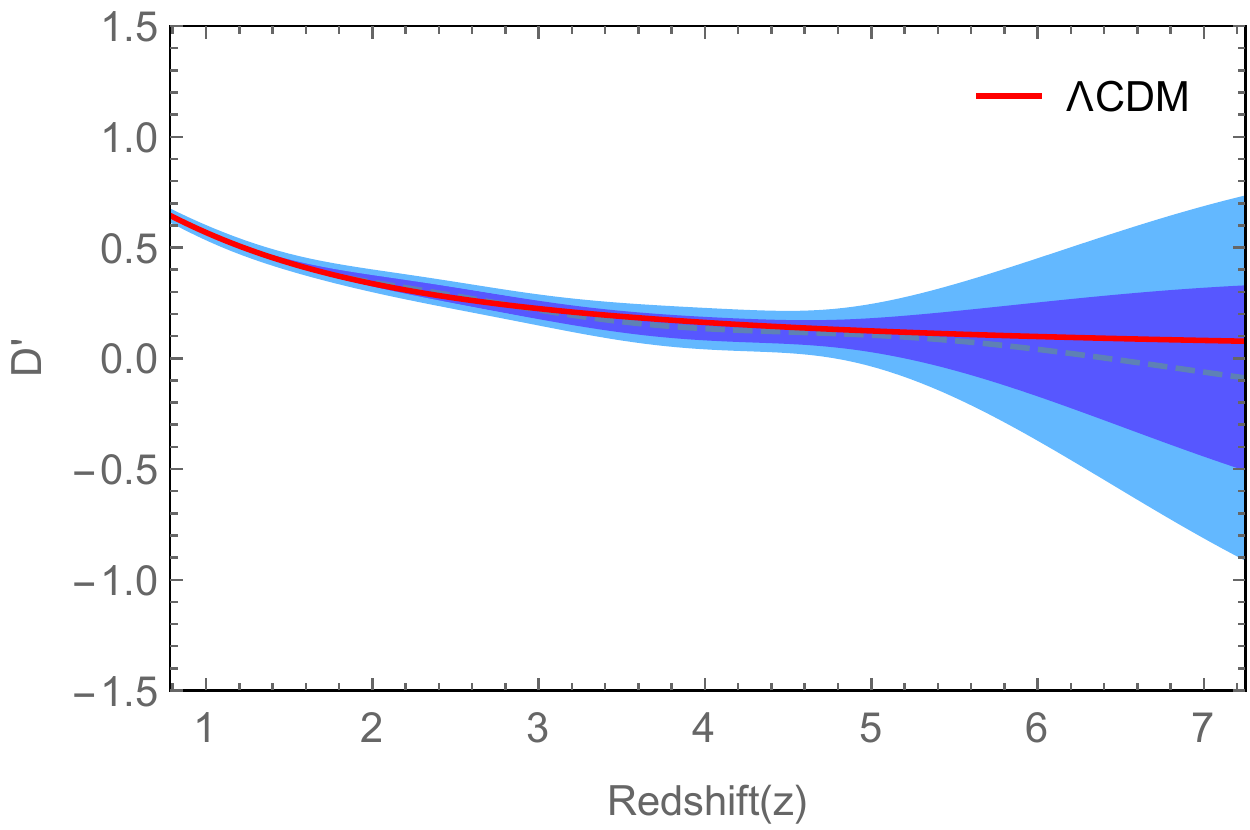}
	\includegraphics[width=.31\textwidth]{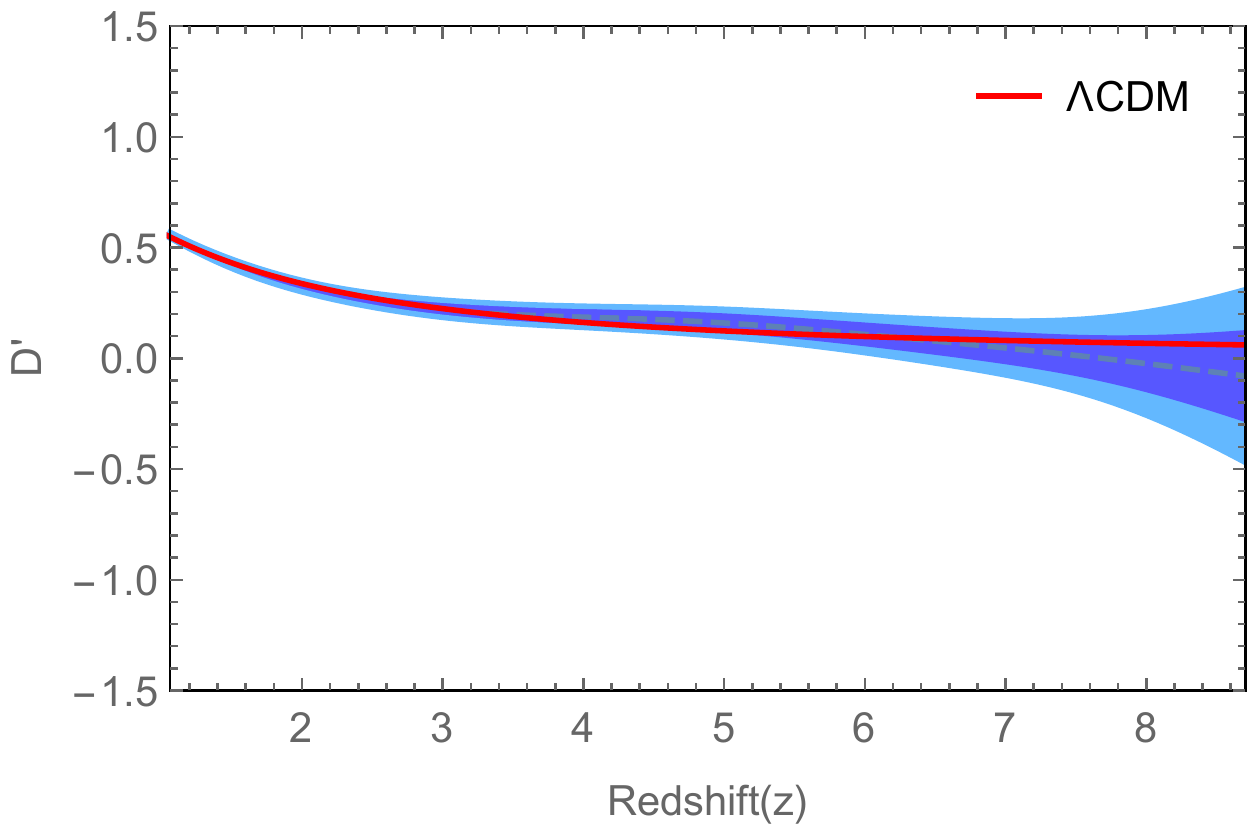}
	\includegraphics[width=.31\textwidth]{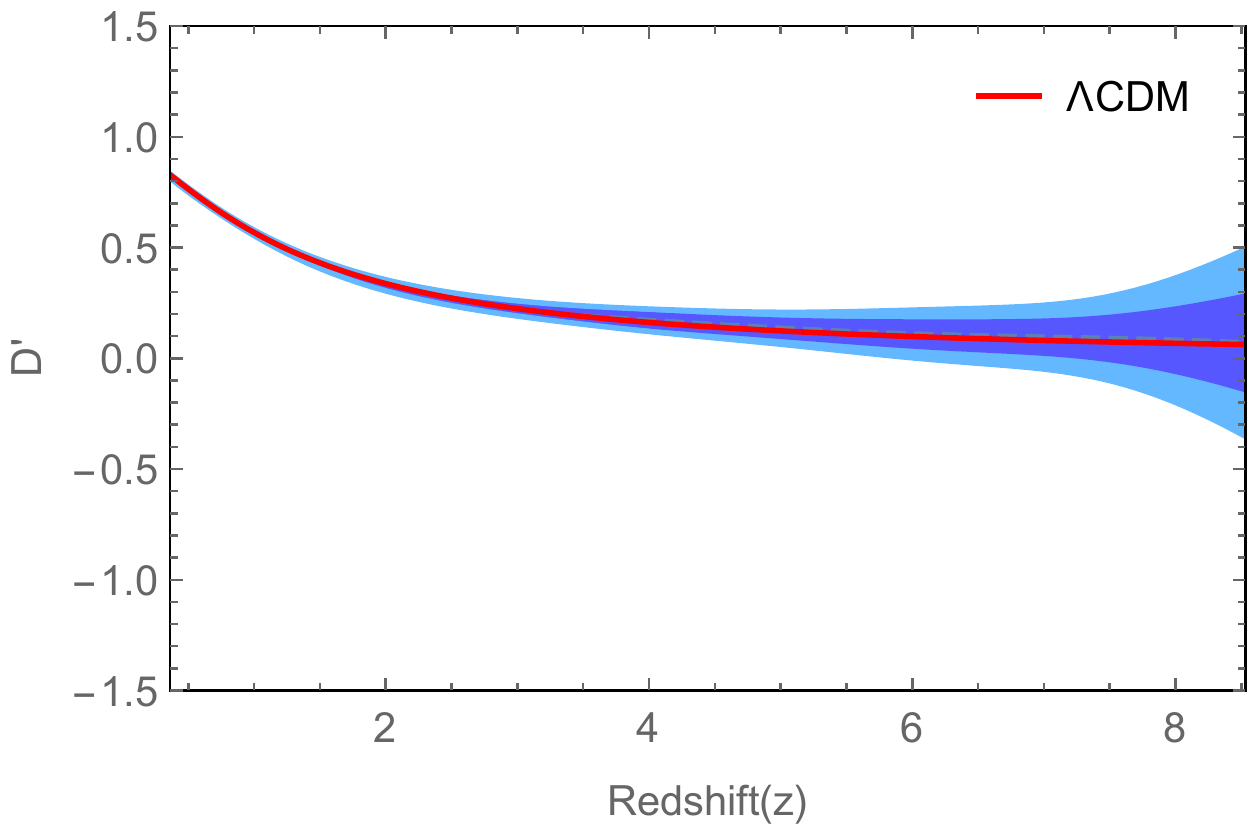}\\
	\includegraphics[width=.31\textwidth]{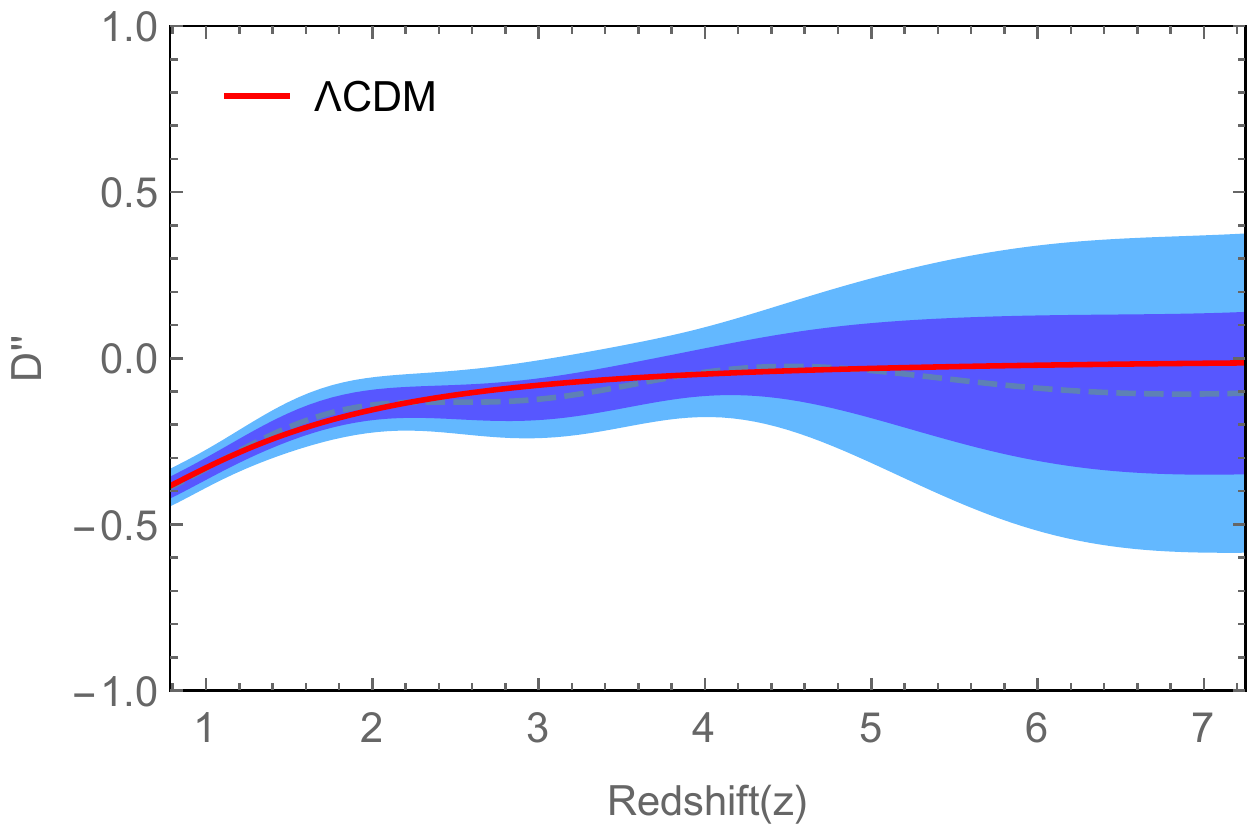}
	\includegraphics[width=.31\textwidth]{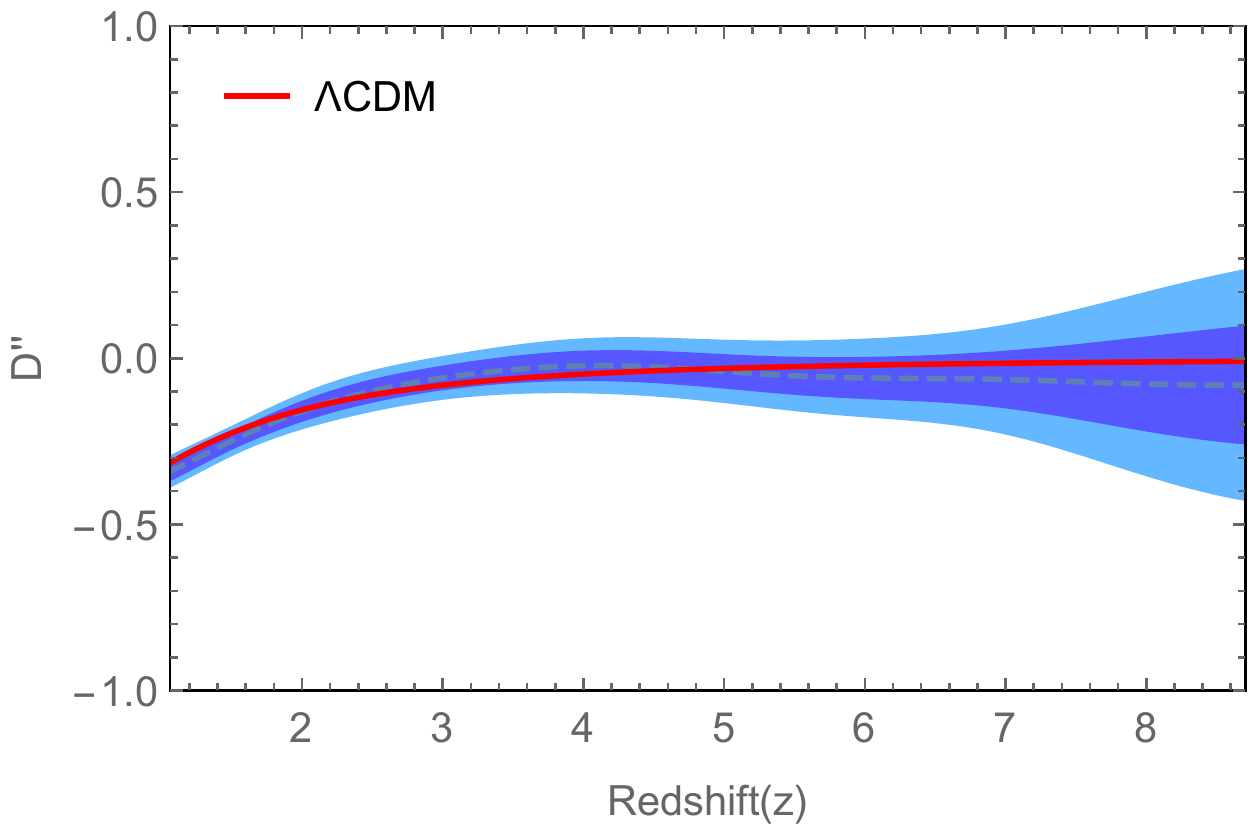}
	\includegraphics[width=.31\textwidth]{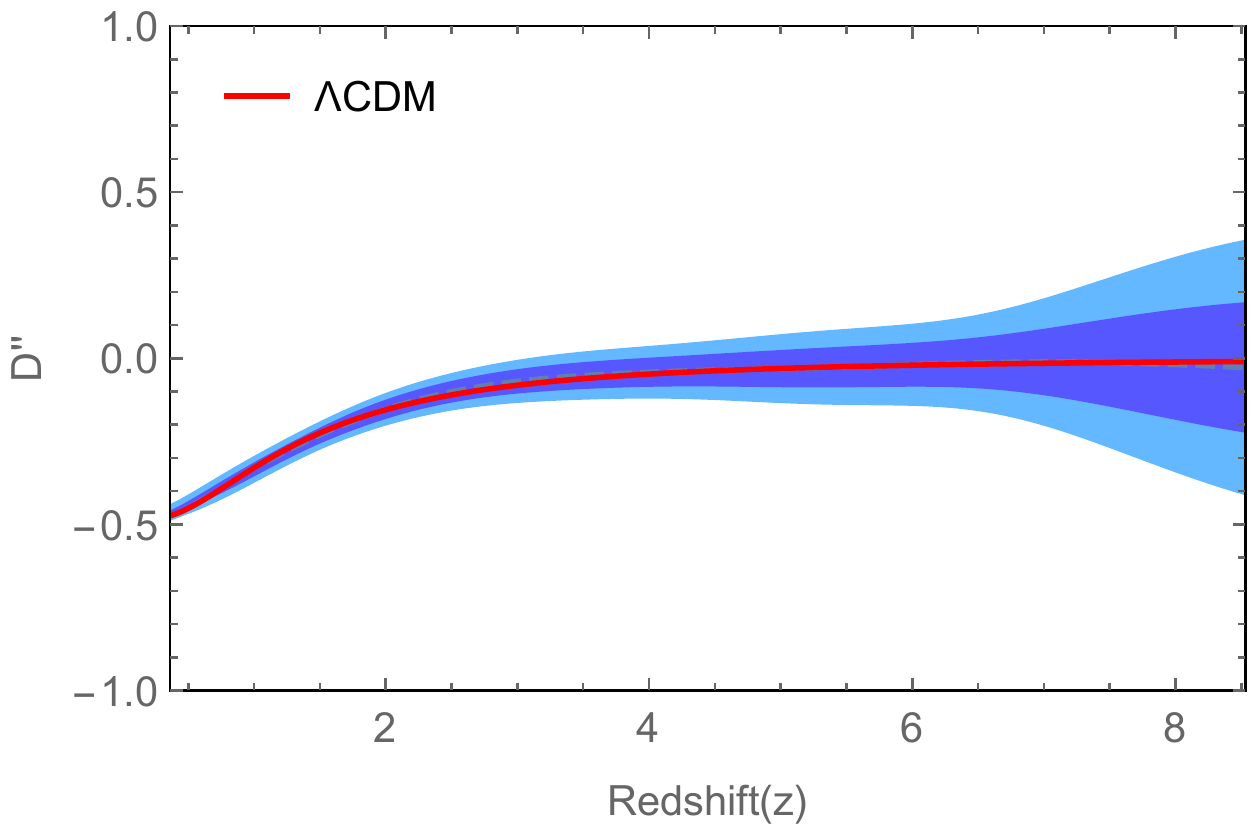}\\
	\includegraphics[width=.31\textwidth]{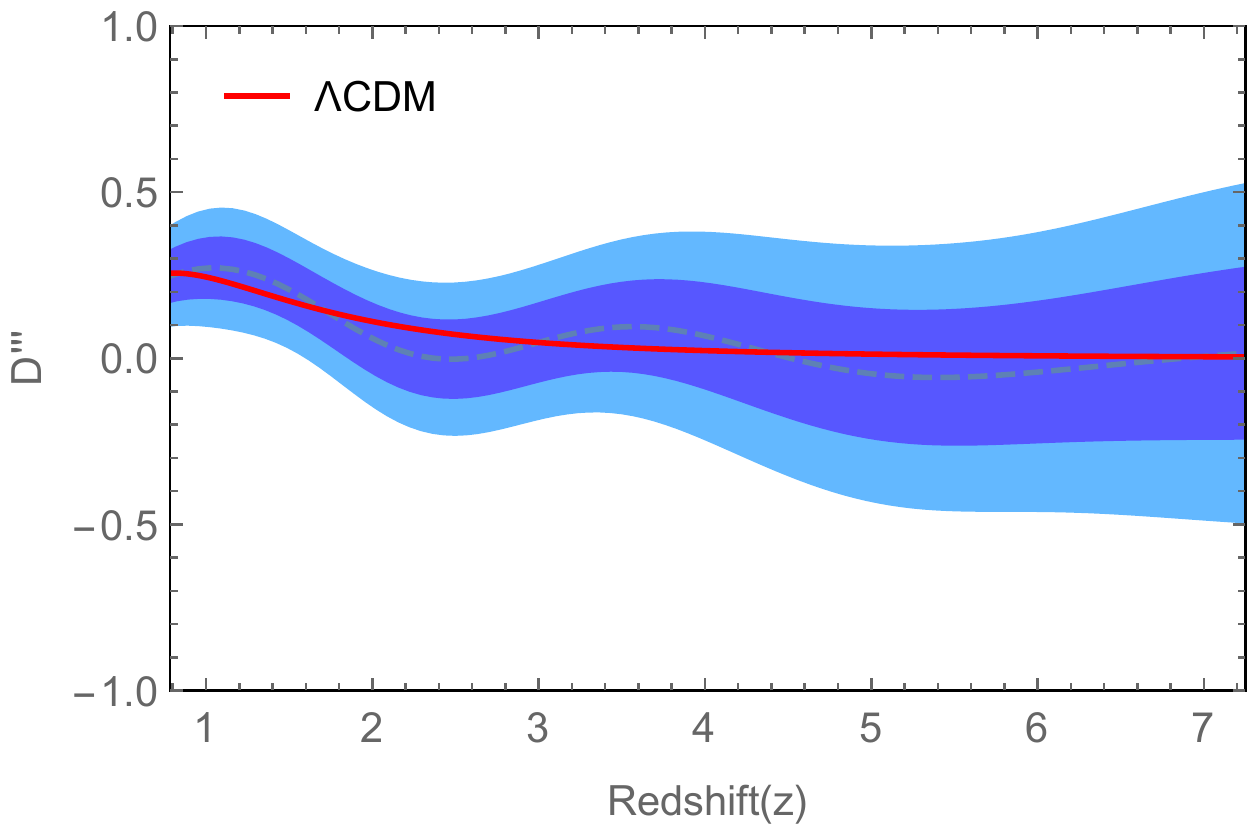}
	\includegraphics[width=.31\textwidth]{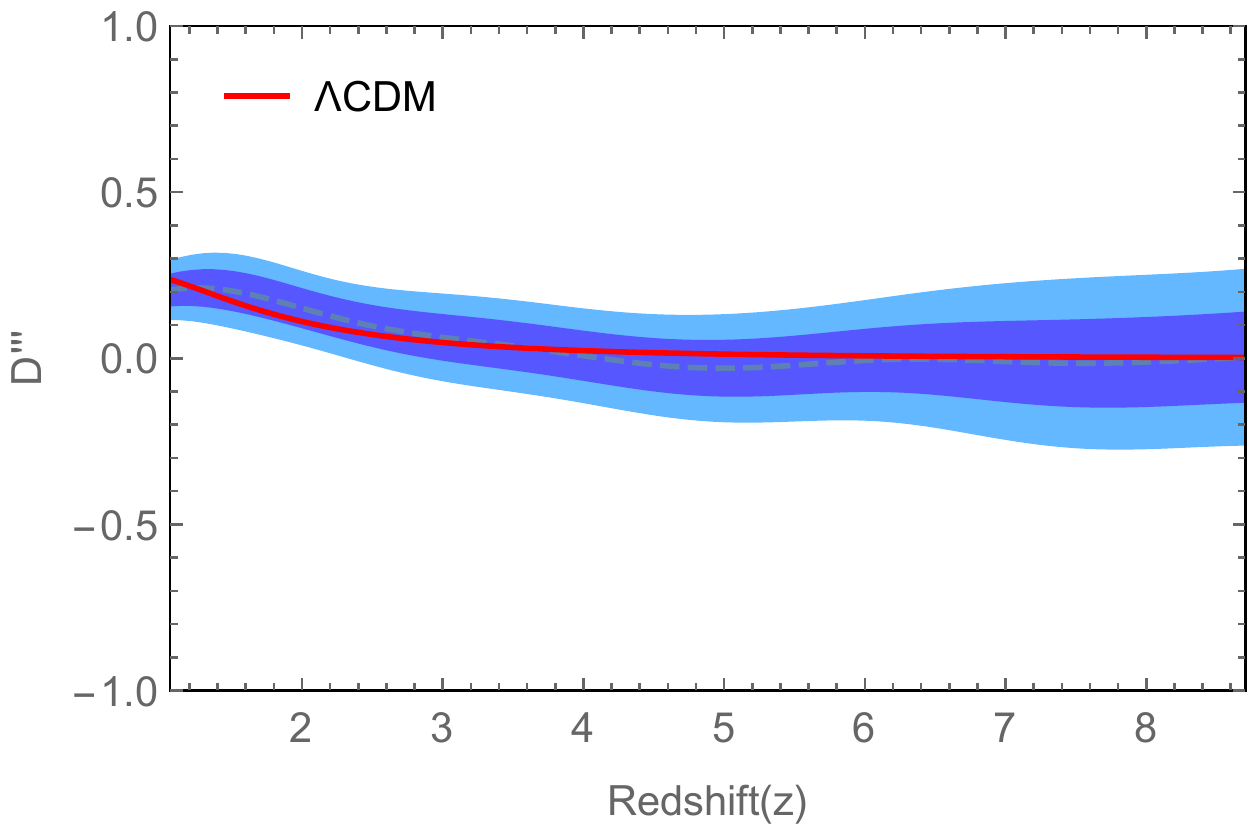}
	\includegraphics[width=.31\textwidth]{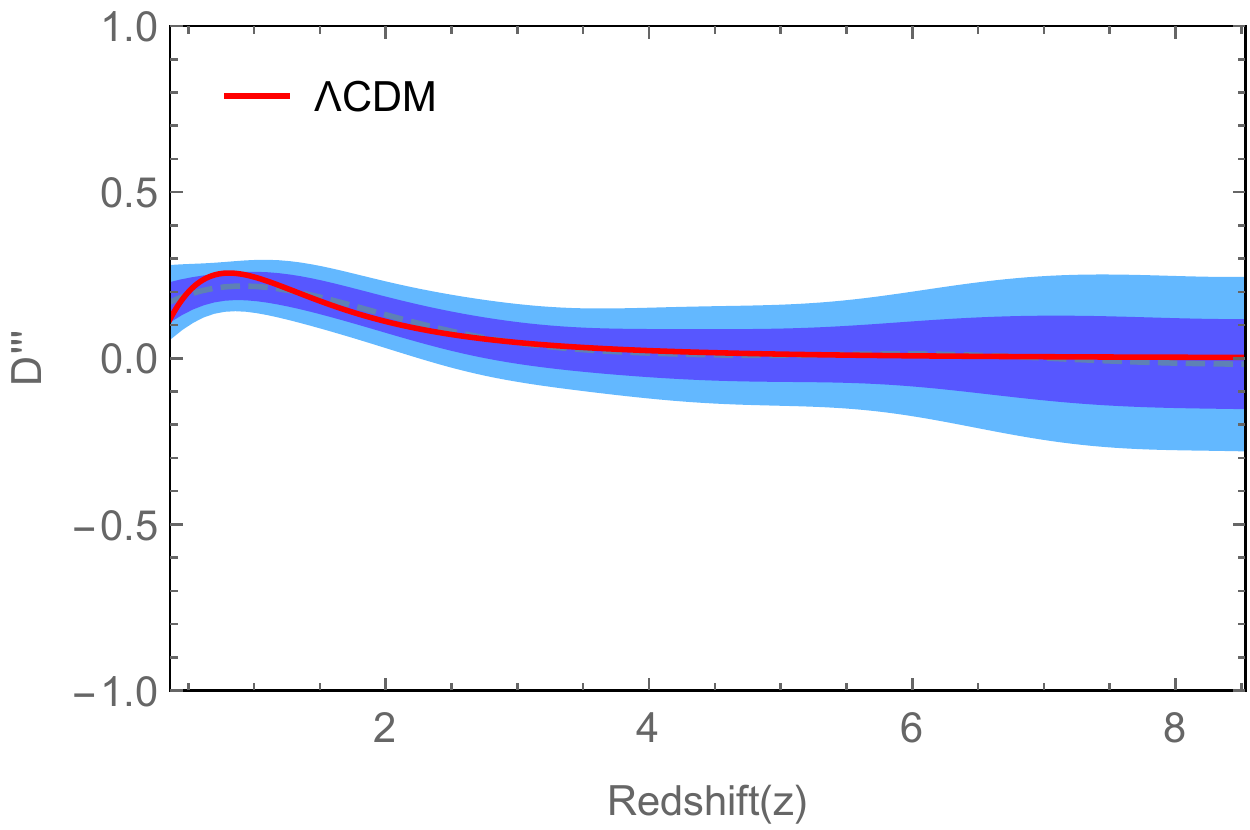}
	\caption{Reconstructions of the derivatives of the distance $D(z)$ using LISA (10 years) alone.
	}
\label{fig:10yDprime_LISA}
\end{figure}

\begin{figure}
	\centering
	\includegraphics[width=.31\textwidth]{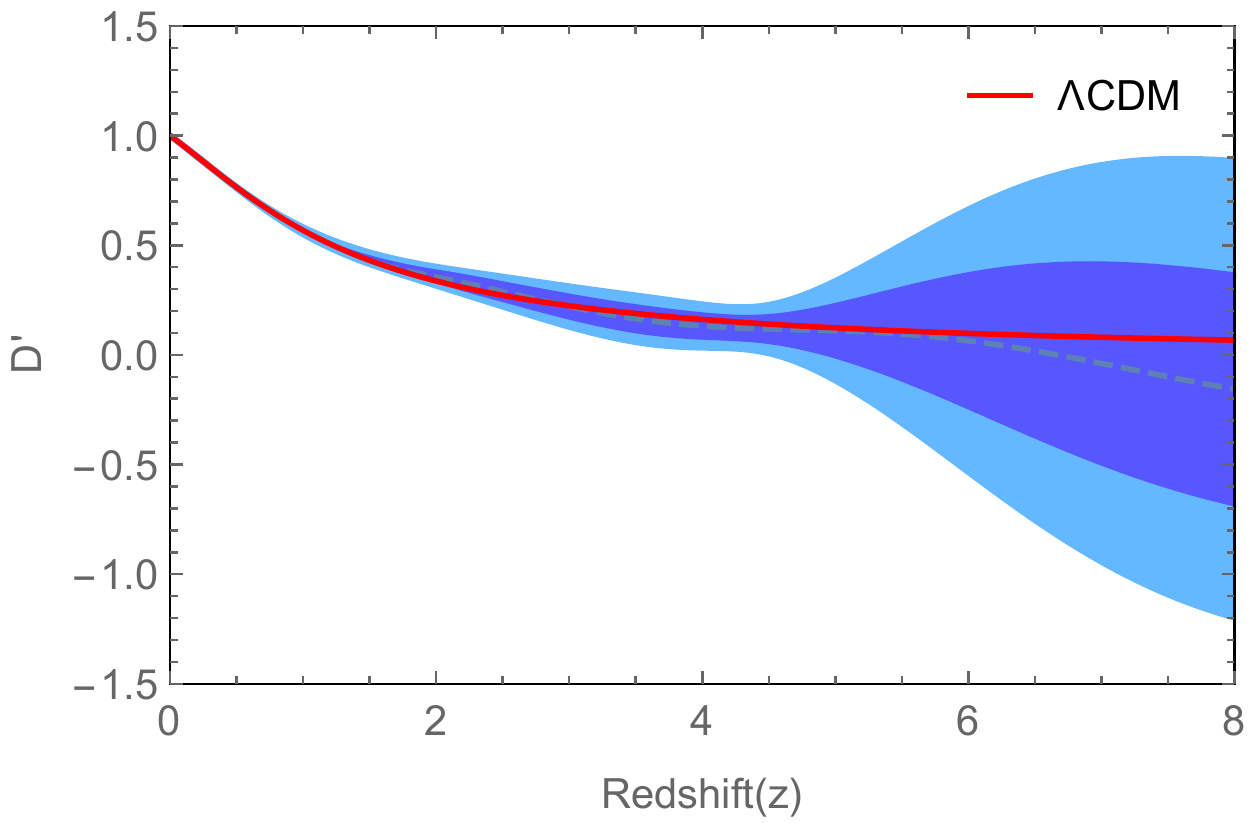}
	\includegraphics[width=.31\textwidth]{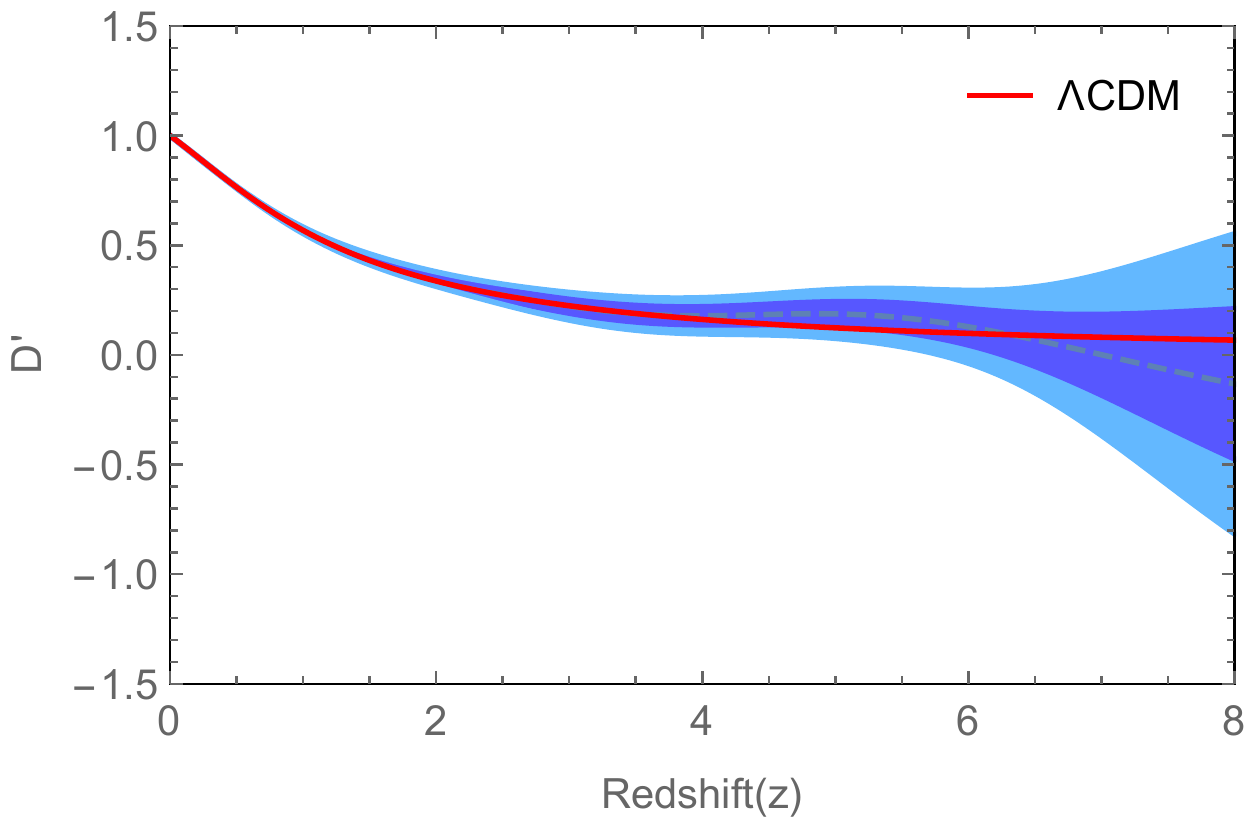}
	\includegraphics[width=.31\textwidth]{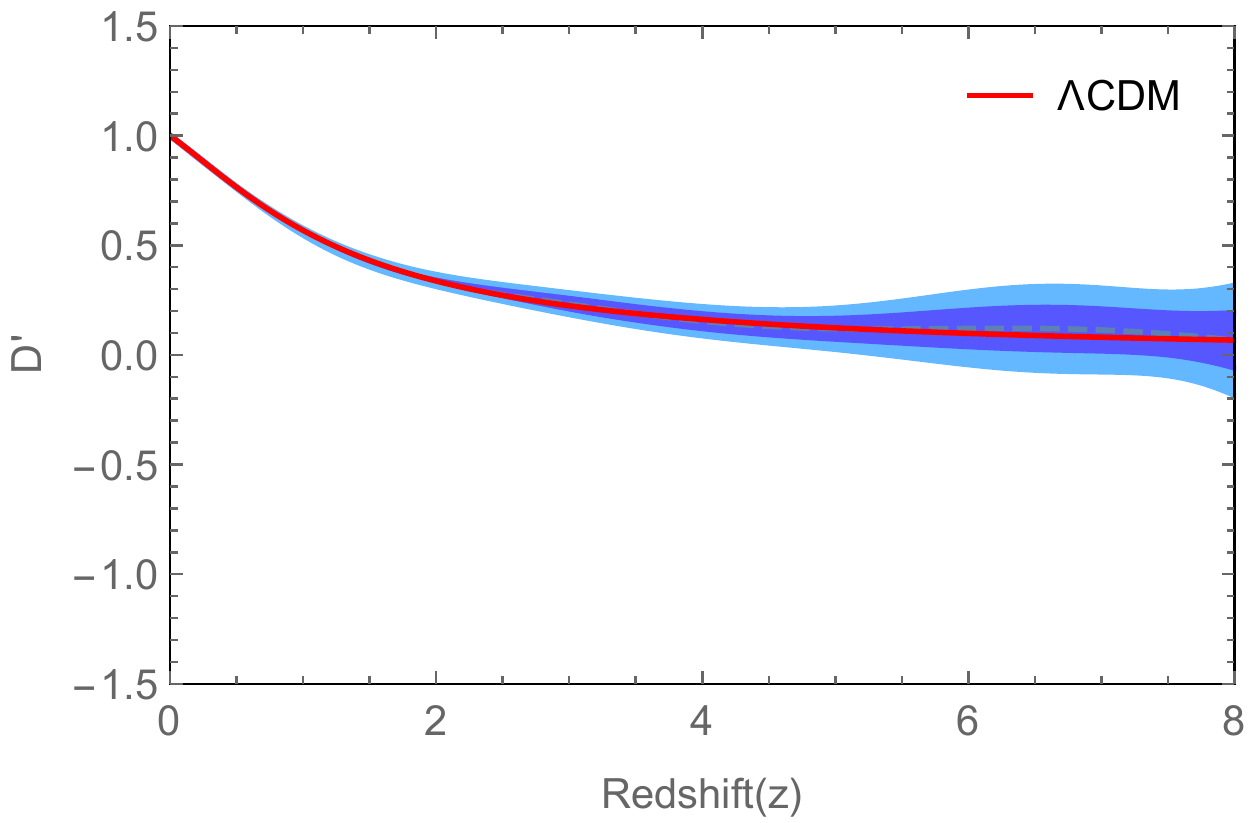}\\
	\includegraphics[width=.31\textwidth]{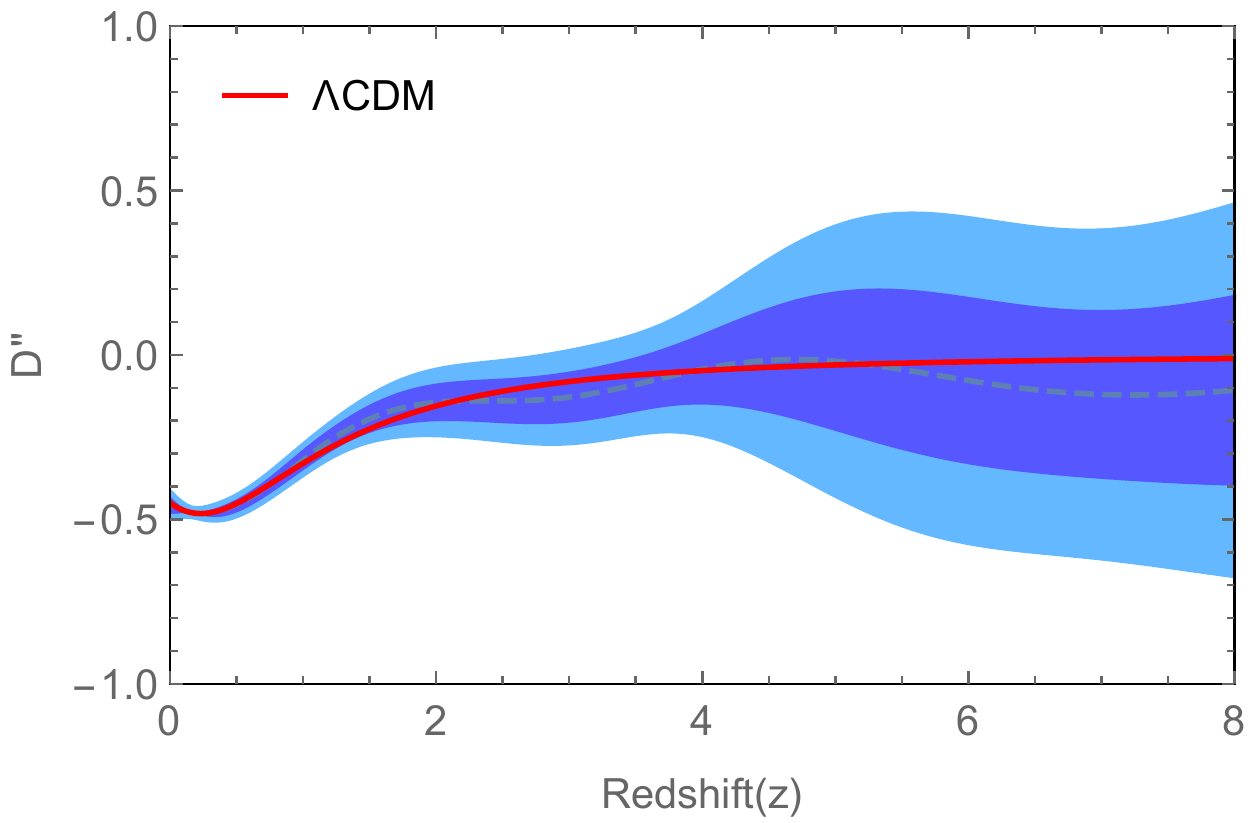}
	\includegraphics[width=.31\textwidth]{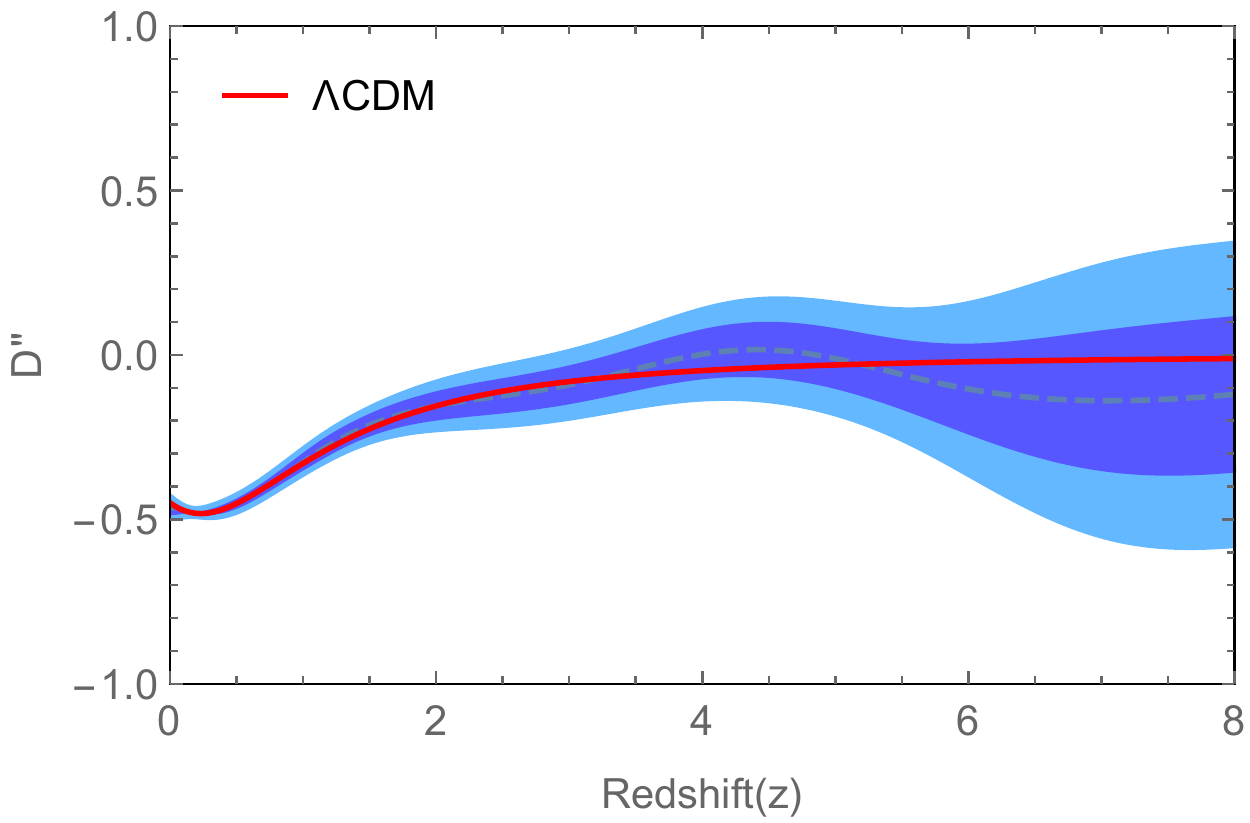}
	\includegraphics[width=.31\textwidth]{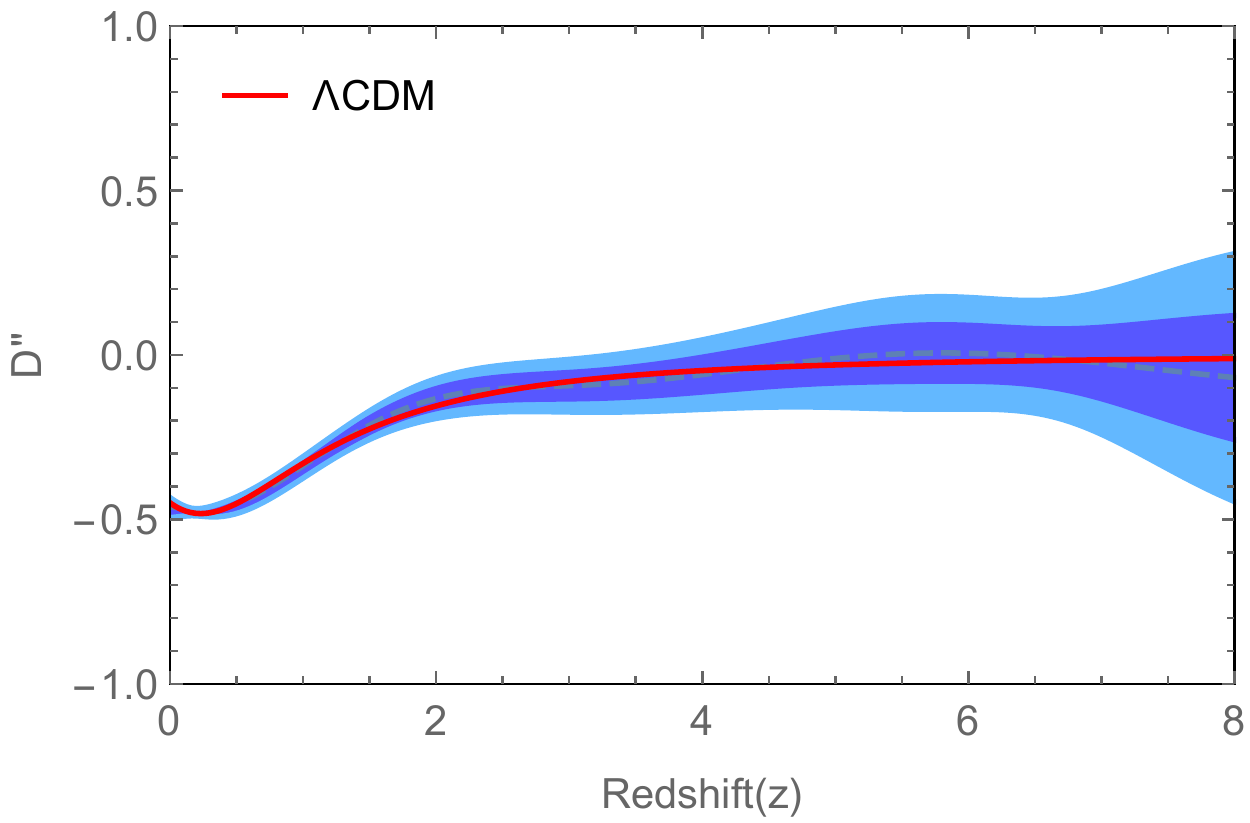}\\
	\includegraphics[width=.31\textwidth]{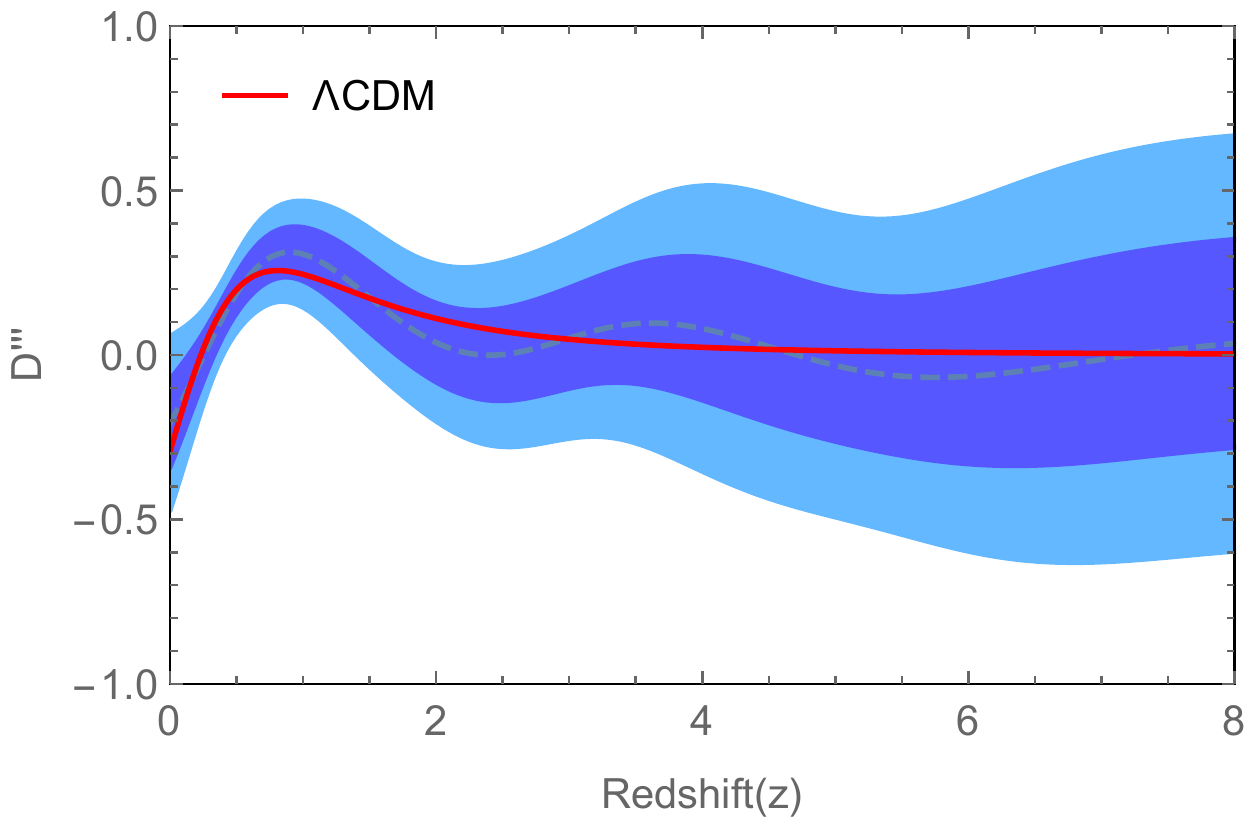}
	\includegraphics[width=.31\textwidth]{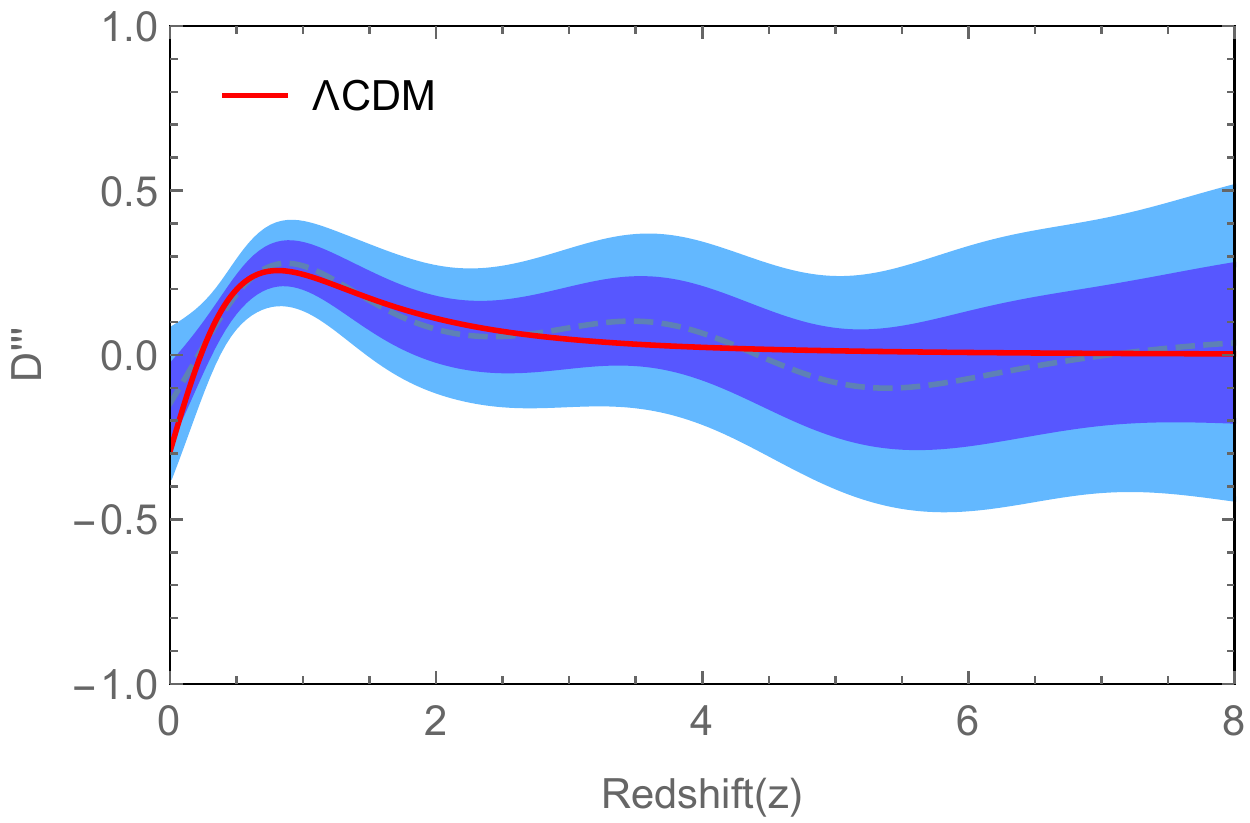}
	\includegraphics[width=.31\textwidth]{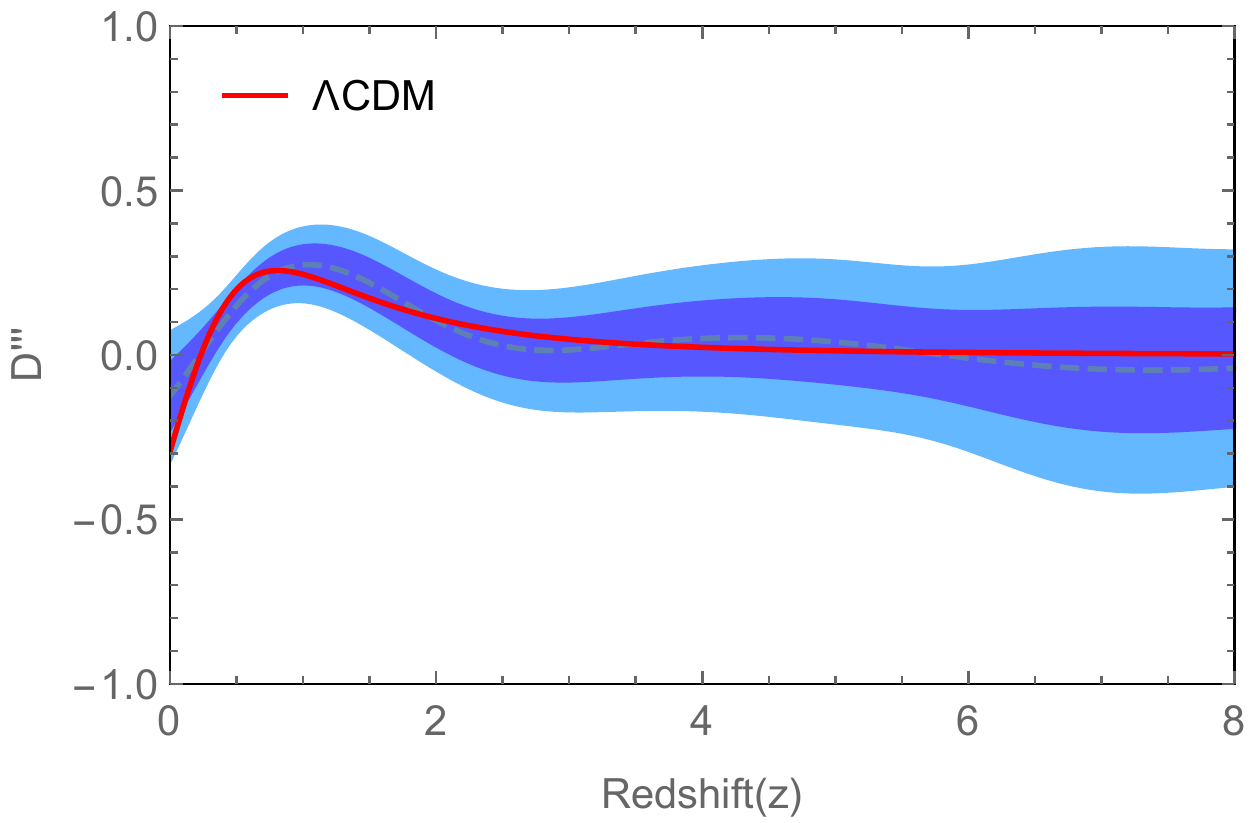}
	\caption{Reconstructions of the derivatives of the distance $D(z)$ using DES$+$LISA (5 years) data.
	}
\label{fig:5yDprime_LISAplusDES}
\end{figure}

\begin{figure}
	\centering
	\includegraphics[width=.31\textwidth]{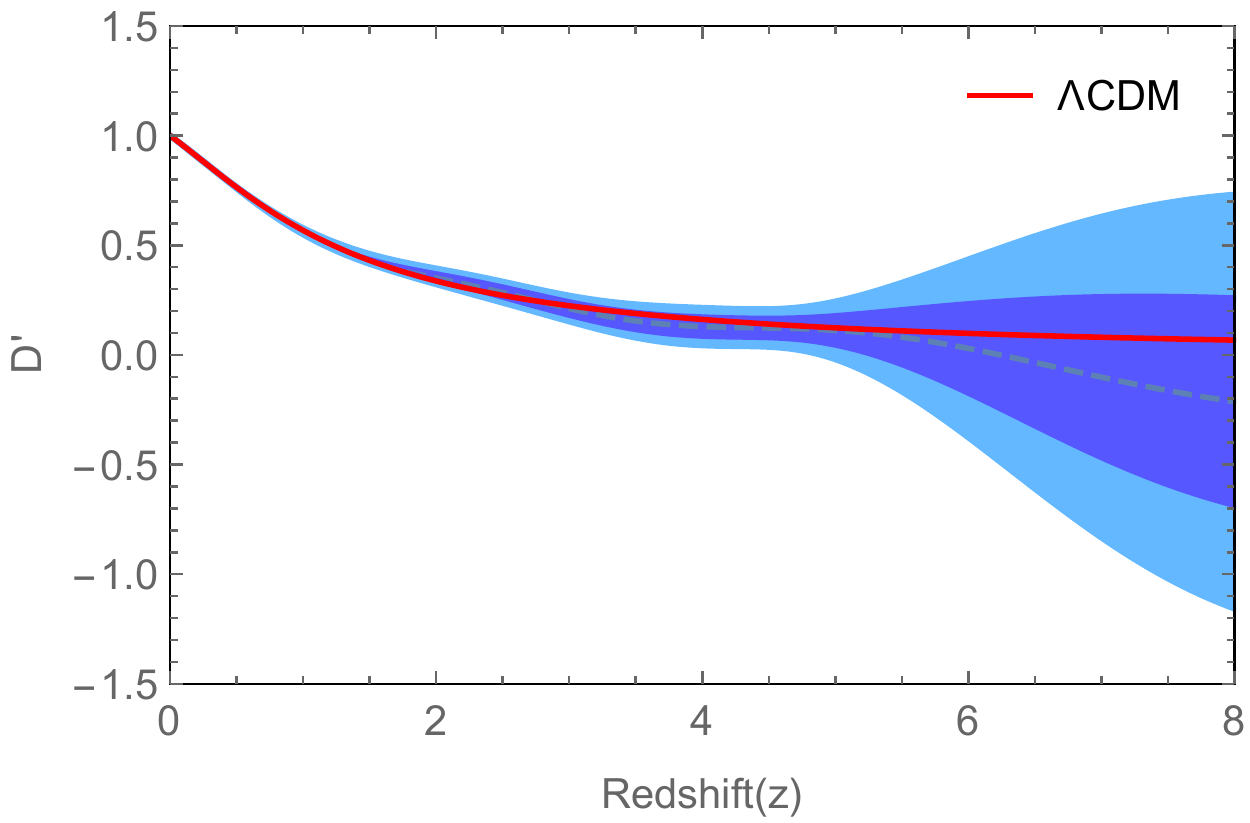}
	\includegraphics[width=.31\textwidth]{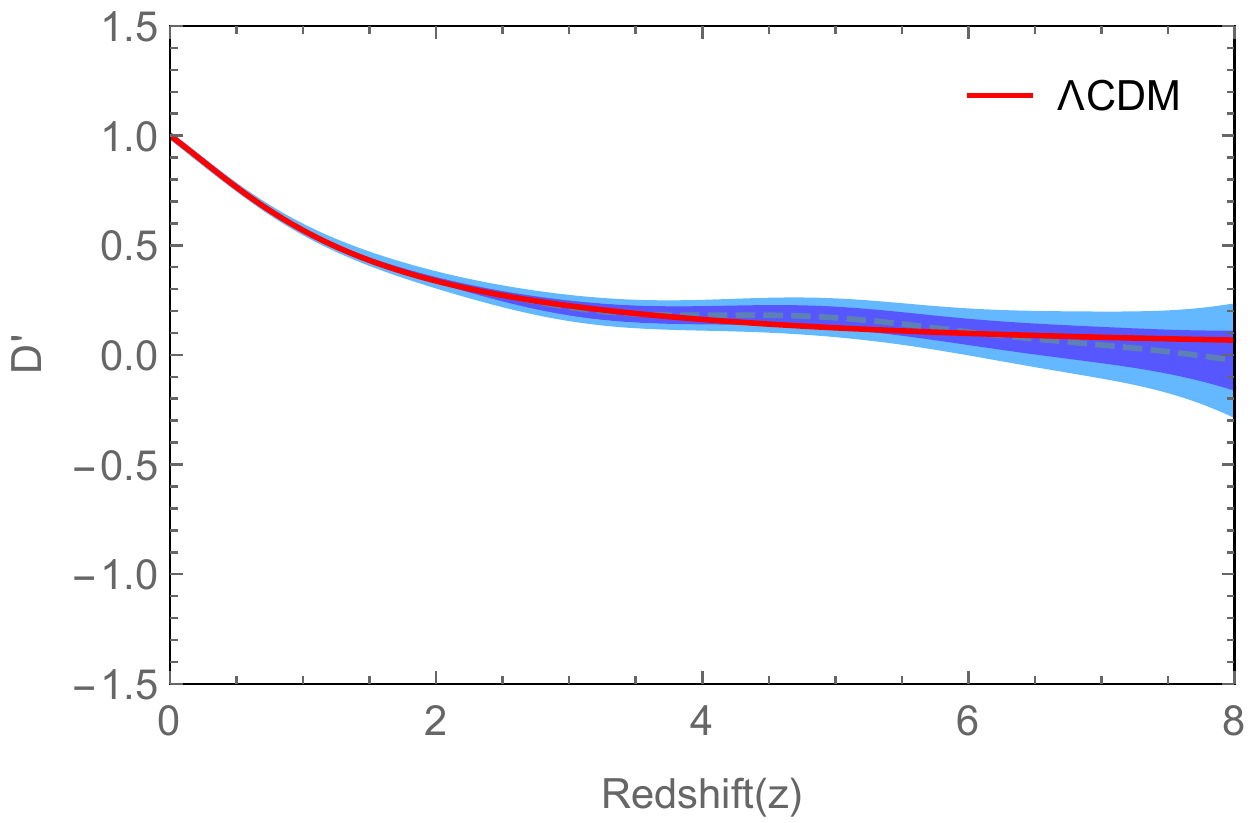}
	\includegraphics[width=.31\textwidth]{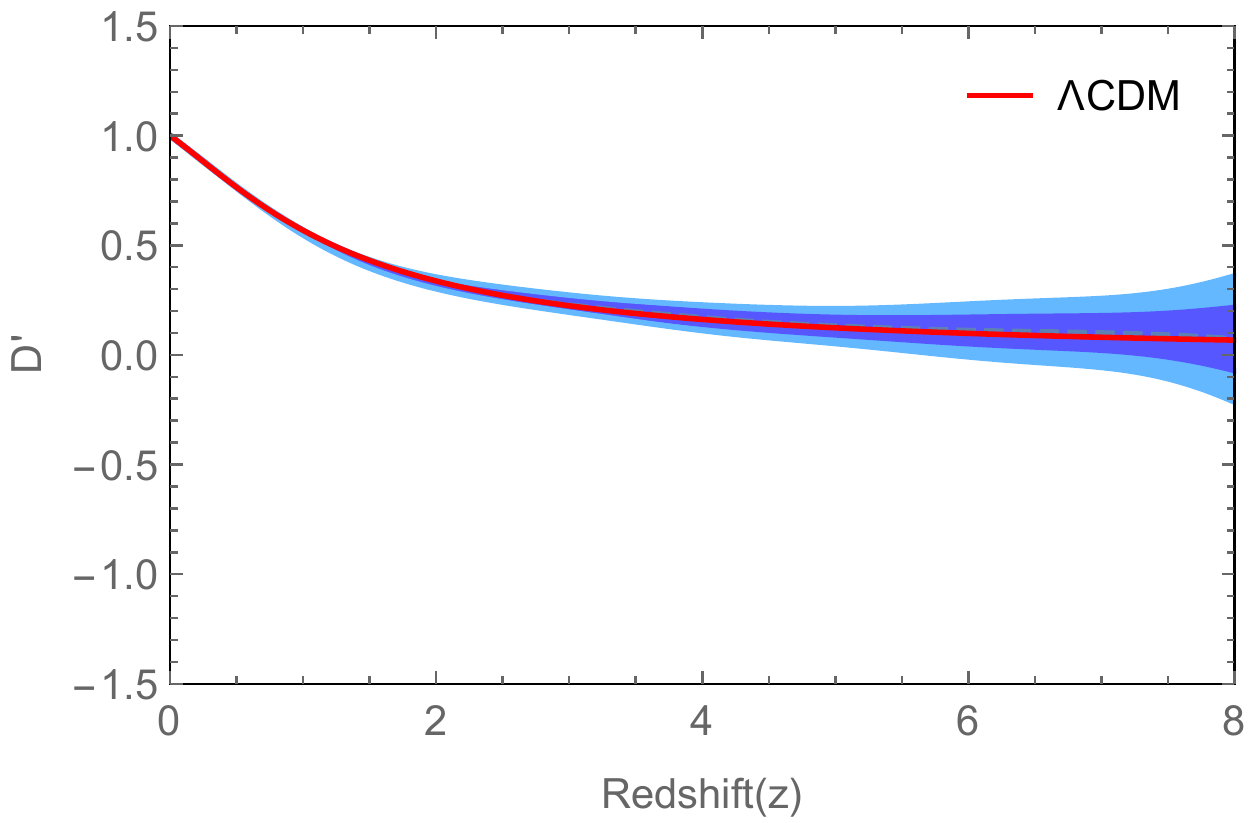}\\
	\includegraphics[width=.31\textwidth]{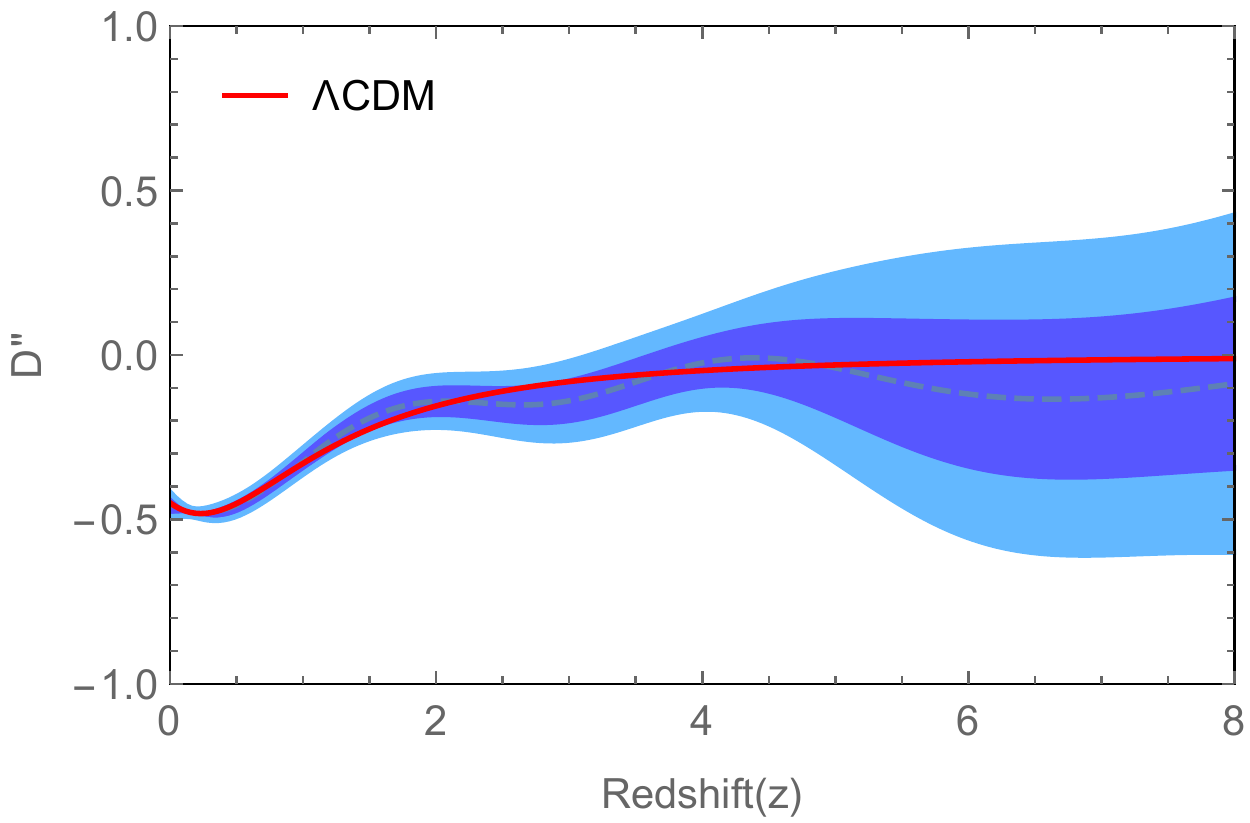}
	\includegraphics[width=.31\textwidth]{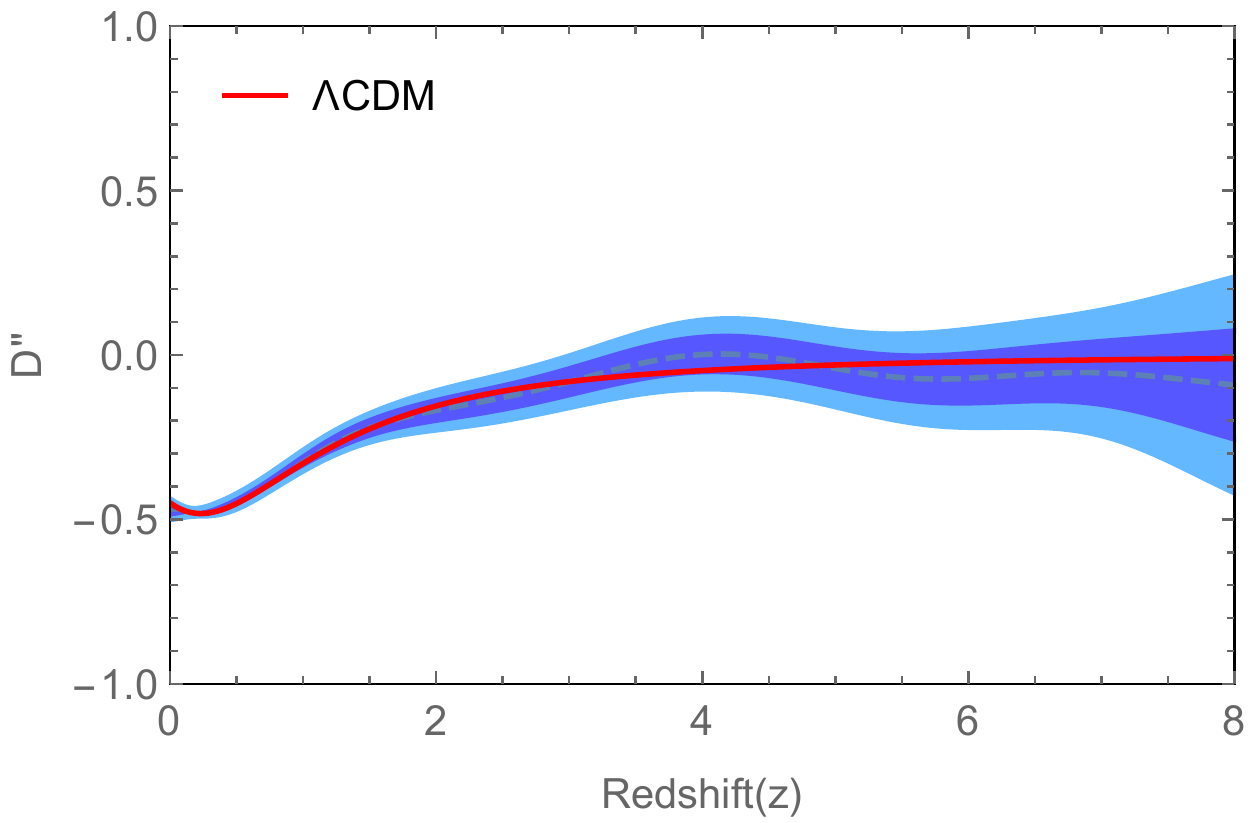}
	\includegraphics[width=.31\textwidth]{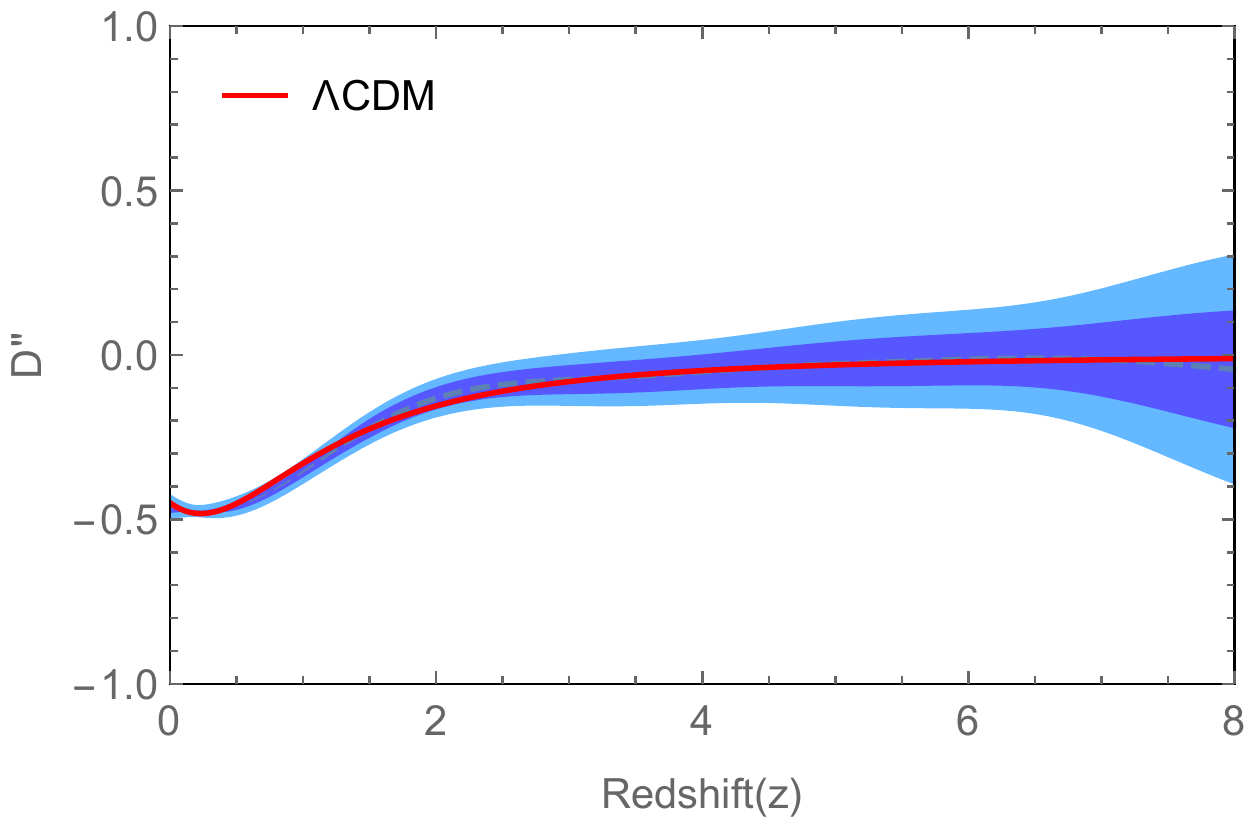}\\
	\includegraphics[width=.31\textwidth]{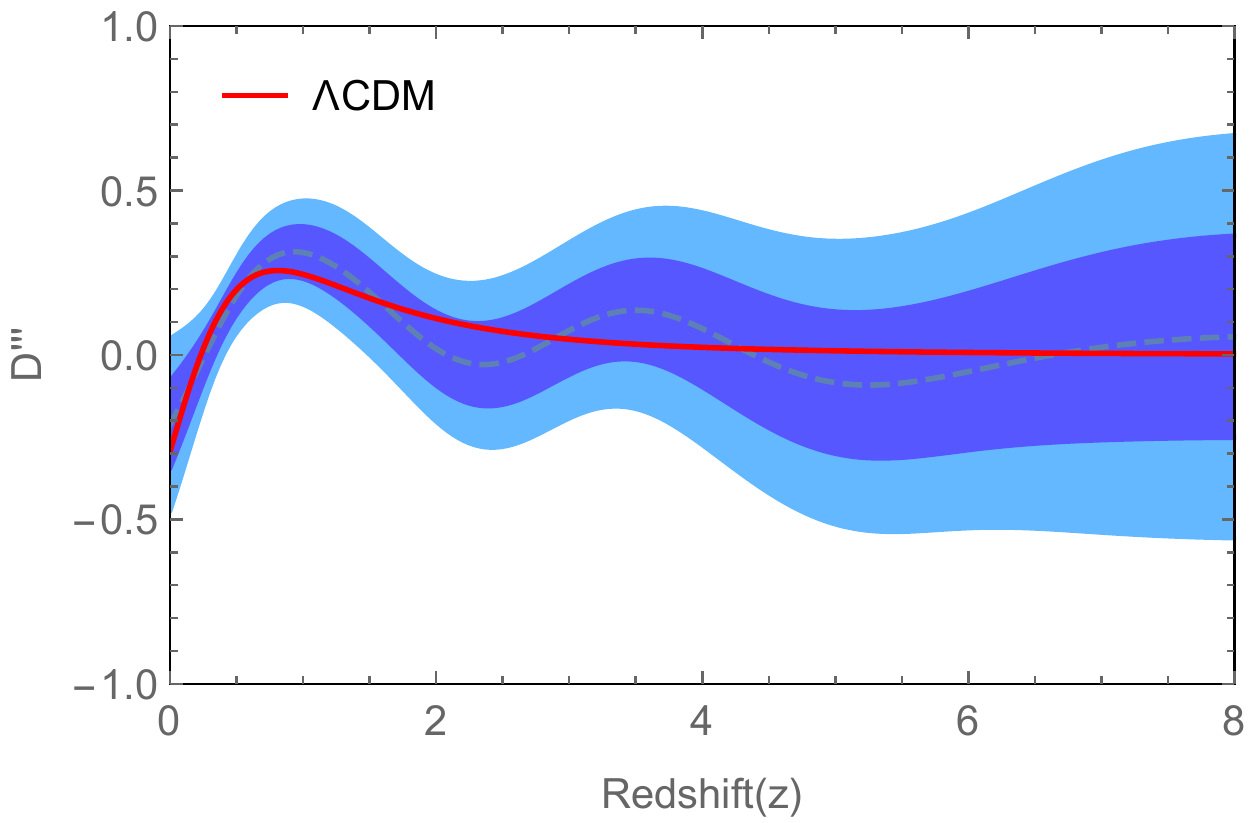}
	\includegraphics[width=.31\textwidth]{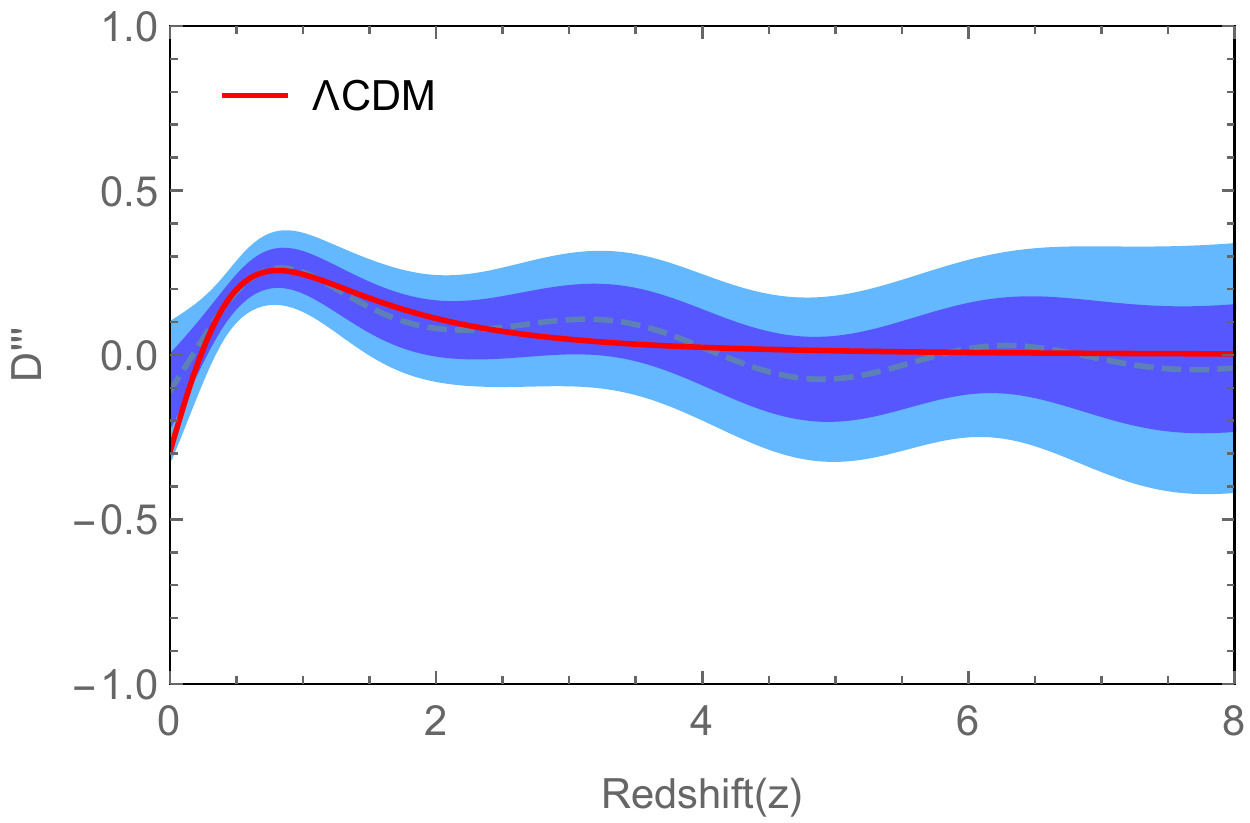}
	\includegraphics[width=.31\textwidth]{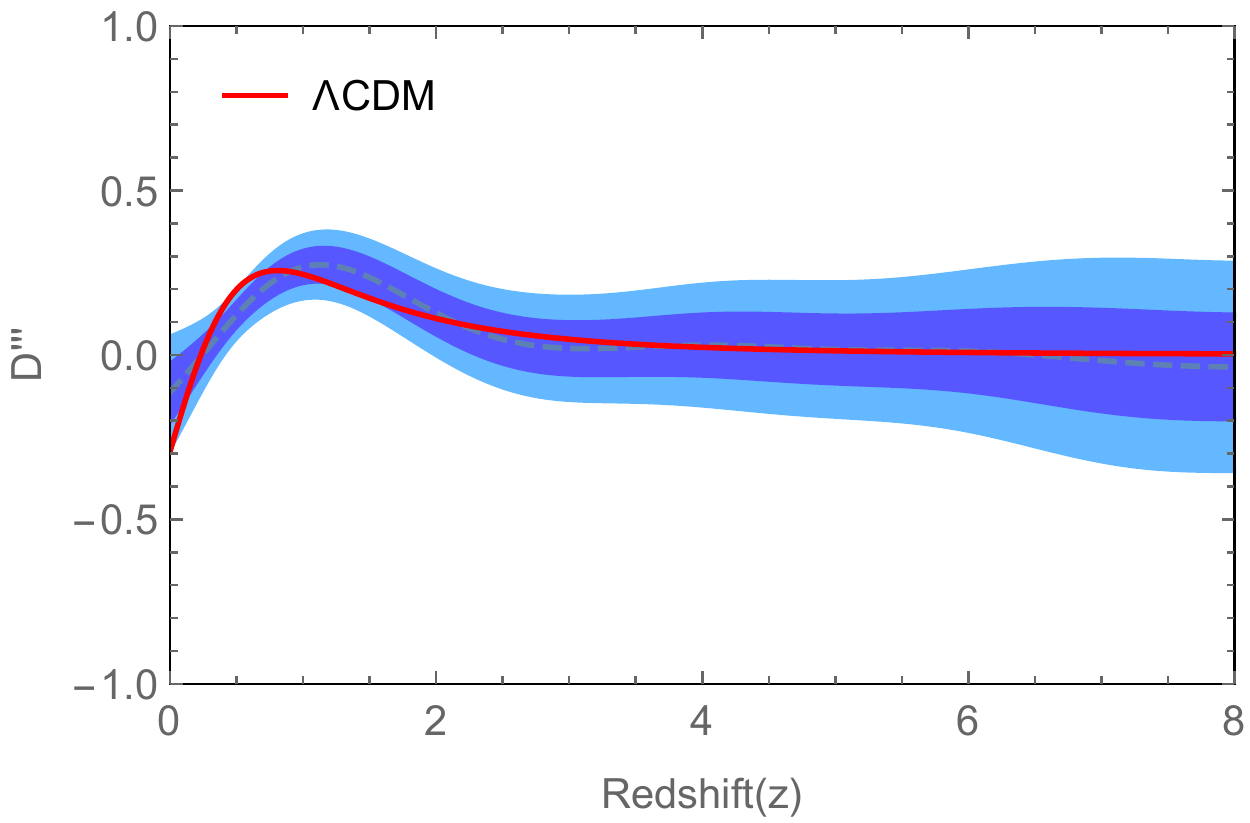}
	\caption{Reconstructions of the derivatives of the distance $D(z)$ using DES$+$LISA (10 years).
	}
\label{fig:10yDprime_LISAplusDES}
\end{figure}

\end{document}